\documentclass[aps,prx,twocolumn,footinbib,floatfix,superscriptaddress]{revtex4-2}
\usepackage{graphicx}
\usepackage{indentfirst}
\usepackage{braket}
\usepackage{float}
\usepackage{amsmath}
\usepackage{physics}
\usepackage{amssymb}
\usepackage{CJK}
\usepackage{esint}
\usepackage{color}
\usepackage[T1]{fontenc}
\usepackage{subfigure}
\usepackage{amsfonts}
\usepackage{footmisc}
\usepackage{scrextend}
\usepackage{multirow}
\usepackage[outdir=./]{epstopdf}
\usepackage{svg}
\usepackage{algorithm}
\usepackage{algpseudocode}

\usepackage[english]{babel}
\usepackage{url}
\usepackage{comment}
\usepackage{array}
\usepackage{subfiles}
\usepackage{tikz}
\usetikzlibrary{shapes,arrows}
\usepackage{mathrsfs}
\usepackage[mathscr]{eucal}
\usepackage[hyperfootnotes=false]{hyperref}
\usepackage{cleveref}
\usepackage{stmaryrd}
\definecolor{darkblue}{rgb}{0,0,0.5}
\hypersetup{
colorlinks=true,
linkcolor=black,
filecolor=blue,
citecolor=darkblue,  
urlcolor=black,
}

\newtheorem{lemma}{Lemma}

\usepackage{trimclip}

\makeatletter
\DeclareRobustCommand{\shortto}{%
  \mathrel{\mathpalette\short@to\relax}%
}

\newcommand{\short@to}[2]{%
  \mkern2mu
  \clipbox{{.5\width} 0 0 0}{$\m@th#1\vphantom{+}{\shortrightarrow}$}%
  }
\makeatother

\newenvironment{proof}[1][Proof]{\noindent\textbf{#1.} }{\ \rule{0.5em}{0.5em}}

\newcommand\argmin{\mathop{\mathrm{argmin}}}

\usepackage{pict2e,picture,graphicx}

\makeatletter
\DeclareRobustCommand{\Arrow}[1][]{%
\check@mathfonts
\if\relax\detokenize{#1}\relax
\settowidth{\dimen@}{$\m@th\rightarrow$}%
\else
\setlength{\dimen@}{#1}%
\fi
\sbox\z@{\usefont{U}{lasy}{m}{n}\symbol{41}}%
\begin{picture}(\dimen@,\ht\z@)
\roundcap
\put(\dimexpr\dimen@-.7\wd\z@,0){\usebox\z@}
\put(0,\fontdimen22\textfont2){\line(1,0){\dimen@}}
\end{picture}%
}
\makeatother
\newcommand{\veryshortrightarrow}{\hspace{.2mm}\scalebox{.8}{\Arrow[.1cm]}\hspace{.2mm}}

\def\be{\begin{equation}}
\def\ee{\end{equation}}
\def\ba{\begin{eqnarray}}
\def\ea{\end{eqnarray}}
\def\bal{\begin{equation}\begin{aligned}}
\def\eal{\end{aligned}\end{equation}}

\def\bp{\begin{pmatrix}}
\def\ep{\end{pmatrix}}

\usepackage{bm}

\newcommand{\calF}{{\cal F}}

\newcommand{\calM}{{\cal M}}
\newcommand{\calN}{{\cal N}}

\newcommand{\calR}{{\cal R}}

\newcommand{\1}{^{(1)}}

\newcommand{\state}[1]{\ketbra{#1}{#1}}

\newcommand{\hw}[1]{{{\textcolor{black}{#1}}}}

\begin{document}
\title{
Fulfilling entanglement's optimal advantage via converting correlation to coherence
}
\author{Haowei Shi}
\thanks{These two authors contributed equally.}
\affiliation{
Ming Hsieh Department of Electrical and Computer Engineering, University of Southern California, Los
Angeles, California 90089, USA
}
\affiliation{
James C. Wyant College of Optical Sciences, University of Arizona, Tucson, Arizona 85721, USA
}
\affiliation{
Department of Electrical and Computer Engineering, University of Arizona, Tucson, Arizona 85721, USA
}

\author{Bingzhi Zhang}
\thanks{These two authors contributed equally.}
\affiliation{
Ming Hsieh Department of Electrical and Computer Engineering, University of Southern California, Los
Angeles, California 90089, USA
}
\affiliation{
Department of Physics, University of Arizona, Tucson, AZ 85721, USA
}
\affiliation{
Department of Electrical and Computer Engineering, University of Arizona, Tucson, Arizona 85721, USA
}

\author{Quntao Zhuang}
\email{qzhuang@usc.edu}
\affiliation{
Ming Hsieh Department of Electrical and Computer Engineering, University of Southern California, Los
Angeles, California 90089, USA
}
\affiliation{
Department of Electrical and Computer Engineering, University of Arizona, Tucson, Arizona 85721, USA
}
\affiliation{
James C. Wyant College of Optical Sciences, University of Arizona, Tucson, Arizona 85721, USA
}

\begin{abstract}
Entanglement boosts performance limits in sensing and communication, and surprisingly the advantage over classical protocols can be even larger in presence of entanglement-breaking noise. However, to maximally fulfill such advantages requires an optimal measurement design, a challenging task as information is encoded in the feeble quantum correlation after entanglement is destroyed by loss and noise. For this reason, the optimal measurement design is still elusive for various entanglement-enhanced protocols long after their debut. We propose a conversion module to capture and transform the quantum correlation to coherent quadrature displacement, which enables the optimal receiver design for a wide range of entanglement-enhanced protocols, including quantum illumination, phase estimation, classical communication, and arbitrary thermal-loss channel pattern classification. 
Via heterodyne and passive linear optics, the conversion module maps the multi-mode quantum detection problem to the semi-classical detection problem of a single-mode noisy coherent state, so that explicit measurements can be constructed to achieve the optimal performance.
Our module provides a paradigm of processing noisy quantum correlations for near-term implementation. 
\end{abstract}

\date{\today}

\maketitle

\tableofcontents

\section{Introduction}
Quantum entanglement not only refreshes our understanding of the world but also brings unprecedented power in sensing~\cite{giovannetti2006,giovannetti2011advances,sidhu2020geometric,lawrie2019quantum,toth2014quantum,pirandola2018advances,degen2017quantum,zhang2021dqs} and communication~\cite{gisin2007quantum,kimble2008quantum,wilde2013quantum,wehner2018quantum}. 
Entanglement is fragile---noise can easily destroy entanglement.
Surprisingly, by evaluating information-theoretical limits, people find that benefits from entanglement survive in various applications such as target detection (quantum illumination, QI)~\cite{tan2008quantum,zhuang2021quantum,zhuang2022ultimate} and classical communication~\cite{Bennett2002}. In fact, the quantum advantage relative to the classical protocols thrives on the large noise in these applications.
However, it is hard to actually design and build systems to fulfill such benefits optimally, as it requires extracting information that is delicately hidden in the remaining quantum correlations from entanglement's destruction. Indeed, till now the experimental demonstrations of these protocols are far from optimal~\cite{hao2021entanglement,zhang2013,assouly2022demonstration} and optimal measurement schemes are either challenging~\cite{zhuang2017optimum} or completely unknown.

\begin{figure*}
    \centering
    \includegraphics[width=0.85\textwidth]{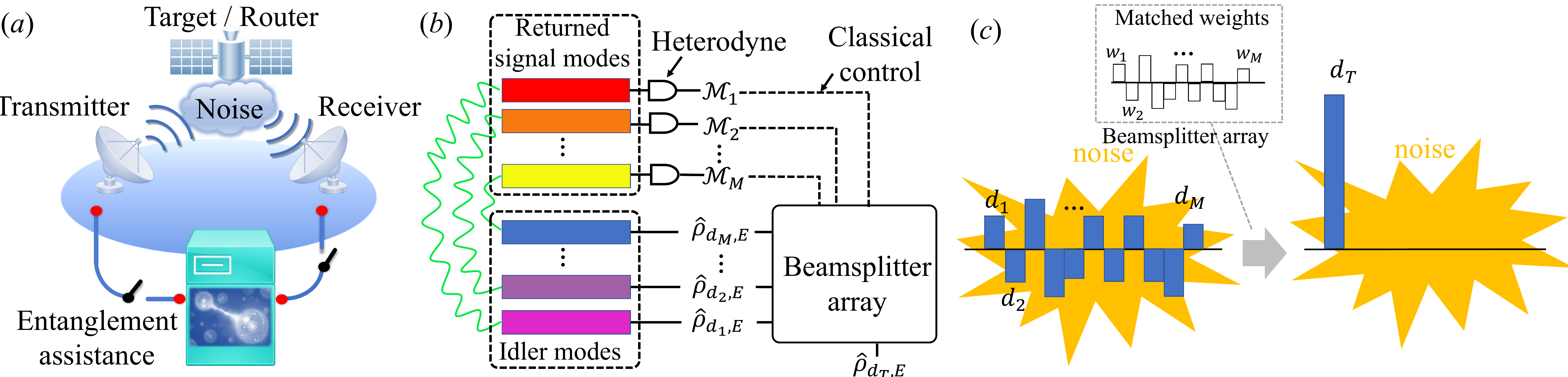}
    \caption{Schematic of (a) an entanglement-assisted/classical (switches on/off) sensing or communication protocol and
    (b) the $\rm C\veryshortrightarrow D$ conversion module that transforms correlation between $2M$ return-idler modes to a single mode of noisy coherent state. (c) Schematic of the beamsplitter mode transform to combine the complex mean fields $d_1,\cdots,d_M$ to a single mean field $d_T$ (indicated by the blue patterns), while the noise (indicated by the orange pattern) is not added due to independence.
    }
    \label{fig:conversion} 
\end{figure*}

We resolve this open problem and fulfill entanglement's optimal advantage in a surprisingly neat fashion---with a conversion module from intermodal correlation to intramodal coherence. 
We prove that such a correlation-to-displacement (`$\rm C\veryshortrightarrow D$') conversion preserves all information of interest and therefore enables the optimal performance in a long list of entanglement-assisted (EA) sensing and communication protocols---QI target detection, phase sensing and classical communication.
Moreover, the conversion module enables exact performance analyses and extends quantum advantages to the non-asymptotic region, unexplored due to the limitations of asymptotic tools~\cite{Audenaert2007,Pirandola2008,nussbaum2011asymptotic,zhuang2017optimum,li2016discriminating}. It also enables a proof of a folklore that a six-decibel advantage in the error exponent exists in an arbitrary thermal-loss channel pattern classification problem. The conversion module reduces the multi-mode quantum measurement problem to the detection of a single-mode noisy coherent state. Combining our conversion module with a standard optimal coherent-state measurement~\cite{dolinar_processing_1973,guha2011structured}, we provide explicit optimal measurement designs, relying only on linear optics and photon detection.  
Our $\rm C\veryshortrightarrow D$ conversion module has broad applications and brings new insights into how quantum correlations can be processed.

\section{Results}
\subsection{Entanglement-assisted protocol} 
In optical sensing and communication, light propagation can often be modeled as an overall quantum phase-shift thermal-loss channel $\Phi_{\kappa, \theta}$~\cite{weedbrook2012gaussian}: upon an input mode described by the annihilation operator $\hat{a}_{S}$, the output mode
\be 
\hat{a}_{R}=e^{i\theta}\sqrt{\kappa}\hat{a}_{S}+\sqrt{1-\kappa}\hat{a}_{B},
\label{input_output_main}
\ee 
where $\kappa$ is the transmissivity, $\theta$ is the phase shift, and $\hat{a}_{B}$ models the thermal background with mean photon number $N_B$ per mode. 

As shown in Fig~\ref{fig:conversion}, in a sensing protocol, $\hat{a}_{S}$ describes the probe signal sent out from the transmitter to interact with a subject, and then unavoidably encounters noise $\hat{a}_{B}$, before finally get detected. For example, in an ideal target detection scenario~\cite{tan2008quantum,zhuang2021quantum}, a present target reflects a $\kappa$ portion of the signals back to the receiver, corresponding to the channel $\Phi_{\kappa, 0}$, assuming a known reflection phase (which is set to zero without loss of generality); while when the target is absent, only noise can be received, and the channel is modeled as $\Phi_{0, 0}$ with zero reflection. In a phase sensing scenario~\cite{Escher_2011} that models bio-sensing, ranging~\cite{zhuang2022ultimate} and gravitational-wave detection~\cite{abbott2016observation}, time of flight in different media leads to an unknown phase shift and therefore we consider the estimation of the parameter $\theta$ of channel $\Phi_{\kappa, \theta}$. 
Similarly, in a phase-shift-keying (PSK)~\cite{shi2020practical} communication protocol, one encodes a classical message $\theta$ to the phase of the signal $e^{i\theta}\hat{a}_{S}$, and then a $\kappa$ portion of the signal is received in mixture with noise. From sender's encoding to the receiver, light propagation is modeled as an overall channel $\Phi_{\kappa, \theta}$. In all the above sensing and communication scenarios, the information of interest is in the transmissivity $\kappa$ and phase $\theta$ of the channel.

In a classical strategy, one directly measures the received mode $\hat{a}_R$ from inputting a signal mode $\hat{a}_S$.
To boost the sensing and communication performance, an EA strategy entangles the initial signal with an ancilla, which is jointly measured with the received signal.
In particular, we consider $M$ signal-idler pairs $\{\hat{a}_{S_m},\hat{a}_{I_m}\}_{m=1}^M$, where each pair is in a two-mode squeezed-vacuum (TMSV) state with mean photon number $N_S$~\cite{weedbrook2012gaussian}, known to be optimal in these applications~\cite{nair2020fundamental,shi2020practical}. 
TMSV is a zero-mean two-mode Gaussian state with the covariance matrix (See Appendix~\ref{sec:GS} for details on Gaussian state formalism)
\begin{equation}
V_{SI} =
\begin{pmatrix}
(2N_S+1) \mathbb{I} & 2\sqrt{N_S(N_S+1)}\mathbb{Z}\\
2\sqrt{N_S(N_S+1)}\mathbb{Z} & (2N_S+1)\mathbb{I}
\end{pmatrix},
\label{eq:CM_SI_main}
\end{equation}
where $\mathbb{Z}$ is the Pauli-Z matrix and $\mathbb{I}$ is $2\times 2$ identity. 

While the signals are sent through the channel $\Phi_{\kappa,\theta}$, the idlers are stored or pre-shared to the receiver side, leading to $M$ return-idler pairs $\{\hat{a}_{R_m},\hat{a}_{I_m}\}_{m=1}^M$. For the input-output relation in Eq.~\eqref{input_output_main}, the covariance matrix of the return and the idler is then
\begin{equation}
V_{RI} = \begin{pmatrix}
(2\kappa N_S+2N_B+1)\mathbb{I} & 2\sqrt{\kappa N_S(N_S+1)}\bm{R}\mathbb{Z}\\
2\sqrt{\kappa N_S(N_S+1)}\mathbb{Z}\bm{R}^T & (2N_S+1) \mathbb{I}
\end{pmatrix},
\label{eq:CM_RI_main}
\end{equation}
where $\bm{R} = \cos\theta \mathbb{I}-i \sin\theta \mathbb{Y}$ and $\mathbb{Y}$ is the Pauli-Y matrix. In the sensing and communication applications of interest, information is encoded in either the transmissivity $\kappa$ or the phase $\theta$ in the return-idler states.
We note that both $\kappa$ and $\theta$ are embedded in the phase-sensitive cross-correlation 
$ 
\expval{\hat{a}_{R_m} \hat{a}_{I_m}}=e^{i \theta} C_p
$
with the amplitude $C_p\equiv \sqrt{\kappa    N_S \left(  N_S+1\right)}$. At the same time, when $N_S$ is small, the amplitude of the correlation $C_p \propto \sqrt{N_S}$ in an EA protocol. As a comparison, for a classical sensing protocol with a coherent-state probe of the same brightness $N_S$ and strong local oscillator as the reference, the correlation $\propto N_S$ and is therefore much smaller when $N_S$ is small. In this regard, the crucial part of a measurement design to fulfil entanglement's benefit is to detect the phase-sensitive cross correlation.

\subsection{Correlation-to-displacement conversion} 
Now we design a module to convert phase-sensitive cross-correlation between $M$ signal-idler pairs to the complex displacement amplitude of a single-mode coherent state. Through this module, the quantum problem of receiver design is mapped to a semi-classical problem of coherent state processing.

Given the return-idler pairs, $\{\hat{a}_{R_m},\hat{a}_{I_m}\}_{m=1}^M$,
as shown in Fig.~\ref{fig:conversion}(b), the module first performs individual heterodyne measurement on each $\hat a_{R_m}$, producing the complex measurement result $\calM_m $, which obeys a circularly-symmetric complex Gaussian distribution with variance 
\be 
v_\calM\equiv (N_B + \kappa N_S+1)/2.
\ee 
Here $N_B/(1-\kappa)$ is the mean photon number of noise $\hat{a}_{B}$~\footnote{Similar to previous works~\cite{tan2008quantum,zhuang2017optimum,zhuang2021quantum}, we have chosen the $1-\kappa$ factor so that passive signature are not present. }.
Conditioned on the output ${\calM_m}$~\cite{genoni2016conditional,weedbrook2012gaussian}, each $\hat a_{I_m}$ is in a displaced thermal state $\hat{\rho}_{d_m, E}$, with mean 
$d_m=(C_p/2v_\calM) e^{ i \theta }{\calM_m^*}$ and thermal noise photon number
\be 
E={N_S \left(N_B+1-\kappa \right)}/{ 2v_\calM}\le N_S
.
\ee 
A succinct summary of the derivation can be found in Appendix~\ref{method}, while the full details can be found in Appendix~\ref{sec:details_C2D}. To gain an intuitive understanding on the order of magnitudes of the displacement and noise, we consider the noisy ($N_B\gg1$) and low brightness region ($N_S\ll1$) in which EA sensing and communication thrive. In this case, the typical amplitude of the measurement outcome $|\calM_m|\sim \sqrt{v_\calM} \sim \sqrt{N_B}$, the order of magnitude of the mean $|d_m|\sim C_p/\sqrt{v_\calM}\sim \sqrt{\kappa N_S/N_B}$. Therefore, $|d_m|^2\sim \kappa N_S/N_B$ is typically much smaller than the noise $E\sim N_S$---the non-zero mean of the idlers are embedded in significant noise, as indicated in Fig.~\ref{fig:conversion}(c). 

\begin{figure}[t]
    \centering
    \includegraphics[width=0.45\textwidth]{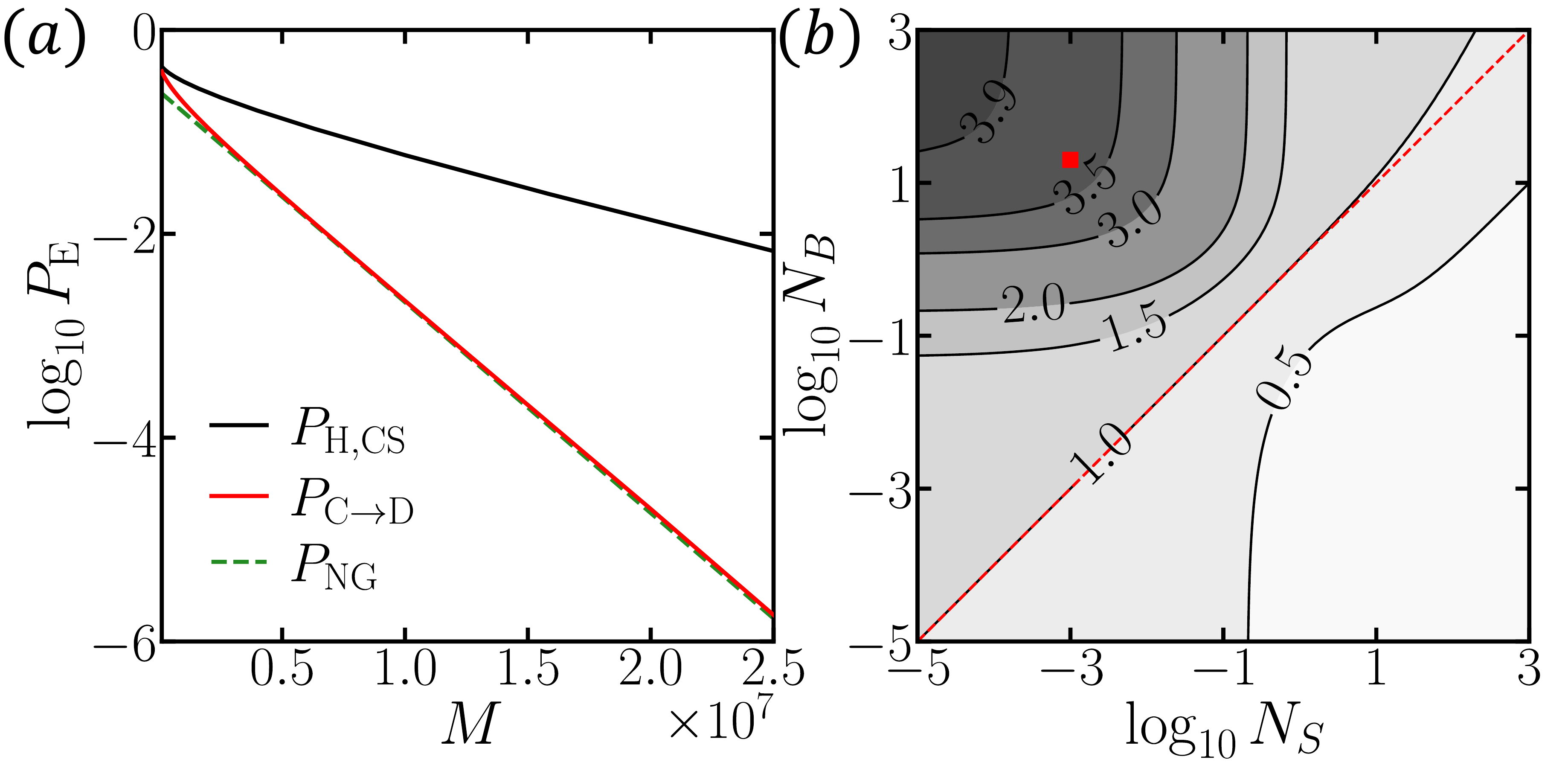}
    \caption{QI in the asymptotic region. (a) Error probability versus number of copies $M$ with $N_S=0.001, N_B = 20$. (b) Error exponent ratio $r_{{\rm C}\veryshortrightarrow{\rm D}}/r_{\rm CS}$ versus $N_S, N_B$. Red dashed line indicates $N_S=N_B$.  The red dot indicates the parameters chosen in (a). In both cases $\kappa = 0.01$. 
    }
    \label{fig:illumination_exponent}
\end{figure}

To extract the information with sufficient signal-to-noise ratio, as indicated in Fig.~\ref{fig:conversion}(c), we perform a beamsplitter array with proper weights to constructive combine the mean field of the idler modes, which are known from the measurement results; while the noise are not accumulating due to the independence between noise among different idler modes. 
With the weights of the beamsplitter array $w_m= d_m^*/|d_T|$ chosen for each input idler mode, all $M$ idler modes have the mean field combined into a single mode $\sum_{m=1}^M w_m \hat a_{I_m}$ in state $\hat{\rho}_{d_T,E}$, with mean 
$
d_T=|d_T|e^{i\theta}
$
and thermal noise $E$. 
Here the amplitude square 
$
|d_T|^2=\sum_{m=1}^M |d_m|^2
$
satisfies the $\chi^2$ distribution of $2M$ degrees of freedom 
$ 
P_{\kappa}^{(M)}(x)\propto \left({x}/{\xi}\right)^{M-1}e^{-x/(2\xi)},
$
with mean $2M\xi$ and variance $4M\xi^2$, where 
$ 
\xi\equiv C_p^2/4v_\calM\,.
$
As the mean accumulates with number of modes $M$, while the noise stays constant, the signal is therefore much stronger than noise when the number of modes are large, as indicated in Fig.~\ref{fig:conversion}(c).

A few points are worth addressing before the performance analyses. First, the experimental realization of the beamsplitter array depends on the specific protocol. For time-domain modes, it can be realized by a single beamsplitter that adjusts its ratio to combine a stored mode with each incoming mode. Similar feed-forward with phase and amplitude modulation has been realized in previous experiments realizing the Dolinar receiver for coherent-state discrimination~\cite{cook2007}. The device level realization of such a conversion module are subject to future study, for example see recent works along this direction~\cite{angeletti2023,reichert2023}. For frequency modes, it can be potentially realized by an integrated four-wave mixing process~\cite{otterstrom2021,li2016,mcguinness2010}. Second, although heterodyne detection on TMSV has been conceptually utilized in the security proof of quantum key distribution~\cite{grosshans2002,garcia2006,navascues2006}, the adaptive manipulation of the conditional quantum state, for the sensing and communication purpose, in such a noisy environment has never been considered.

\subsection{Performance limits}
We now analyze the performance limit of the ${\rm C}\veryshortrightarrow{\rm D}$ conversion module in various sensing and communication applications, while we defer the semi-classical coherent-state processing to later part. 
We will use the label `${\rm C}\veryshortrightarrow{\rm D}$' for quantities involving the correlation-to-displacement conversion module. 

\subsubsection{Quantum illumination}
QI for target detection considers the discrimination between two channels $\Phi_{0,0}$ and $\Phi_{\kappa,0}$. In this case, the conversion module produces two displaced thermal state $\hat{\rho}_{0,N_S}$ (target absent) and $\hat{\rho}_{\sqrt{x},E}$ (target present),
where $x\sim P_{\kappa}^{(M)}(\cdot)$ obeys the $\chi^2$ distribution. This leads to the error probability performance limit
\be 
P_{{\rm C}\veryshortrightarrow{\rm D}}=\int dx P_{\kappa}^{(M)}(x) P_{\rm H}(\hat{\rho}_{0,N_S},\hat{\rho}_{\sqrt{x},E}),
\label{eq:average_pe}
\ee 
where $P_{\rm H}(\hat{\rho},\hat{\sigma})$ is the Helstrom limit of error probability in state discrimination between $\hat{\rho}$ and $\hat{\sigma}$ with equal prior probability~\cite{Helstrom1969, Helstrom_1967,Helstrom_1976} (See Appendix~\ref{app:QCB}).

\begin{figure}[t]
    \centering
    \includegraphics[width=0.45\textwidth]{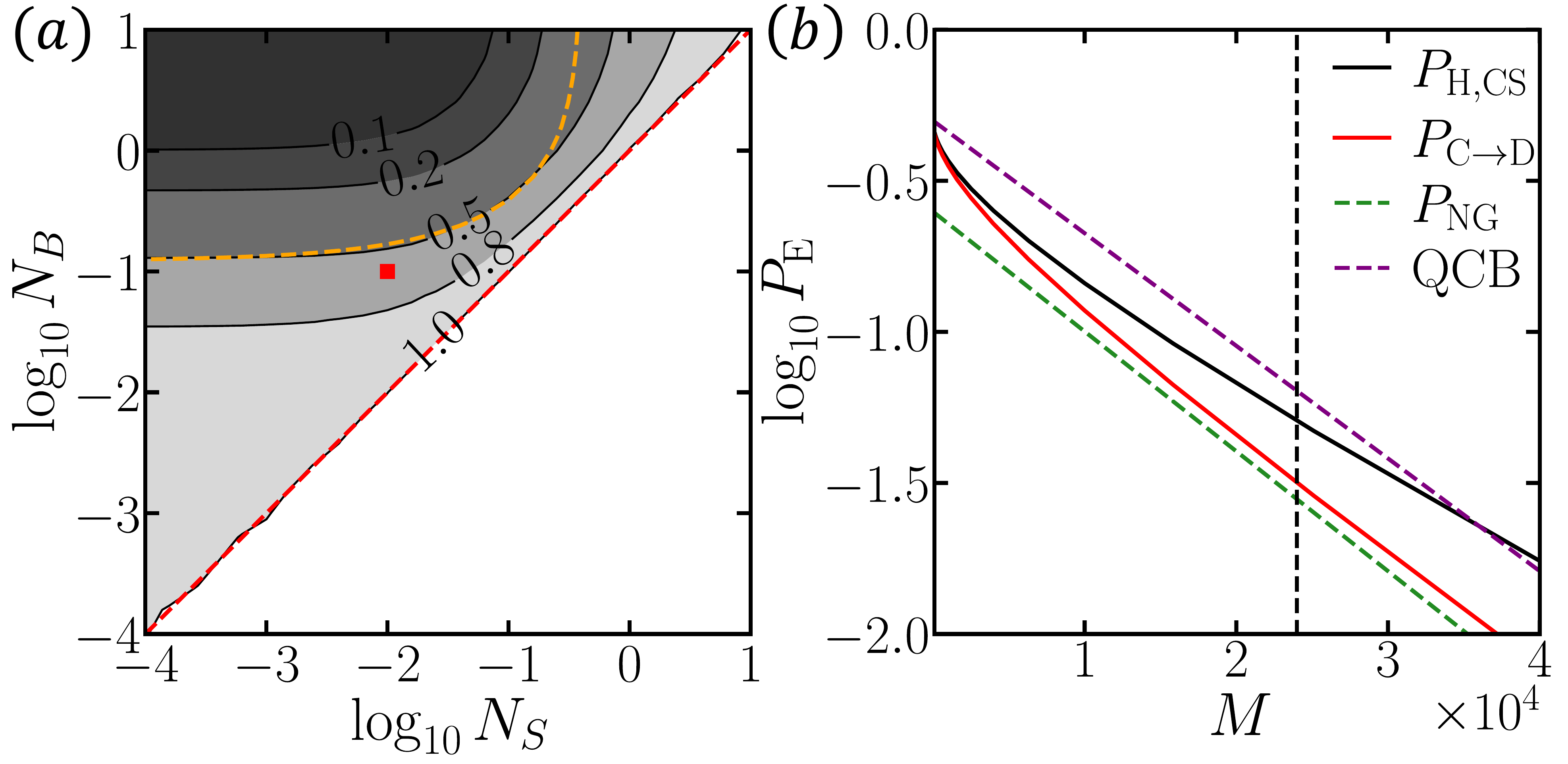}
    \caption{QI in the non-asymptotic region.  (a) Error probability ratio $P_{{\rm C}\veryshortrightarrow {\rm D}}/P_{\rm H,  CS}$ versus $N_S, N_B$, $M$ is chosen to fix $P_{\rm H,  CS}=0.05$. The red dot indicates the choice of $N_S, N_B$ in (b). The red dashed line indicates $N_S= N_B$ and the orange line indicates QCB equaling $P_{\rm H,  CS}$. (b) Error probability versus the number of copies $M$ with $N_S=0.01, N_B=0.1$. The vertical line indicates the $M$ values such that $P_{\rm H,  CS}=0.05$. In both cases, $\kappa=0.01$.}
    \label{fig:illumination_peratio}
\end{figure}


To compare with the ultimate performance, we evaluate the Nair-Gu (NG) lower bound $P_{\rm NG}$~\cite{nair2020fundamental} on the error probability applicable to any source of illumination (See Appendix~\ref{sec:review}). To benchmark for entanglement advantage, we also consider the Helstrom limit of the optimal classical scheme based on coherent states~\cite{tan2008quantum}, $P_{\rm H,CS}=P_{\rm H}(\hat{\rho}_{0,N_B}, \hat{\rho}_{\sqrt{\kappa M N_S},N_B})$.

We begin with the asymptotic limit of low brightness $N_S\ll1$ and low reflectivity $\kappa\ll 1$ considered in prior works~\cite{tan2008quantum,zhuang2017optimum}, where $M$ is large to guarantee a decent signal-to-noise ratio. At this limit, we can approximate $\hat{\rho}_{\sqrt{x},E}$ as a coherent state and $\hat{\rho}_{0,N_S}$ as vacuum; Therefore, the Helstrom limit $P_{\rm H}(\hat{\rho}_{0,N_S},\hat{\rho}_{\sqrt{x},E})\simeq e^{-x}/4$ and Eq.~\eqref{eq:average_pe} leads to
\be 
P_{{\rm C}\veryshortrightarrow{\rm D}}\simeq \frac{1}{4}(1+2\xi)^{-M}\simeq \frac{1}{4}\exp{-M r_{{\rm C}\veryshortrightarrow{\rm D}}},
\label{eq:illumination_ph_approx}
\ee 
which saturates the lower bound $P_{\rm NG}$ (See Appendix~\ref{sec:review}) with the error exponent
$
r_{{\rm C}\veryshortrightarrow{\rm D}}=2\xi.
$ 
In fact, one can easily check that the optimality holds as long as $N_S\ll1$ and $\kappa\ll 1+N_B$. We verify this optimality in Fig.~\ref{fig:illumination_exponent}(a), where a close agreement is seen between $P_{{\rm C}\veryshortrightarrow{\rm D}}$ (red) and $P_{\rm NG}$ (green). At the same time, huge advantage over the classical limit $P_{\rm H,CS}$ (black) can be observed.

Now we examine the error exponent more closely.
In general, when $\xi \ll1$ (e.g., due to $\kappa\ll1$) we can obtain a lower bound on the error exponent,
$
r_{{\rm C}\veryshortrightarrow{\rm D}}\ge
2\xi (\sqrt{N_S+1}-\sqrt{N_S})^2,
$
while the coherent state error exponent
$ 
r_{\rm CS}=\kappa N_S(\sqrt{N_B+1}-\sqrt{N_B})^2
$~\cite{tan2008quantum} (See Appendix~\ref{sec:review}).
We can show that the entanglement advantage exists as long as the signal brightness is smaller than the noise brightness, i.e., $N_S\le N_B$, as also confirmed in Fig.~\ref{fig:illumination_exponent}(b) via plotting $r_{{\rm C}\veryshortrightarrow{\rm D}} / r_{\rm CS}$.


Finally, we emphasize that Eq.~\eqref{eq:average_pe} provides an efficiently calculable and achievable error-probability lower bound for QI, in contrast to the asymptotically tight quantum Chernoff (upper) bound (QCB)~\cite{Audenaert2007,Pirandola2008,tan2008quantum}. As a consequence, Eq.~\eqref{eq:average_pe} allows the exploration of QI's advantage in the non-asymptotic region. As shown in Fig.~\ref{fig:illumination_peratio}(a), when the classical Helstrom limit is fixed at $P_{\rm H,CS}=0.05$, the ratio $P_{{\rm C}\veryshortrightarrow{\rm D}}/P_{H,\rm CS}\le 1$ when $N_S\le N_B$ (above the red dashed line). However, QCB can only show quantum advantage in a strictly smaller region above the orange dashed curve. We also pick a set of parameters $N_S, N_B$ to explicitly plot the error probability versus the number of copies $M$ in Fig.~\ref{fig:illumination_peratio}(b)---when $M$ is small, QCB fails to identify quantum advantage, while our Eq.~\eqref{eq:average_pe} shows advantage.

\begin{figure}[t]
    \centering
    \includegraphics[width=0.45\textwidth]{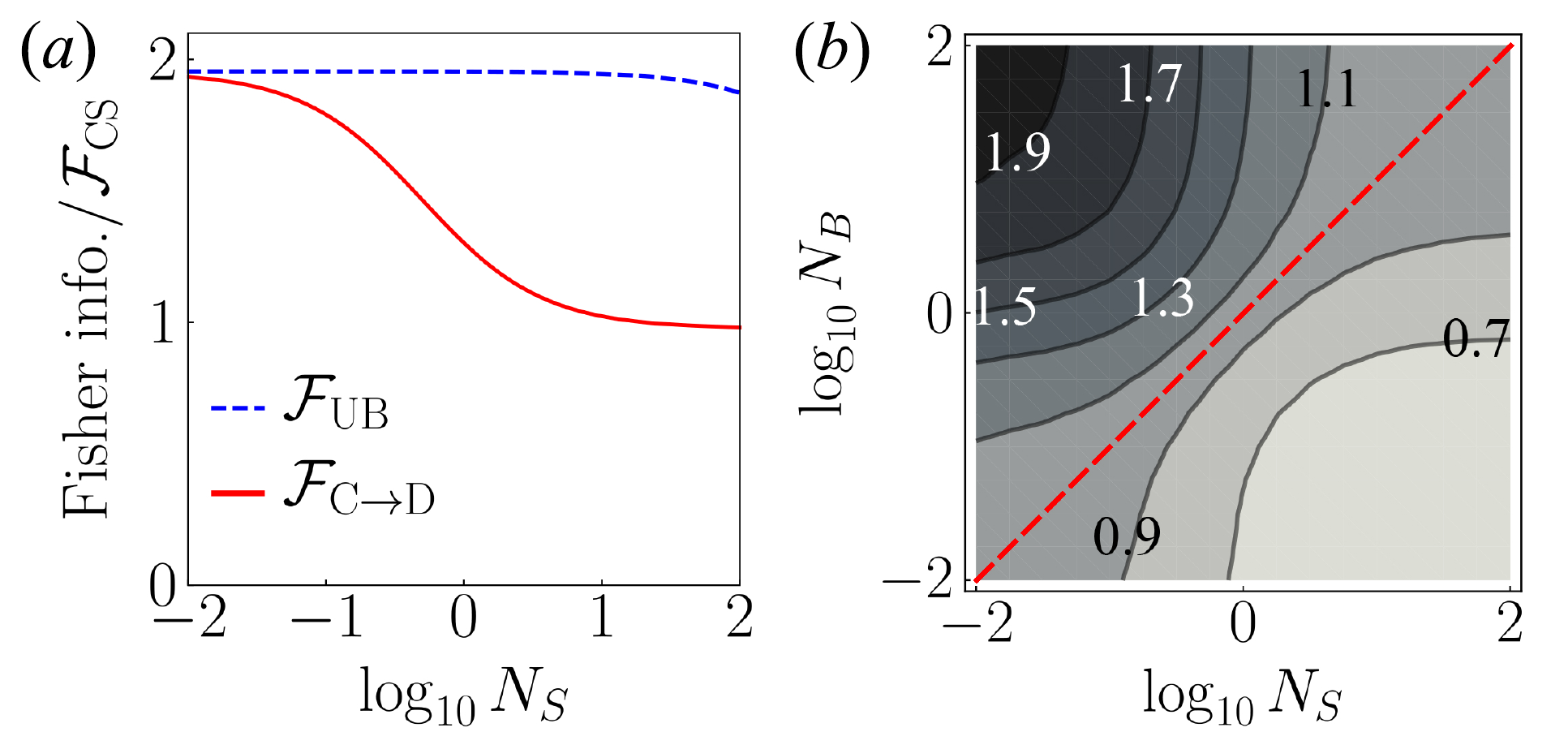}
    \caption{Quantum Fisher information normalized by the coherent-state limit $\calF_{\rm CS}$. (a) The conversion module (red) in comparison the EA upper bound (blue dashed), with $N_B=20$. (b) Contour of $\calF_{\rm C\veryshortrightarrow D}/\calF_{\rm CS}$ versus $N_S,N_B$. The red dashed diagonal line indicates $N_S=N_B/(1-\kappa)$. In both cases, $\kappa=0.01$.}
    \label{fig:qfi_limits_NbNs}
\end{figure}

\subsubsection{Quantum phase estimation}
For quantum phase estimation, we aim at estimating the phase shift $\theta$ of the channel $\Phi_{\kappa, \theta}$ described in Eq.~\eqref{input_output_main}. After the conversion module, we obtain a displaced thermal state $\hat{\rho}_{e^{i\theta}\sqrt{x},E}$, where $x\sim P_{\kappa}^{(M)}(\cdot)$. The variance of unbiased estimators has an asymptotically tight lower bound $\delta\theta^2= 1/\calF$, with $\calF$ being the quantum Fisher information (QFI). The overall QFI enabled by the conversion module can be obtained as (See Appendix~\ref{sec:Fisher})
\be 
\calF_{{\rm C}\veryshortrightarrow{\rm D}}=\frac{4 M\kappa  N_S \left(N_S+1\right)}{1+N_B+N_S \left(2 N_B+2-\kappa \right)}.
\label{eq:qfi_c2d}
\ee
We compare $\calF_{{\rm C}\veryshortrightarrow{\rm D}}$ with the ultimate upper bound of Fisher information $\calF_{\rm UB}$ derived in Ref.~\cite{gagatsos2017bounding} (See Appendix~\ref{sec:review}. 
At low brightness $N_S\ll1$, we find \hw{$\calF_{{\rm C}\veryshortrightarrow{\rm D}}\simeq \left[1-\kappa/(1+N_B)\right]\calF_{\rm UB}$} and if reflectivity $\kappa$ is low, we can further show that $\calF_{{\rm C}\veryshortrightarrow{\rm D}}\simeq \calF_{\rm UB}$ achieves the optimum.

\begin{figure}[t]
    \centering
    \includegraphics[width=0.45\textwidth]{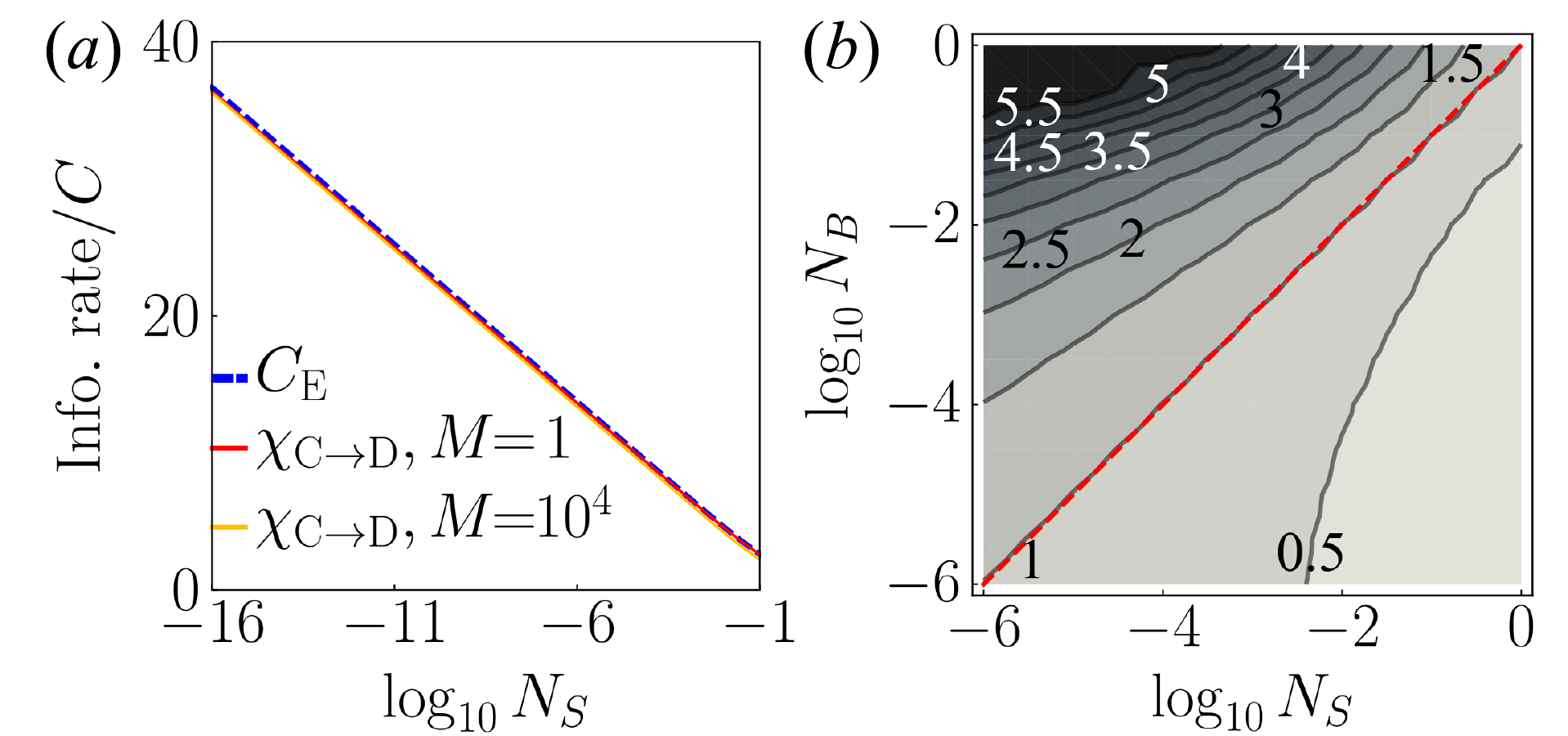}
    \caption{Information rates normalized by the unassisted capacity $C$.  (a) $\chi_{{\rm C}\veryshortrightarrow {\rm D}}$ under repetition coding order $M=1$ (red) and $M=10^4$ (yellow), compared with the ultimate EA capacity (blue dashed). Here the three lines almost overlap. $N_B=100$. (b) Contour of $\chi_{{\rm C}\veryshortrightarrow {\rm D}}/C$ versus $N_S,N_B$. The red dashed diagonal line indicates $N_S=N_B$. In both cases, $\kappa=0.01$. }
    \label{fig:EACOMM_c2d_limits}
\end{figure}

Now we show the quantum advantage by comparing with the classical limit using coherent-state sources
$
\calF_{\rm CS}={4M\kappa N_S}/{(1+2N_B)}\,.
$ 
When the noise $N_B\gg1$, the optimal performance of $\calF_{{\rm C}\veryshortrightarrow{\rm D}}$ has a factor of two advantage over $\calF_{\rm CS}$, as verified in Fig.~\ref{fig:qfi_limits_NbNs} (a).
Comparing $\calF_{{\rm C}\veryshortrightarrow{\rm D}}$ and $\calF_{\rm CS}$, we can show quantum advantage ($\calF_{{\rm C}\veryshortrightarrow{\rm D}}\ge \calF_{\rm CS}$) as long as 
$ 
N_S\le N_B/(1-\kappa),
$
as verified in Fig.~\ref{fig:qfi_limits_NbNs}(b).

\subsubsection{Entanglement assisted communication}
Consider PSK with repetitions, where $M$ signal modes are modulated by the same phase $\theta$ uniformly randomly chosen from $[0,2\pi)$. Below we present the results, while details can be found in Appendix~\ref{sec:holevo}.
The output of the conversion module is in state $\hat{\rho}_{\sqrt{x}e^{i\theta},E}$, where $x\sim P_{\kappa}^{(M)}(\cdot)$. At this point, the achievable information rate per symbol from the output state is
\be 
\chi_{{\rm C}\veryshortrightarrow {\rm D}}=\frac{1}{M}\int dx P_{\kappa}^{(M)}(x) \chi\left(\{\hat{\rho}_{\sqrt{x}e^{i\theta},E}\}\right),
\label{Holevo_module}
\ee 
where $\chi\left(\{\hat{\rho}_{\sqrt{x}e^{i\theta},E}\}\right)$ is the Holevo information~\cite{holevo1973bounds,wilde2013quantum} of the corresponding state ensemble.
Due to the uniform phase modulation and the Gaussian nature of each state $\hat{\rho}_{\sqrt{x}e^{i\theta},E}$, Eq.~\eqref{Holevo_module} can be efficiently evaluated.

To compare with the ultimate performance, we consider the EA classical capacity $C_E$~\cite{Bennett2002}. At the same time, to understand the advantage over the classical schemes, we also compare with the classical capacity without assistance $C$~\cite{giovannetti2014ultimate}. In Fig.~\ref{fig:EACOMM_c2d_limits}(a), we see that $\chi_{{\rm C}\veryshortrightarrow {\rm D}}$ approaches $C_{\rm E}$, therefore verifying the optimality of the conversion module to fulfill the EA advantage in communication.
Indeed, at the limit of low brightness, $N_S\to 0$, we obtain 
$
\chi_{{\rm C}\veryshortrightarrow {\rm D}}
   \sim \kappa  N_S \ln(1/N_S)/{(N_B+1)}\,,
$
which achieves the scaling of the EA capacity. The same optimality result also holds for the binary PSK modulation. 

To fully understand the advantage enabled by the conversion module, we plot the ratio $\chi_{{\rm C}\veryshortrightarrow {\rm D}}/C$ in Fig.~\ref{fig:EACOMM_c2d_limits}(b) versus $N_S,N_B$. When $N_S\ll1, N_B\gg1$, we indeed see the huge advantage; Moreover, we find that entanglement's benefit can be identified in a large region when $N_S\le N_B$, similar to the previous cases. Note that the $N_S\ll1$ region is relevant to covert communication, where the brightness is low to avoid the revelation of communication attempts~\cite{bash2015quantum,shi2020entanglement}.

\begin{figure}[t]
    \centering
    \includegraphics[width=0.45\textwidth]{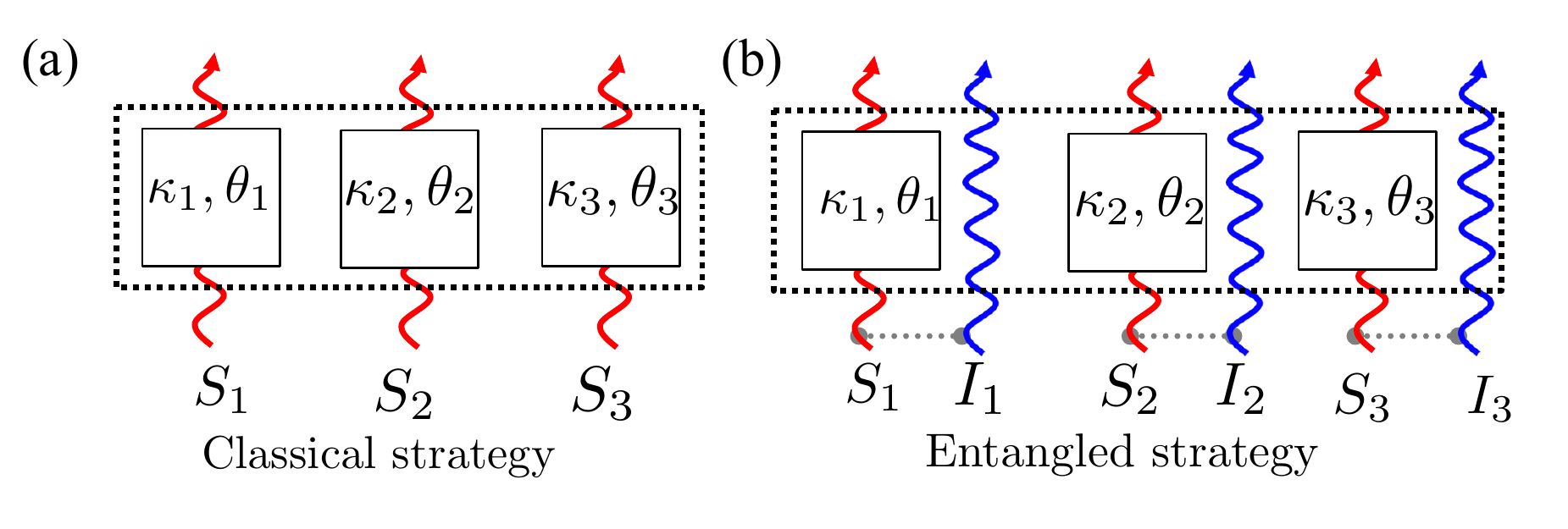}
    \caption{Schematic of quantum sensing over a composite channel in a quantum channel pattern classification problem. (a) Classical strategy. (b) Entangled strategy.}
    \label{fig:scheme_pattern}
\end{figure}



\begin{figure*}
    \centering
    \includegraphics[width=0.75\textwidth]{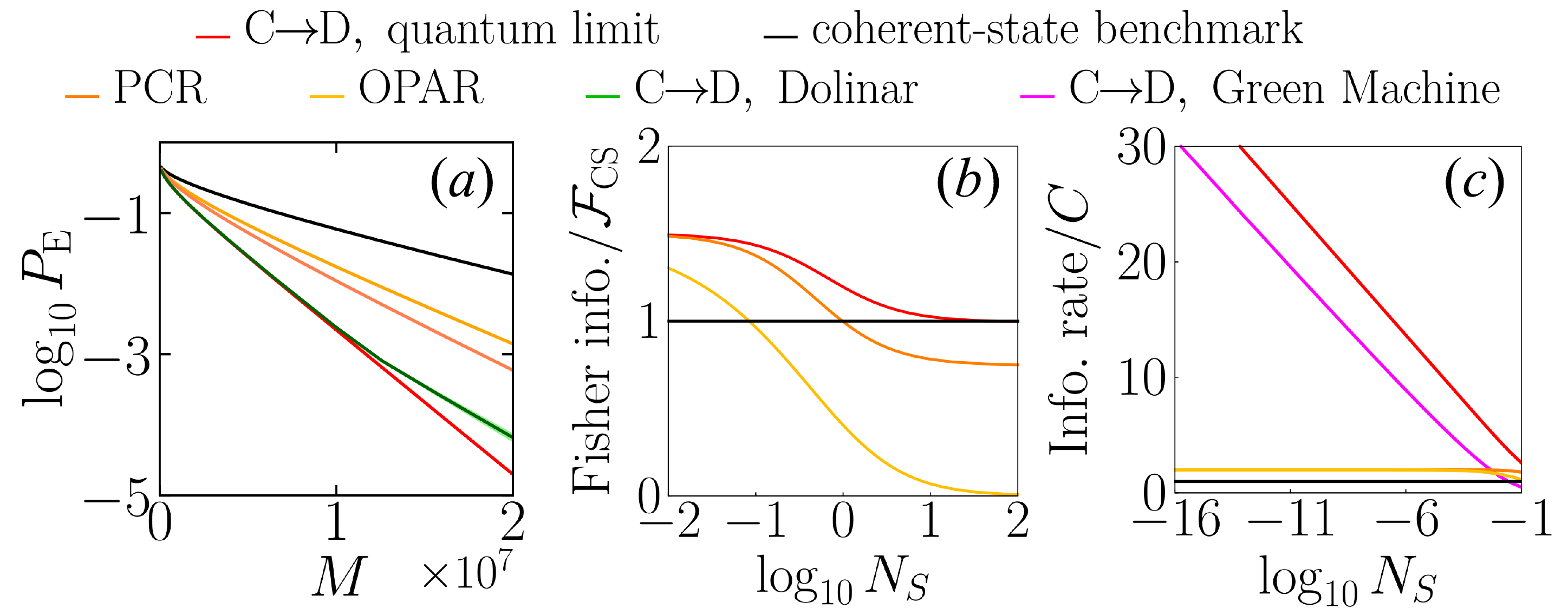}
    \caption{ (a) QI error probability. The limit (red) of the $\rm C\veryshortrightarrow D$ conversion module and its practical performance achieved by Dolinar receiver (green) are shown in comparison with PCR(orange), OPAR(yellow), and the classical homodyne receiver using coherent state (black). The light green area shows the numerical precision. $N_S=0.001$, $N_B=20$ and $\kappa=0.01$.
    (b) Fisher information per probe for phase estimation, normalized by $\calF_{\rm CS}$. The QFI of $\rm C\veryshortrightarrow D$ module, which is achieved by homodyne measurement (red), in comparison with performance of PCR (orange), OPAR (yellow), and the classical coherent-state limit (black). $N_B=1, \kappa=0.98$.
    (c) The information rate per symbol, normalized by the unassisted capacity $C$. The quantum limit $\chi_{{\rm C}\veryshortrightarrow{\rm D}}$ (red) and its practical performance achieved by Green machine (magenta) are shown in comparison with the performances of PCR (orange), OPAR (yellow), and the unassisted capacity (black). 
    $N_B=100,\kappa=0.01$. }
    \label{fig:receivers}
\end{figure*}

\subsubsection{Quantum channel Pattern classification}
So far, we have considered the sensing of a single phase-shift thermal-loss channel $\Phi_{\kappa, \theta}$. In general, complex sensing problems often involve composite channels with $M$ different sub-channels,
$
\Phi_{\bm \kappa, \bm \theta}=\otimes_{m=1}^M \Phi_{\kappa_m, \theta_m},
$
where the vector notation $\bm \kappa=\{\kappa_m\}_{m=1}^M$, $\bm \theta=\{\theta_m\}_{m=1}^M$ and we have assumed that the noise background $N_B$ is identical across all sub-channels. Fig.~\ref{fig:scheme_pattern} shows an example of quantum sensing over composite channels with $M=3$ sub-channels. Previous works on quantum channel position-finding~\cite{zhuang2020entanglement}, barcode recognition~\cite{banchi2020quantum}, quantum ranging~\cite{zhuang2020entanglement,zhuang2022ultimate}, and absorption spectrum recognition~\cite{shi2020entanglement} can all be considered as special cases of this composite channel.

As indicated in Fig.~\ref{fig:scheme_pattern}, in a classical strategy, one in general sends in coherent states or their mixtures to probe the composite channel; While in an entanglement-enhanced scheme, one sends in the signal of the TMSV state for probing and store the idler for entanglement-assistance. For hypothesis testing between general composite channels, we are able to prove a universal error exponent advantage from entanglement.

{\it Theorem 1.---}
{ In the high noise $N_B\gg1$ and low signal brightness $N_S\ll1$ limit, entanglement from two-mode squeezed vacuum enables a factor of four (six decibel) error-exponent advantage over classical sources of coherent states for the discrimination between multiple general} composite thermal-loss channels.

The proof directly utilizes the ${\rm C}\veryshortrightarrow{\rm D}$ conversion module, and is achievable with the module plus optimal discrimination between multiple coherent states (See Appendix~\ref{general_pattern} for the full proof). This result immediately implies that the conversion module is also optimal in the quantum channel position finding problem~\cite{zhuang2020entanglement}.

\subsection{Completing the measurement design} 
With the conversion module, the detection of cross-correlation in EA scenarios is reduced to the detection of single-mode semi-classical coherent states, where measurement designs have been extensively explored theoretically and experimentally~\cite{dolinar_processing_1973,tsujino2011quantum,chen2012optical,becerra2013experimental,becerra2015photon,ferdinand2017multi,burenkov2018quantum,izumi2020experimental,guha2011structured}. Below, we present some examples based on only linear optics and photon detection to complete the optimal measurement design. 
We will also benchmark with practical schemes based on optical parametric amplifier receivers (OPAR) or phase conjugation receivers (PCR)~\cite{Guha2009,shi2020practical,hao2021entanglement} (See Appendix~\ref{app:OPA_PCR} for a review of OPAR and PCR).

For QI target detection, as we explained earlier, the conversion module outputs coherent states with low noise at the $N_S\ll 1$ limit; therefore, the well-known Dolinar receiver~\cite{dolinar_processing_1973} based on linear optics and photon counting saturates the Helstrom limit and completes the optimum measurement design (See Appendix~\ref{app:receivers_summary}). Dolinar receiver utilizes an adaptive control of displacement operations and photon counting. We perform numerical simulations to obtain Dolinar receiver's performance in presence of noise. In Fig.~\ref{fig:receivers}(a), we evaluate the performance of the Dolinar receiver combined with a ${\rm C}\veryshortrightarrow{\rm D}$ conversion module (green), which indeed achieves the optimal error probability (red). Some discrepancy can be found when $M$ is too large, due to the small noise $E\simeq N_S$ being significant at low error probability. As expected, OPAR (yellow) and PCR (orange) give worse performance, although still better than the coherent-state homodyne scheme (black)~\cite{Guha2009}.

For phase estimation, ${\rm C}\veryshortrightarrow{\rm D}$ conversion produces a weakly thermal coherent state, whose phase is being estimated. In this regard, a simple homodyne detection on the ${\rm C}\veryshortrightarrow{\rm D}$ conversion output achieves the QFI in Eq.~\eqref{eq:qfi_c2d} and completes the measurement design (See Appendix~\ref{sec:Fisher}). The Fisher information can be obtained from the measurement statistics. Although OPAR and PCR are also asymptotically optimal~\cite{shi2020practical}, we note that the ${\rm C}\veryshortrightarrow{\rm D}$ scheme has larger Fisher information in the non-asymptotic region, especially when $\kappa$ is close to unity and $N_B$ is small, as can be verified in Fig.~\ref{fig:receivers}(b).

For EA communication, after the conversion module, the rest of the measurement design problem reduces to achieving the Holevo information among an ensemble of noisy coherent state~\cite{guha2011structured,wilde2012explicit}. To enable a near-term measurement design, we consider binary PSK combined with the Hadamard code and Green machine~\cite{guha2011structured, guha2020infinite} (See Appendix~\ref{sec:holevo}). The performance is shown in Fig.~\ref{fig:receivers}(c) (magenta), which achieves the optimal $\ln (1/N_S)$ scaling of $\chi_{\rm C\veryshortrightarrow D}$, while only relying on linear optics and photon counting. Note that the constant factor off here is not due to the conversion module, which achieves the optimal as shown in Fig.~\ref{fig:EACOMM_c2d_limits}(a); instead, the Green machine has room for improvement.


\section{Discussions}
Now we address feasibility of experimental realizations. In terms of microwave QI, a recent experiment has eventually realized a $20\%$ advantage in the error exponent, utilizing the sub-optimal OPAR~\cite{assouly2022demonstration}. The $\rm C\veryshortrightarrow D$ module is practical in that it does not require the return signals and the stored idlers to interact---one heterodyne detects the returned signals and perform (potentially adaptive) photon counting on all idler modes conditioned on measurement results, leading to tremendous simplification in experimental realizations~\cite{assouly2022demonstration}. A device-level experimental design of the actual receiver system based on the conversion module is an important future direction, as recently explored in Refs.~\cite{angeletti2023,reichert2023} since the appearance of the pre-print version of this paper on arxiv~\cite{shi2022fulfilling}. Lastly, we point out that although we have assumed a Gaussian channel model with no phase noise, the conversion module can also operate in the presence of non-Gaussian phase noise to enable entanglement's advantage~\cite{zhuang2021quantum-enabled,chen2022}.


\begin{acknowledgements}
This project is supported by the NSF CAREER Award CCF-2142882 and Office of Naval Research Grant No. N00014-23-1-2296. QZ also acknowledges support from Defense Advanced Research Projects Agency (DARPA) under Young Faculty Award (YFA) Grant No. N660012014029, National Science Foundation (NSF) Engineering Research Center for Quantum Networks Grant No. 1941583, and support from Raytheon Missiles and Defense (Mark J. Meisner) during the final completion of the paper.
\end{acknowledgements}

\appendix 
\section{Derivation on correlation-to-displacement conversion.}
\label{method}

By performing the heterodyne measurement on the returned mode, it is mapped to a coherent state with mean $\overline{\bm x}_\Pi\equiv (q_\Pi, p_\Pi)$ and identity covariance matrix, $V_{\Pi}=\mathbb{I}$. 
In general, for input state $\hat \rho_{RI}$ of a single pair of return and idler, conditioned on the heterodyne measurement result $(q_\Pi, p_\Pi)$, we produce an output state
\be 
\hat{\rho}_I= \frac{\expval{q_\Pi+i p_\Pi|\hat \rho_{RI}| q_\Pi+i p_\Pi}}{\tr \left[\expval{q_\Pi+i p_\Pi|\hat \rho_{RI}| q_\Pi+i p_\Pi}\right]},
\ee 
where $\ket{q_\Pi+i p_\Pi}$ is a coherent state with amplitude $q_\Pi+i p_\Pi$.

From conditional Gaussian map analyses~\cite{genoni2016conditional} (See Appendix~\ref{conditional_distribution}), the idler mode's covariance matrix and mean, and the distribution of measurement outcome are
\begin{subequations}
\begin{align}
V_I^\prime &= \left(2\frac{(1-\kappa+N_B)N_S}{\kappa N_S+N_B+1}+1\right)\mathbb{I}, \label{quadrature_conditional_output_CM}\\
\overline{\bm x}_I^\prime &= \frac{\sqrt{\kappa N_S(N_S+1)}}{\kappa N_S+N_B+1}\begin{pmatrix}
 \cos\theta q_\Pi+\sin\theta p_\Pi \\
 +\sin\theta q_\Pi-\cos\theta p_\Pi
\end{pmatrix}, \label{quadrature_conditional_output_mean} \\
p(\overline{\bm x}_\Pi) &= \frac{e^{-\frac{|\overline{\bm x}_\Pi|^2}{4(\kappa N_S+N_B+1)}}}{4\pi(\kappa N_S+N_B+1)}. \label{quadrature_conditional_output_p}
\end{align}
\end{subequations}
One can directly realize that the idler mode is in a displaced thermal state with mean $\overline{\bm x}^\prime_I$ and thermal photon number $E\equiv (1-\kappa+N_B)N_S/(\kappa N_S+N_B+1)$, as stated in the main text.

Using the distribution of quadratures Eq.~\eqref{quadrature_conditional_output_p}, we can obtain the distribution of the complex heterodyne readout on the $m$th returned mode
${\calM_m}=(q_{R_m}+ip_{R_m})/2$ as
\be 
p({\calM_m}) = \frac{e^{-\frac{|{\calM_m}|^2}{(\kappa N_S+N_B+1)}}}{\pi(\kappa N_S+N_B+1)}.
\label{eq:p_M}
\ee 
At the same time, the complex displacement of idler conditioned on the measurement result is 
\begin{align}
d_{m} &= \frac{\sqrt{\kappa N_S(N_S+1)}}{\kappa N_S+N_B+1}e^{i\theta}{\calM_m}^*
\end{align}
where $\calM_m^*$ denotes the complex conjugate.

Through the change of variables, one can write the total displacement amplitude square 
\be 
|d_T|^2\equiv \sum_{m=1}^M |d_m|^2 = \xi \sum_{i=1}^{2M} z_i^2,
\ee 
with 
\be 
\xi=\kappa N_S(N_S+1)/2(N_B+\kappa N_S+1)
\label{xi_definition}
\ee  
and $z_i \sim \calN(0,1)$ being a standard normal random variable, and thus we obtain the $\chi^2$ distribution of $|d_T|^2$
\be 
P_{\kappa}^{(M)}(x)\sim \frac{1}{\xi}\left(\frac{x}{\xi}\right)^{M-1}e^{-x/(2\xi)},
\label{p_overall}
\ee 
with mean $2M\xi$ and variance $4M\xi^2$, where 
$ 
\xi\equiv C_p^2/4v_\calM\,.
$

\section{Preliminary}
\label{sec:preliminary}

In this section, we define Gaussian states and general Gaussian measurements. Then, we review measurement statistics of general Gaussian measurements on general Gaussian states. Finally, we review ultimate limits of quantum state discrimination and quantum parameter estimation.

\subsection{Gaussian states}
\label{sec:GS}

We consider a system of $K$ modes, described by the annihilation operators $\{\hat{a}_\ell, 1\le \ell \le K\}$ satisfying the canonical commutation relation $[\hat{a}_\ell, \hat{a}_{k}^\dagger]=\delta_{\ell k}$. We define the momentum and position quadratures~\cite{weedbrook2012gaussian} as
\begin{align}
\hat{q}_\ell=\hat{a}_\ell+\hat{a}_\ell^\dagger \,,
\hat{p}_\ell=-i(\hat{a}_\ell-\hat{a}_\ell^\dagger ).
\end{align}
Now we can introduce the vector operator
\be 
\hat{\bm x}=(\hat{q}_1,\hat{p}_1, \cdots, \hat{q}_{K},\hat{p}_{K})^T,
\ee 
which satisfies the commutation relation
\be 
[\hat{x}_l, \hat{x}_k]=2i{\bm \Omega}_{lk}. 
\ee 
Here ${\bm \Omega}=\bigoplus_{k=1}^{K} i \mathbb{Y}$ with $\mathbb{Y}$ being the Pauli-Y matrix. For a quantum state $\hat{\rho}$, one can define the mean and the covariance matrix as
\begin{align}
&\overline{\bm{x}}\equiv \expval{\hat{\bm x}}
\\
&
V_{kl} = \frac{1}{2}\expval{\{ \hat{x}_k-\expval{\hat{x}_k}, \hat{x}_l-\expval{\hat{x}_l} \}},
\end{align}
where $\{\hat{a},\hat{b}\}\equiv\hat{a}\hat{b}+\hat{b}\hat{a}$ is the anti-commutator and $\expval{\hat{A}}=\Tr\left(\hat{A} \hat{\rho}\right)$ for any operator $\hat{A}$.

Gaussian states are entirely characterized by the mean and covariance matrix. In this paper, we consider entanglement from two-mode squeezed vacuum (TMSV), which is a zero-mean two-mode Gaussian state with the covariance matrix
\begin{equation}
V_{SI} =
\begin{pmatrix}
(2N_S+1) \mathbb{I} & 2\sqrt{N_S(N_S+1)}\mathbb{Z}\\
2\sqrt{N_S(N_S+1)}\mathbb{Z} & (2N_S+1)\mathbb{I}
\end{pmatrix},
\label{eq:CM_SI}
\end{equation}
where $\mathbb{Z}$ is the Pauli-Z matrix and $\mathbb{I}$ is $2\times 2$ identity. 

The TMSV state is still Gaussian after the signal mode is transmitted through a bosonic Gaussian channel. For the input-output relation in Eq.~\eqref{input_output_main} of the main text, the covariance matrix of the return and the idler is
\begin{equation}
V_{RI} = \begin{pmatrix}
(2\kappa N_S+2N_B+1)\mathbb{I} & 2\sqrt{\kappa N_S(N_S+1)}\bm{R}\mathbb{Z}\\
2\sqrt{\kappa N_S(N_S+1)}\mathbb{Z}\bm{R}^T & (2N_S+1) \mathbb{I}
\end{pmatrix},
\label{eq:CM_RI}
\end{equation}
where $\bm{R} = \begin{pmatrix}
 \cos\theta & -\sin\theta\\
 \sin\theta & \cos\theta
\end{pmatrix}$.

\subsection{General-dyne measurement statistics}
\label{conditional_distribution}

Consider a bipartite system, $A$ with $K_A$ modes and $B$ with $K_B$ modes, in a Gaussian state characterized by the mean and covariance matrix
\begin{align}
&\overline{\bm{x}}\equiv (\overline{\bm{x}_A}, \overline{\bm{x}_B})^T,
\\
&
V = \begin{pmatrix}
 V_A & V_{AB}\\ V_{AB}^T & V_B
\end{pmatrix}.
\end{align}
A general-dyne measurement on subsystem $B$ can be described by a projective measurement with a set of pure Gaussian states $\{\ket{\psi_\Pi}\}$, each with mean $\overline{\bm{x}_\Pi}$ and covariance matrix $V_\Pi$ where $\Pi$ denotes the corresponding Gaussian measurement. Under a general-dyne measurement, the remaining subsystem results in a Gaussian state conditioned on the measurement outcome following a Gaussian distribution, as shown in Ref.~\cite{genoni2016conditional}. Below, we provide a brief derivation of the measurement results and output Gaussian states.

Following the Fourier-Weyl relation, any quantum state $\hat{\rho}$ can be written in the basis of displacement operator as
\begin{equation}
    \hat{\rho} = \frac{1}{(2\pi)^{K_A+K_B}}\int_{\mathbb{R}^{2(K_A+K_B)}}d\bm{z} \chi(\bm{z})\hat{D}(-\bm{z})
\end{equation} 
where the displacement operator $\hat{D}(\bm \xi)\equiv \exp\left(i \hat{\bm x}^T \bm \Omega \bm \xi\right)$ and satisfys orthogonal relation $\tr\left[\hat{D}(\bm \xi) \hat{D}(\bm \xi^\prime)\right]=\pi^{K_A+K_B}\delta(\bm \xi+\bm \xi^\prime)$. For Gaussian states, the Wigner characteristic function has a Gaussian form
\begin{equation}
    \chi(\bm{z})=\Tr{\hat{\rho} D(\bm{z})}=\exp\left[-\frac{1}{2}\bm{z}^T\left(\Omega V\Omega^T\right)\bm{z}-i\left(\Omega\overline{\bm{x}}\right)^T\bm{z} \right].
\end{equation}
From the state overlap between $\hat{\rho}$ and $\ket{\psi_\Pi}$, we have
\begin{widetext}
\begin{equation}
\begin{split}
    \bra{\psi_\Pi}\hat{\rho}\ket{\psi_\Pi} &= \frac{1}{(2\pi)^{K_A+K_B}}\int {\rm d}\bm{z} e^{-\frac{1}{2}\bm{z}^T\Omega V \Omega^T \bm{z}-i(\Omega\overline{\bm x})^T\bm{z}}\bra{\psi_\Pi}\hat{D}(-\bm z)\ket{\psi_\Pi}\\
    &= \frac{1}{(2\pi)^{K_A+K_B}}\int {\rm d}\bm{z} e^{-\frac{1}{2}\bm{z}^T\Omega V \Omega^T \bm{z}-i(\Omega\overline{\bm x})^T\bm{z}}
    \times e^{-\frac{1}{2}\bm{z}_B^T\Omega V_\Pi \Omega^T \bm{z}_B+i(\Omega\overline{\bm x}_\Pi)^T\bm{z}_B}\hat{D}\left(-\bm z_A\right)\\
    &= \frac{1}{(2\pi)^{K_A+K_B}}\int {\rm d}\bm{z}^{\prime}_A e^{-\frac{1}{2}\bm{z}^{\prime T}_A V_A \bm{z}^{\prime}_A-i\overline{\bm x}_A^T\bm{z}^{\prime}_A} \hat{D}(-\Omega\bm{z}^{\prime}_A) \int {\rm d}\bm{z}'_B e^{-\frac{1}{2}\bm{z}^{\prime T}_B(V_B+V_\Pi)\bm{z}'_B+\bm{z}^{\prime T}_B\left[-V_{AB}^T\bm{z}^{\prime}_A+i(\overline{\bm x}_B-\overline{\bm x}_\Pi)\right]}\\
    &= \frac{1}{(2\pi)^{K_A}\sqrt{\det(V_B+V_\Pi)}}\int {\rm d}\bm{z}_A e^{-\frac{1}{2}\bm{z}_A^T V_A \bm{z}_A-i\overline{\bm x}_A^T\bm{z}_A} \hat{D}(-\Omega\bm{z}_A) e^{\frac{1}{2}\left[-V_{AB}^T\bm{z}_A+i(\overline{\bm x}_\Pi-\overline{\bm x}_B)\right]^T\frac{1}{V_B+V_\Pi}\left[-V_{AB}^T\bm{z}_A+i(\overline{\bm x}_\Pi-\overline{\bm x}_B)\right]}\\
    &= \frac{e^{-\frac{1}{2}(\overline{\bm x}_\Pi-\overline{\bm x}_B)^T\frac{1}{V_B+V_\Pi}(\overline{\bm x}_\Pi-\overline{\bm x}_B)}}{(2\pi)^{K_A}\sqrt{\det(V_B+V_\Pi)}} \int {\rm d}\bm{z}_A e^{-\frac{1}{2}\bm{z}_A^T \left(V_A-V_{AB}\frac{1}{V_B+V_\Pi}V_{AB}^T\right)\bm{z}_A} e^{-i\bm{z}_A^T\left(\overline{\bm x}_A+V_{AB}\frac{1}{V_B+V_\Pi}(\overline{\bm x}_\Pi-\overline{\bm x}_B) \right)}\hat{D}(-\Omega\bm{z}_A),
\end{split}
\label{projection_state}
\end{equation}
\end{widetext}
where we have denoted $\bm z=(\bm z_A, \bm z_B)$ in the second line; in the third line we replaced the variable $(\bm z_A', \bm z_B') =\Omega \bm z$; in the fourth line, we apply the formula of integral $\int {\rm d}\bm{z} e^{\bm{z}^T A \bm{z}+\bm{z}^T\bm{b}} = {\pi^{K_A}}e^{\bm{b}^TA^{-1}\bm{b}/4}/\sqrt{\det(A)}$ and omitted prime for simplicity. From the above projection, the mean and covariance matrix of the unmeasured subsystem $A$, and the measurement outcome distribution are directly found as
\begin{subequations}
\begin{align}
   V_A^\prime &= V_A-V_{AB}\frac{1}{V_B+V_\Pi}V_{AB}^T \label{eq:cond_cm},\\
   \overline{\bm x}_A^\prime &= \overline{\bm x}_A+V_{AB}\frac{1}{V_B+V_\Pi}(\overline{\bm x}_\Pi-\overline{\bm x}_B), \label{eq:cond_x}\\
   p(\overline{\bm x}_\Pi) &= \frac{ e^{-\frac{1}{2}(\overline{\bm x}_\Pi-\overline{\bm x}_B)^T\frac{1}{V_B+V_\Pi}(\overline{\bm x}_\Pi-\overline{\bm x}_B)}}{(2\pi)^{K_B}\sqrt{\det(V_B+V_\Pi)}}. \label{eq:cond_p}
\end{align}
\label{conditional_output}
\end{subequations}

\subsection{Helstrom limit and Quantum Chernoff bound}
\label{app:QCB}

In general, given $m_s$ quantum states $\{p_j, \hat{\rho}_j\}_{j=1}^{m_s}$ with prior probability $p_j$ for each state, there exists a well-known lower bound of error probability in discrimination---the Helstrom limit~\cite{Helstrom1969, Helstrom_1967,Helstrom_1976}, 
\begin{equation}
    P_{\rm H}(\{\hat{\rho}_j\}_{j=1}^{m_s}) = 1 - \max_{\{\hat{\Pi}_j\}} \sum_{j=1}^{m_s} p_j \Tr{\hat{\Pi}_j \hat{\rho}_j}
\end{equation}
where $\{\hat{\Pi}_j\}$ is a set of POVM operators whose $j$th element corresponds to state $\hat{\rho}_j$ and $\sum_j \hat{\Pi}_j = \hat{\mathbb{I}}$. In the case of $m_s=2$ with equal prior probability $p_1=p_2=1/2$, the Helstrom limit is
\begin{equation}
    P_{\rm H}(\hat{\rho}_1, \hat{\rho}_2) = \frac{1}{2}\left(1-\frac{1}{2}\Tr{|\hat{\rho}_1 - \hat{\rho}_2|}\right)
\end{equation}
and for pure states $\ket{\psi_1}, \ket{\psi_2}$, it can be further simplified to $P_{\rm H} = \left(1-\sqrt{1-|\bra{\psi_1}\ket{\psi_2}|^2}\right)/2$. 

In general, the Helstrom limit is hard to evaluate. A useful upper bound of the the Helstrom limit is the quantum Chernoff bound (QCB)~\cite{Audenaert2007}. For the binary state discrimination between M identical copies of states, $\hat{\rho}_0^{\otimes M}$ and $\hat{\rho_1}^{\otimes M}$, we have
\begin{equation}
    P_{\rm H}(\hat{\rho}_0^{\otimes M}, \hat{\rho_1}^{\otimes M})\le P_{\rm QCB} =\frac{1}{2}\left({\rm inf}_{s\in[0,1]}Q_s\right)^M,
\end{equation}
with $Q_s(\hat{\rho}_1,\hat{\rho}_2) = \Tr{\hat{\rho}_1^s\hat{\rho}_2^{1-s}}$.

QCB can be efficiently evaluated for Gaussian states~\cite{Pirandola2008}.
For two $K$-mode Gaussian states $\{\hat{\rho}_h\}_{h=1}^2$ with mean quadrature $\overline{\bm x}_h$ and covariance matrix $V_h$, one can find the symplectic decomposition of the covariance matrix as $V_h = S_hV_h^{\oplus}S_h^T$ where $V_h^{\oplus} = \oplus_{j=1}^K \nu_j^{(h)}\mathbb{I}$ with the symplectic spectrum $\{\nu_j^{(h)}\}_{j=1}^K$~\cite{weedbrook2012gaussian}.
In this case, the QCB can be evaluated via 
\begin{equation}
    Q_s = \overline{Q}_s\exp{-\frac{1}{2}{\bm d}^T\left(\tilde{V}_1(s)+\tilde{V}_2(1-s)\right)^{-1}{\bm d}}
\end{equation}
where ${\bm d}=\overline{\bm x}_1 - \overline{\bm x}_2$ and $\overline{Q}_s$ is defined as
\begin{equation}
    \overline{Q}_s = \frac{2^K\prod_{j=1}^n G_s(\nu_j^{(1)})G_{1-s}(\nu_j^{(2)})}{\sqrt{\det\left[\tilde{V}_1(s)+\tilde{V}_2(1-s)\right]}}
\end{equation}
with $\tilde{V}_h(s)=S_h\left[\oplus_{j=1}^K \Lambda_s(\nu_j^{(h)})\mathbb{I}\right]S_h^T$. Here two known functions are introduced as
\begin{subequations}
\begin{align}
    G_p(\nu) &\equiv \frac{2^p}{(\nu+1)^p-(\nu-1)^p},
    \\
    \Lambda_p(\nu) &\equiv \frac{(\nu+1)^p+(\nu-1)^p}{(\nu+1)^p-(\nu-1)^p}.
\end{align}
\end{subequations}

\subsection{Quantum Fisher information for Gaussian states}
We will utilize the formula of quantum Fisher information (QFI) of Gaussian states proposed in Ref.~\cite{gao2014bounds}, summarized as the following. To be consistent with Ref.~\cite{gao2014bounds}, we will adopt a different definition of mean and covariance matrix compared with Appendix~\ref{sec:GS}. For an arbitrary $n$-mode Gaussian state, define a vector of annihilation operators $\hat{\bm a}=[\hat a_1,\hat a_1^\dagger, \ldots, \hat a_n,\hat a_n^\dagger]$. The state has mean 
$
\bm d\equiv \expval{\hat{\bm a}}\,,
$
and covariance matrix 
$
\hw{\Sigma_{\mu\nu} \equiv \frac{1}{2}\expval{(\hat{a}_\mu-d_\mu)(\hat{a}_{\nu}- d_\nu)+(\hat{a}_\nu-d_\nu)(\hat{a}_{\mu}- d_\mu)} }\,.
$
The commutation relation is $[\hat{a}_i,\hat{a}_j]=\Omega_{ij}$, where ${\bm \Omega}=\bigoplus_{k=1}^{K} i \mathbb{Y}$ with $\mathbb{Y}$ being the Pauli-Y matrix.
Given the mean $\bm d$ and covariance matrix $\Sigma$, Ref.~\cite{gao2014bounds} states the formula of the Gaussian-state QFI as
\be 
\calF=\frac{1}{2} {\mathscr R}^{-1}_{\alpha\beta,\mu\nu} \partial_\theta\Sigma_{\alpha\beta}\partial_\theta\Sigma_{\mu\nu}+\Sigma^{-1}_{\mu\nu}\partial_\theta d_\mu \partial_\theta d_\nu,
\label{eq:QFI_Gaussian_supp}
\ee
where ${\mathscr R}\equiv\Sigma\otimes \Sigma+\Omega\otimes\Omega/4$. Here $\theta$ can be an arbitrary parameter, while we focus on the estimation of the signal phase $\theta$ in this paper.

\section{Review of known results}
\label{sec:review}

In this section, we review known results mentioned in the main text and/or utilized in Appendix~\ref{sec:details_C2D}.

\subsection{Nair-Gu lower bound $P_{\rm NG}$ on error probability}

Nair and Gu derived a lower bound on the error probability of QI~\cite{nair2020fundamental} target detection applicable to $M$ probes with mean photon number $N_S$ assisted by arbitrary form of entanglement,
\begin{equation}
    P_{\rm E}\ge P_{\rm NG} \equiv \frac{1}{4}\exp\left[-\beta M N_S\right], 
    \label{eq:nair_gu}
\end{equation}
where $\beta \equiv -\log(1-\kappa/(N_B+1))$. We adopt this lower bound for comparison in quantum illumination target detection.

\subsection{Upper bound $\calF_{\rm UB}$ on phase sensing Fisher information}

Ref.~\cite{gagatsos2017bounding} derived an ultimate upper bound of Fisher information in noisy phase estimation applicable to any form of entangled input 
\bal 
&\calF_{\rm UB}=\\
&\hw{\frac{4M\kappa N_S\left(\kappa N_S+\left(1-\kappa\right)N_B'+1\right)}{\left(1-\kappa\right)\!\left[\kappa N_S \left(2N_B'+1\right)-\!\kappa N_B'\left(N_B'+1\right)+\!\left(N_B'+1\right)^2\right]}\,,}
\label{eq:qfi_UB}
\eal 
\hw{where $N_B'=N_B/(1-\kappa)$ is the mean photon number of the thermal state at the environment mode in the Stinespring representation of the channel \cite{gagatsos2017bounding}.}

\subsection{Comparing with `no-go' results}

Here, we discuss the relationship of our results on quantum illumination to some no-go type of results. In general, local operations and classical communication (LOCC) strategy on each copy is not optimal for the discrimination between a pair of identical copies of states~\cite{LOCC_NO_GO,cheng2021discrimination}. For finite $N_S$, our approach is not LOCC between copies because it requires jointly detection of the $M$ idler modes, e.g. via a beamsplitter transform and then joint photodetection on output. At the $N_S\ll1$ limit, the conversion module produces pure coherent states, and then LOCC in the form of Dolinar receiver can be adopted directly without the beamsplitter on all idlers to achieve the optimal. However, this does not contradict Refs.~\cite{LOCC_NO_GO,cheng2021discrimination} as they do not preclude special cases of optimal mixed-state discrimination to be achievable by LOCC between copies.

Another relevant paper is Ref.~\cite{bradshaw2017overarching}, however, there they consider first measuring the idler and then show that quantum advantage is gone; in our work, we are measuring the signal and then perform operations on the idlers conditioned on the measurement results, which is entirely different from Ref.~\cite{bradshaw2017overarching}.

\subsection{Review of classical communication capacity. }

For the channel $\Phi_{\kappa,\theta}$ described by Eq.~\eqref{input_output_main} of the main text, the classical capacity with energy constraint $\expval{\hat a^\dagger \hat a}\le N_S$, without entanglement assistance, is known as~\cite{giovannetti2004}
\be 
C=g(\kappa N_S+N_B)-g(N_B)\,.
\ee
Here, $g(n)=(n+1)\log_2(n+1)-n \log_2 n$ is the entropy of a thermal state with mean photon number $n$. 
When $N_S\to 0$, we can expand $C$ to the leading order as
\be 
C=\kappa N_S\log_2(1+\frac{1}{N_B})+O(N_S^2)\,.
\ee
In the noisy scenario $N_B \gg 1$, the classical capacity $C\simeq \kappa N_S/\ln(2)N_B$ is saturated by a heterodyne or a homodyne receiver~\cite{shi2020practical}. 

\begin{figure}
    \centering
    \includegraphics[width=0.25\textwidth]{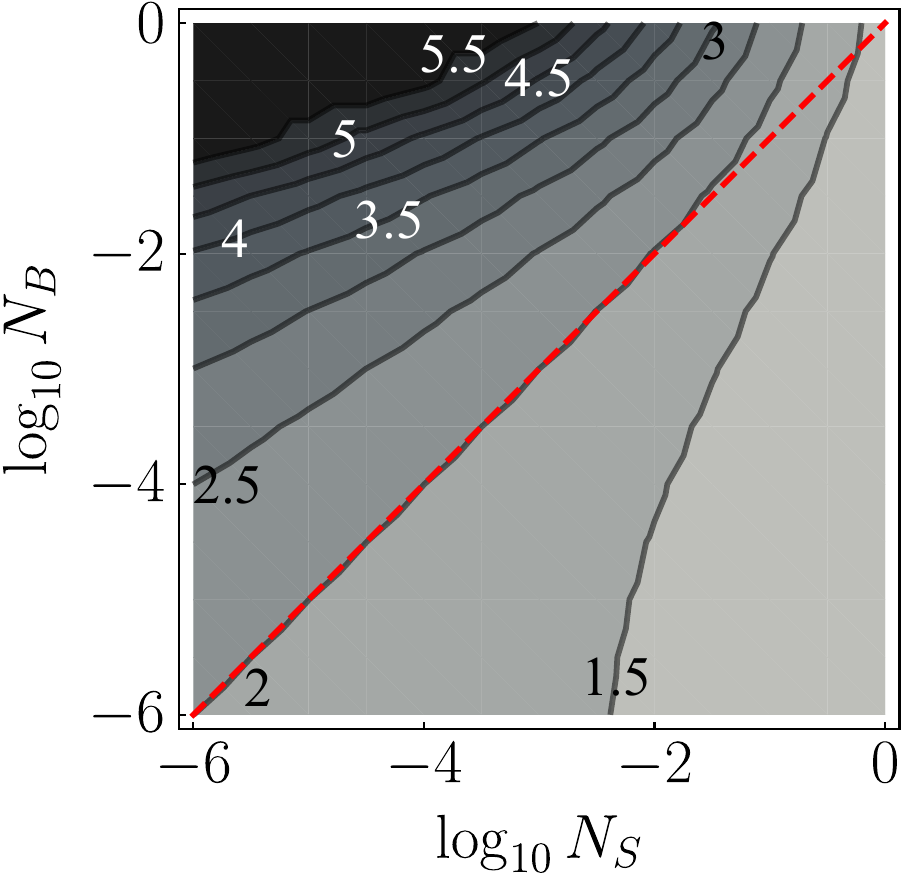}
    \caption{The ultimate EA capacity over lossy channel under various $N_B,N_S$, normalized by the unassisted capacity $C$. The red dashed diagonal line indicates $N_S=N_B$. Channel transmissivity $\kappa=0.01$.}
    \label{fig:EACOMM_NbNs_CE}
\end{figure}

Entanglement assistance boosts the communication capacity to \cite{Bennett2002}
\be
C_E=g(N_S)+g(N_S^\prime)-g(A_+)-g(A_-),
\label{CE_exact}
\ee 
where 
$A_\pm=(D-1\pm(N_S^\prime-N_S))/2$, $N_S^\prime=\kappa N_S+N_B$ and $D=\sqrt{(N_S+N_S^\prime+1)^2-4\kappa N_S(N_S+1)}$. 
When $N_S\to 0$, the leading order can be obtained as
\bal 
C_E&= \frac{\kappa N_S}{N_B+1} \left[\log_2 \left(\frac{1}{N_B N_S \left(N_B-\kappa+1\right)}\right)+\calR
   \right]\!+\!O\left(N_S^2\right)\\
   &= \frac{\kappa N_S\log_2(1/N_S)}{N_B+1}+O(N_S)\,,
\label{CE_expansion}
\eal 
where $\calR=\big[\left(N_B+1\right) \log_2 \left(N_B-\kappa +1\right)+\kappa +\left(-N_B+2 \kappa -1\right) \log_2
   \left(N_B+1\right)\big]/\kappa$ is independent of $N_S$.
We see a diverging advantage $
C_E/C\sim \ln\left(1/N_S\right),
$ in the $ N_S\ll1$ limit. Remarkably, such an advantage is not limited to the region $N_B\gg 1, N_S\ll1$ considered in Ref.~\cite{Bennett2002,shi2020practical}, as shown in Fig.~\ref{fig:EACOMM_NbNs_CE}. In the main text such an extension of advantageous region is shown to hold for our conversion module as well.
The EA capacity is known to be achieved by the Holevo information of phase encoding on TMSV at this limit~\cite{shi2020practical}.

\subsection{Review of Green machine for achieving optimal scaling of EA communication}
\label{app:GM}

\begin{figure*}[tbp]
    \centering
    \includegraphics[width=0.9\textwidth]{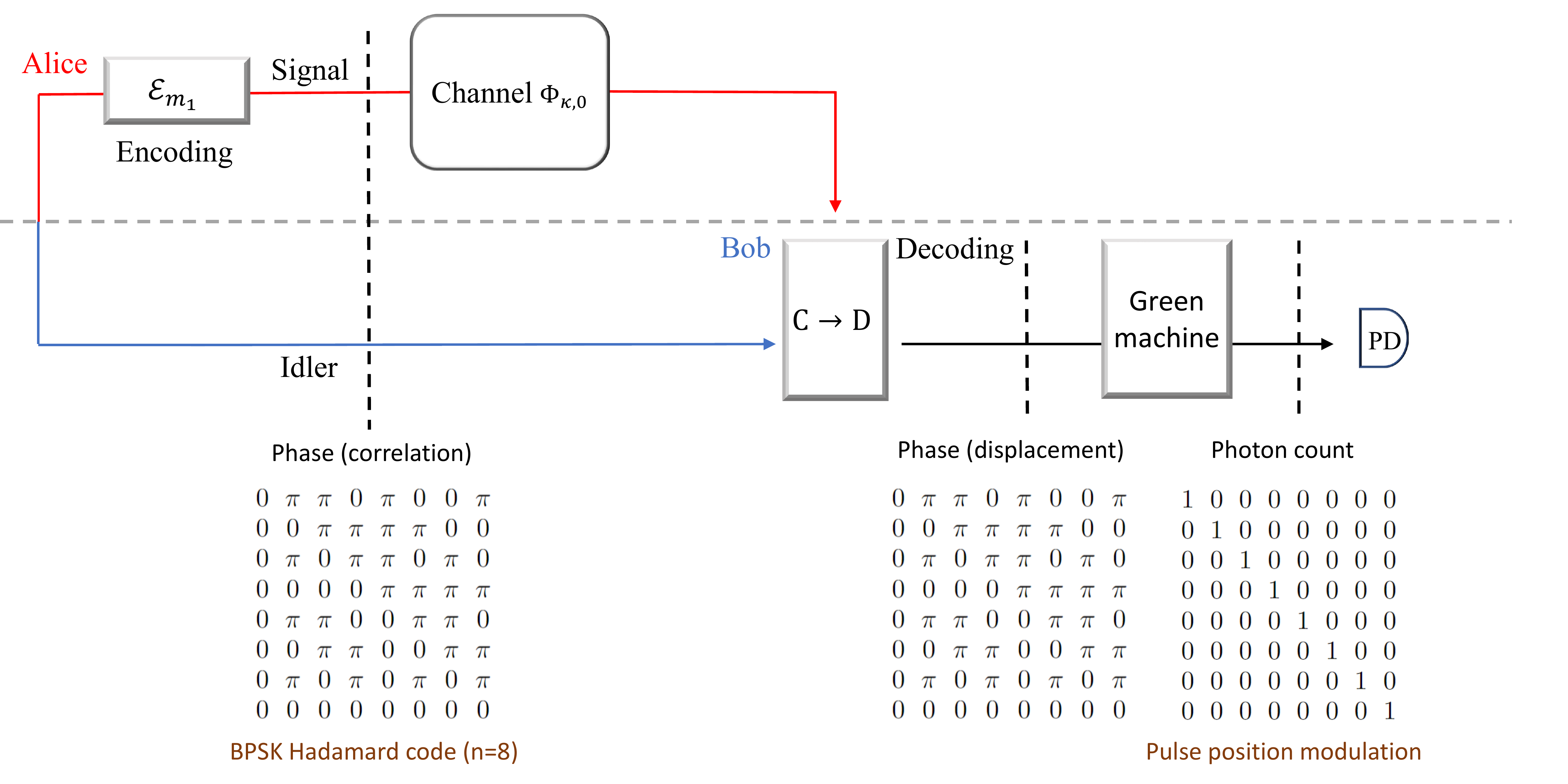}
    \caption{Entanglement-assisted communication protocol using BPSK Hadamard code and Green machine. The Green machine, a beamsplitter array defined in Eq.~\eqref{eq:GM_BS}, converts the BPSK Hadamard code in phase to pulse position modulation in photon count.}
    \label{fig:GM}
\end{figure*}

\begin{figure}
    \centering
    \includegraphics[width=0.25\textwidth]{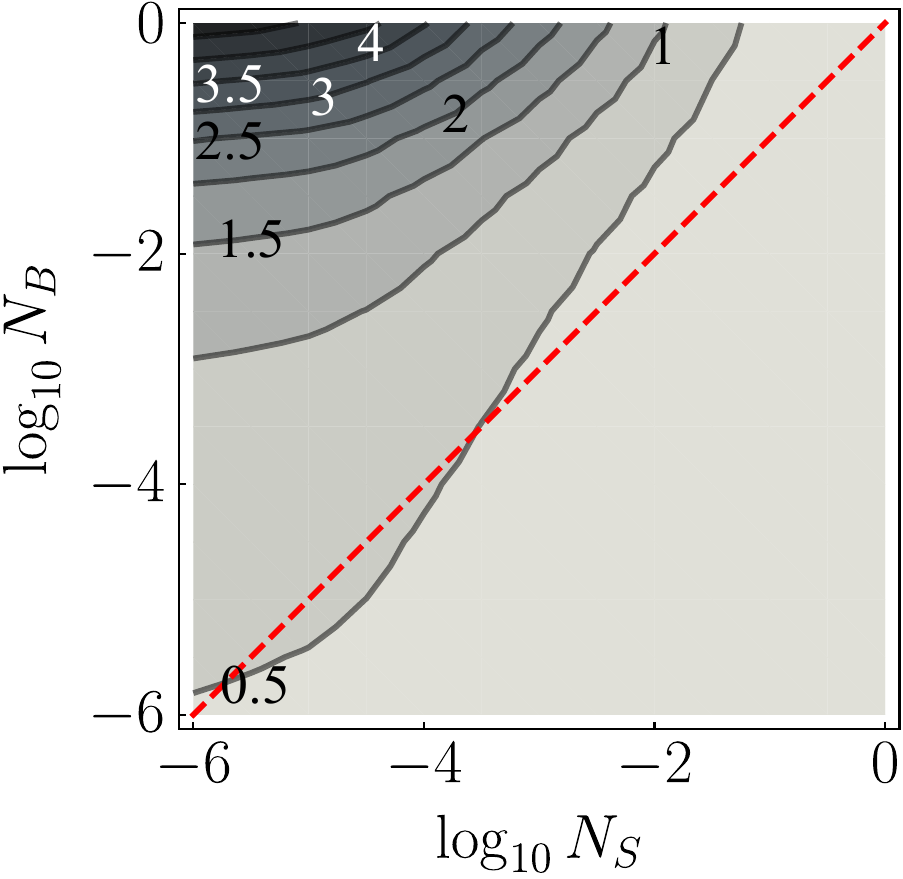}
    \caption{The information rate per symbol of the Green machine using BPSK-encoded TMSV state under various $N_B,N_S$, normalized by the unassisted capacity $C$. Plotted with numerical optimized repetition $M$ and asymptotically optimal codeword length $n$ Eq.~\eqref{eq:optn}. The black diagonal line indicates $N_S=N_B$. $\kappa=0.01$. }
    \label{fig:EACOMM_NsNb_GM}
\end{figure}

The `Green machine' is a receiver that attains a rate very close to the classical communication capacity with phase modulation of coherent
states~\cite{guha2011structured}. Here we show that a design concatenating it with our $\rm C\veryshortrightarrow D$ module (see Fig.~\ref{fig:EACOMM_NsNb_GM} for a schematic plot) achieves the same scaling of the ultimate EA capacity.
The sender jointly encodes a block of $n$ signal modes by BPSK modulation according to an $n$-codeword Hadamard code, where $n$ is a power of 2 and each codeword contains $n$ symbols $\{\theta_k\}_{k=1}^n$. Fig.~\ref{fig:GM} shows an example with $n=8$. The information rate can be further improved by repetitive encoding over $M$ i.i.d. copies of signal modes with identical symbols, where $M$ is to be optimized. After an $n$-block of signals goes through the channel $\Phi_{\kappa,0}$, the receiver obtains an $n$-block of returned signals. Then the receiver applies the $\rm C\veryshortrightarrow D$ conversion module to the $n$-block of return-idler pairs, which yields $n$ displaced thermal states as defined in Eq.~\eqref{eq:dts_app}, with quadrature phase subject to the $n$-codeword Hadamard code. We note that, the $\rm C\veryshortrightarrow D$ conversion module combines the $M$ copies together, thereby the brightness of the $n$ displaced thermal states is increased such that the thermal background $E$ is negligible and the states resemble coherent states. The $n$ quasi-coherent-state outputs are input into the Green machine. The Green machine consists of a beamsplitter array. Denote the input modes as a vector $\bm {\hat a}=[\hat a_1,\ldots \hat a_n]^T$, with mean $\expval{\bm{\hat a}}\propto [e^{i\theta_1},\ldots,e^{i\theta_n}]^T$, where we have left out the amplitude to focus on the phase. The beamsplitter array fulfills a Bogoliubov transform $\bm {\hat a}\to S\bm {\hat a}$ with the unitary matrix
\be 
S=\frac{1}{\sqrt{n}}
\bp 
1 & 1\\
-1 & 1
\ep ^{\otimes \log_2 n}\,.
\label{eq:GM_BS}
\ee
Finally the receiver makes zero-or-not photon counting on the $n$ output modes of the Green machine individually. At the limit of weak thermal background $E\to 0$, each output mode yields non-zero photon count iff the input modes are constructively interfered by the beamsplitter array.
Concretely, the Green machine converts one of the $n$ codewords of BPSK Hadamard code (in the quadrature phase) into one of the $n$ codewords of coherent-state pulse position modulation (PPM) (in the photon count), i.e. the photons of $n$ input modes are merged into one output mode.

Now we evaluate the performance of the above protocol. The magnitudes of the means at the output modes of the $\rm C\veryshortrightarrow D$ module depend on the squared mean $x$ of the random heterodyne readout, which is subject to the $\chi^2$ distribution $P_\kappa (x)$ as defined in Eq.~\eqref{p_overall}. 
For an $M$-copy, $n$-codeword Green machine, the per-symbol rate given squared mean $x$ is \cite{guha2020infinite}
\begin{widetext}
\bal 
R_{\rm GM}(x)=&
\frac{1}{ M n \ln 2}\left[\left(n-1\right) P_d\left(x\right) \ln \left(\frac{n P_d\left(x\right)}{P_{\rm E}\left(x\right)}\right)-\Bigg(\left(n-1\right) P_d\left(x\right)+P_{\rm E}\left(x\right)\Bigg) \ln \left(\frac{\left(n-1\right) P_d\left(x\right)}{P_{\rm E}\left(x\right)}+1\right)+P_{\rm E}\left(x\right) \ln \left(n\right)\right]
\label{rate_formula}
\eal
\end{widetext}
where $P_d(x)=[1-P_c(x)]P_b(1-P_b)^{n-2}$, $P_{\rm E}(x)=P_c(x)(1-P_b)^{n-1}$ and
\bal 
P_c(x)=1-\frac{e^{-\frac{x}{1+E}}}{1+E}\,,P_b=1-\frac{1}{1+E}\,.
\eal 
Here the thermal background $
E={N_S \left(N_B+1-\kappa \right)}/{ 2v_\calM}\le N_S
$ is defined in the main text.
Thus, the overall rate is
\be 
R_{\rm GM}=\int_{0}^\infty dx P_\kappa (x) R_{\rm GM}(x).
\label{eq:Rgm_exact_supp}
\ee



Observe that $R_{\rm GM}$ depends on $M, n$. We numerically optimize $R_{\rm GM}$ over integer $M$ for each value of $N_S$ in Fig.~\ref{fig:EACOMM_NsNb_GM}, while choosing the block size $n^\star$ to be the asymptotic optimum Eq.~\eqref{eq:optn}. The results in the main text is obtained similarly.

Below, we provide asymptotic analyses to obtain more insights. Note that our numerical results above are evaluated via the exact formula Eq.~\eqref{eq:Rgm_exact_supp} without asymptotic approximations, using only the value of $n^\star$ derived below, and therefore represents an exact achievable rate.
The optimized $M^*\equiv{\rm argmax}_M R_{\rm GM}$ is numerically found $\sim 10^4$, sufficiently large for all parameters being considered. Thus it is fair to invoke the law of large numbers, that the squared mean $x$ converges in probability to the expectation value $\overline x=M\kappa N_S(N_S+1)/(N_B+\kappa N_S+1)$ of the $\chi^2$ distribution. Hence, the rate converges to the value at $x=\overline x$
\be 
R_{\rm GM}\to R_{\rm GM}(\overline x)=R_{\rm GM}\left(\frac{M\kappa N_S(N_S+1)}{N_B+\kappa N_S+1}\right)
\ee 
Below we optimize $R_{\rm GM}$ with respect to $n$, in the asymptotic regime $N_S\to 0$. Let the optimal $n$ be
\be 
n^*\equiv {\rm argmax}_{n\in \calN} R_{\rm GM},
\ee
where $\calN=\{2^k| k\in \{1,2,\ldots\}\}$ is the set of positive powers of 2. 
In this case $P_{\rm E}\gg P_d$, thus the rate in Eq.~\eqref{rate_formula} is dominated by $P_{\rm E}\ln (n)/Mn$.
Then one can obtain the second-order expansion of the rate
\be 
MR_{\rm GM}\simeq \frac{P_{\rm E}\ln(n)}{n  \ln 2}=(u +v n) \log_2(n)+O(N_S^3)
\label{eq:GM_R_1}
\ee
where 
\bal 
u&=\frac{\!- N_S \big[\!-\!\kappa  \left(N_B\!+\!1\right) \left(M\!+\!2 N_S\right) \!+\! \left(N_B\!+\!1\right){}^2 N_S\!+\! \kappa ^2 N_S\big]}{\left(N_B+1\right){}^2}\,,\\
v&=-\frac{\kappa  M N_S^2 \big[2 \left(N_B+1\right)+\kappa  (M-2)\big]}{2 \left(N_B+1\right)^2}\,.
\eal 
By solving ${\rm d} R_{\rm GM}/{\rm d} n=0$, we obtain
\be 
n^*\simeq -\frac{u}{v W(-ue/v)},
\label{eq:optn}
\ee
where $W$ is the principle branch of the Lambert W function which satisfies $W(xe^x)=x$ for $x\ge -1$. Using the relation $\ln W(x)=\ln (x)-W(x)$ and the asymptotic expansion $W(x)=\ln(x)-\ln\ln(x)+O(1)$ as $\ln(x)\to \infty$,
\begin{widetext}
\bal 
\ln n^*\simeq &-1+W(\frac{-ue}{v})\\
=&-\ln \left[\ln \left(-\frac{2 e\left(-\kappa  \left(N_B+1\right) \left(M+2 N_S\right)+\left(N_B+1\right){}^2 N_S+\kappa ^2 N_S\right)}{\kappa  M N_S \left(2 N_B-2 \kappa +\kappa  M+2\right)}\right)\right]\\
&+\ln \left(-\frac{2  \left(-\kappa  \left(N_B+1\right) \left(M+2 N_S\right)+\left(N_B+1\right){}^2 N_S+\kappa ^2 N_S\right)}{\kappa  M N_S \left(2 \left(N_B+1\right)+\kappa  (M-2)\right)}\right).
\label{eq:optlogn}
\eal 
\end{widetext}
Plugging Eqs.~\eqref{eq:optn} and~\eqref{eq:optlogn} in Eq.~\eqref{eq:GM_R_1}, we have the asymptotic rate
\bal 
R_{\rm GM}&= \frac{\kappa  N_S \left(\ln \left(\frac{2 \left(N_B+1\right)}{N_S \left(2 \left(N_B+1\right)+\kappa 
   (M-2)\right)}\right)-1\right)}{(N_B+1) \ln 2}+O\left(N_S^2\right).
\label{eq:GM_R_fin}
\eal 
At the limit of $N_S\to 0$, it achieves the optimal scaling of the ultimate EA capacity $C_{\rm E}$~\cite{Bennett2002}
\be 
R_{\rm GM}=\frac{\kappa N_S}{(N_B+1) \ln 2}(\ln(1/N_S)-O(1))\propto N_S\ln (1/N_S)\,.
\label{eq:GM_R_asym}
\ee
Remarkably,
\be 
\frac{R_{\rm GM}}{C_{\rm E}}= 1- O(\frac{1}{\ln (1/N_S)})
\ee
which goes to 1 as $N_S\to 0$.

\subsection{Review of coherent state discrimination}
\label{app:receivers_summary}

In this section, we summarize coherent-state discrimination, including the Helstrom limit~\cite{Helstrom1969, Helstrom_1967,Helstrom_1976}, homodyne detection, heterodyne detection, Kennedy receiver~\cite{Kennedy_1972} and Dolinar receiver. 

In the following discussion, we consider the case of discrimination between a vacuum state $\ket{0}$ and a coherent state $\ket{\alpha}$, where in general $\alpha=\alpha_R+i\alpha_I$. The noisy version of this discrimination problem is exactly the sub-task necessary to complete the receiver design for the entanglement-assisted applications after the conversion module. 

We begin with the noiseless version, in which case the Helstrom limit of error probability has a closed-form solution
\begin{equation}
    P_{\rm H}(\ket{0},\ket{\alpha}) = \frac{1}{2}\left(1-\sqrt{1-e^{-|\alpha|^2}}\right).
\end{equation}

Now we discuss the performance of homodyne detection.
Homodyne detection consists of measuring a single quadrature of the mode, for example the position quadrature $\hat{q}$. For binary discrimination, we can write out the POVM element as
\begin{equation}
\begin{split}
    \hat{\Pi}_{0,{\rm homo}} &= \int_{-\infty}^B dq \ket{q}\bra{q}\\
    \hat{\Pi}_{1,{\rm homo}} &= \hat{\mathbb{I}} - \hat{\Pi}_{0,{\rm homo}},
\end{split}
\end{equation}
where $B$ determines the decision threshold. With the POVM elements, the error probability using homodyne detection is
\begin{equation}
\begin{split}
    &P_{\rm E, homo} = \min_B\frac{1}{2}\left(\bra{\alpha}\Pi_{0,{\rm homo}}\ket{\alpha} + \bra{0}\Pi_{1,{\rm homo}}\ket{0}\right)\\
    &= \min_B\frac{1}{2}\left(\int_{-\infty}^B dq |\bra{\alpha}\ket{q}|^2 + 1-\int_{-\infty}^B dq |\bra{0}\ket{q}|^2\right)\\
    &= \min_B\frac{1}{2}+\frac{1}{2\sqrt{\hw{2}\pi}}\left(\int_{-\infty}^B dq  e^{-(q-\hw{2}\alpha_R)^2\hw{/2}} -\int_{-\infty}^B dq e^{-q^2\hw{/2}} \right)\\
    &= \min_B\frac{1}{4}\left({\rm Erfc}(\hw{\sqrt 2}\alpha_R-B\hw{/\sqrt{2}})+{\rm Erfc}(B\hw{/\sqrt{2}})\right)\\
    &= \frac{1}{2}{\rm Erfc}\left(\frac{\alpha_R}{\sqrt{2}}\right),
\end{split}
\end{equation}
where in the last line the error probability is minimized at the threshold $B=\alpha_R/\sqrt{2}$. When the amplitude of coherent state $\alpha_R\gg1$, as ${\rm Erfc}(x)\sim e^{-x^2}/\sqrt{\pi}x$, we have $P_{\rm E, homo}\sim \exp\left(-|\alpha_R|^2/2\right)$. 

Similarly, heterodyne detection projects the modes to coherent states with the POVM for a binary discrimination task
\begin{equation}
\begin{split}
    \hat{\Pi}_{0,{\rm het}} &= \frac{1}{\pi}\int_A d\beta \ket{\beta}\bra{\beta}\\
    \hat{\Pi}_{1,{\rm het}} &= \hat{\mathbb{I}} - \hat{\Pi}_{0,{\rm het}},
\end{split}
\end{equation}
where $A$ denotes a decision region in the complex plane (denoted as $x$ and $y$ axes in the following discussion). The error probability applying heterodyne detection is thus 
\begin{equation}
\begin{split}
    P_{\rm E, het} &= \frac{1}{2}\left(\bra{\alpha}\Pi_{0,{\rm het}}\ket{\alpha} + \bra{0}\Pi_{1,{\rm het}}\ket{0}\right)\\
    &= \frac{1}{2}\left(\frac{1}{\pi}\int_A d\beta |\bra{\beta}\ket{\alpha}|^2 + 1-\frac{1}{\pi}\int_A d\beta |\bra{\beta}\ket{0}|^2\right)\\
    &= \frac{1}{2}+\frac{1}{2\pi}\left(\int_A d\beta e^{-|\beta-\alpha|^2} - \int_A d\beta e^{-|\beta|^2}\right).
\end{split}
\end{equation}
Through simple geometry analysis, we can find the region $A$ to achieve the optimal error probability as $A\equiv \{(x,y)|y\le -\frac{\alpha_R}{\alpha_I}x+\frac{|\alpha|^2}{2\alpha_I}$ with boundary denoted as $l_A$ for simplification. Therefore, we have the optimal error probability with heterodyne as
\begin{equation}
\begin{split}
    P_{\rm E, het} &= \frac{1}{2} + \frac{1}{2\hw{\cdot 2\pi}}\int_{-\infty}^{\infty} dx \int_{-\infty}^{l_A} dy e^{-(x-\hw{\sqrt 2}\alpha_R)^2\hw{/2}-(y-\hw{\sqrt 2}\alpha_I)^2\hw{/2}}\\
    &- \frac{1}{2\hw{\cdot 2\pi}}\int_{-\infty}^\infty dx \int_{-\infty}^{l_A} dy e^{-x^2\hw{/2}-y^2\hw{/2}}\\
    &= \frac{1}{2} + \frac{1}{4}\left(1-{\rm Erf}\left(\frac{|\alpha|}{2}\right)\right) - \frac{1}{4}\left(1+{\rm Erf}\left(\frac{|\alpha|}{2}\right)\right)\\
    &= {\rm Erfc}\left(\frac{|\alpha|}{2}\right).
\end{split}
\end{equation}
When $|\alpha|\gg1$, we have $P_{\rm E, het}\sim \exp\left(-|\alpha|^2/4\right)$.

For coherent state discrimination, there exists a well-known nulling receiver called Kennedy receiver~\cite{Kennedy_1972}. For two arbitrary coherent states $\ket{\alpha_1},\ket{\alpha_2}$, the Kennedy receiver performs a displacement $-\alpha_1$, such that one of the state results in vacuum state $\ket{\alpha_1}\rightarrow \ket{0}$ while the other is $\ket{\alpha_2} \rightarrow\ket{\alpha_2-\alpha_1}$. The unknown state is considered to be $\ket{\alpha_1}$ when there is no photon detected and $\ket{\alpha_2}$ otherwise. Therefore the error probability with Kennedy receiver is
\begin{equation}
    P_{\rm E, Kennedy} = \frac{1}{2}|\bra{0}\ket{\alpha_2-\alpha_1}|^2 = \frac{1}{2}e^{-|\alpha_2-\alpha_1|^2},
\end{equation}
since there is no error in predicting the state $\ket{\alpha_1}$. For the case under consideration, $\ket{0}$ versus $\ket{\alpha}$, as one of the state is always in vacuum state, the Kennedy receiver is equivalent to the a direct photon counting with error probability
\begin{equation}
    P_{\rm E, Kennedy} = \frac{1}{2}e^{-|\alpha|^2}
\end{equation}
which is approximately $P_{\rm E, Kennedy} \simeq 2P_{\rm H}$ when the mean photon number $|\alpha|^2\gg 1$.

Note that Kennedy receiver is only sub-optimal in the coherent states discrimination, an adaptive receiver, Dolinar receiver~\cite{dolinar_processing_1973}, has been proposed to approach the Helstrom limit in the noiseless case. The Dolinar receiver splits the input coherent state into $S$ slices and makes a decision in terms of the prior probability of the each slice where displacement and Bayesian updating rule are applied. To help explain the detail, we introduce $h$ to denote the true state, $g$ as the current decision ($h,g\in\{0,1\}$) and $p_h^{(k)}$ as the prior probability for $k$th slice to be state $\hat{\rho}_h$. The number of photons measured from $k$th slice is denoted as $N^{(k)}$ following a distribution $P_N^{(M)}(N^{(k)},g|h)$. We use $p(N^{(k)},g|h)$ to represent the Bayesian conditional probability for obtaining $N^{(k)}$ photons when the $k$th slice is determined to be $\hat{\rho}_g$ while it is actually $\hat{\rho}_h$.
For $\ket{0}$ and $\ket{\alpha}$ with equal prior probability $p_0^{(0)}=p_1^{(1)}$, the Dolinar receiver works as the following.
\begin{algorithm}[H]
\caption{Dolinar receiver}
\begin{algorithmic}
\State $S$, $h$, $\gamma = \sqrt{|\alpha|^2}/2\sqrt{S}$
\State $k \gets 1$, $g \gets None$
\While{$k \le S$}
\State $u^{(k)} = \gamma/\sqrt{1-\exp{-|\alpha|^2(k-1/2)/S}}$
\If{$p_0^{(k)} > p_1^{(k)}$}
    \State $g \gets 0$
\ElsIf{$p_0^{(k)} < p_1^{(k)}$}
    \State $g \gets 1$
\Else
    \State $g \gets \{0,1\}$ with equal probability
\EndIf
\If{$g = 0$}
    \State Perform displacement $-\gamma+u^{(k)}$
\Else
    \State Perform displacement $-\gamma-u^{(k)}$
\EndIf
\State Measure the photon number $N^{(k)}$ with probability $p_N(N^{(k)},g|h)$
\State Update prior probability \\
$p_0^{(k+1)} \gets p_0^{(k)} p(N^{(k)},g|0)/\sum_{h^\prime=0}^1 p_{h^\prime}^{(k)} p(N^{(k)},g|h^\prime)$\\
$p_1^{(k+1)} \gets p_1^{(k)} p(N^{(k)},g|1)/\sum_{h^\prime=0}^1 p_{h^\prime}^{(k)} p(N^{(k)},g|h^\prime)$
\EndWhile
\If{$p_0^{S+1} > p_1^{S+1}$}
    \State $g \gets 0$
\ElsIf{$p_0^{S+1} < p_1^{S+1}$}
    \State $g \gets 1$
\Else
    \State $g \gets \{0,1\}$ with equal probability
\EndIf.\\
\Return $g$
\end{algorithmic}
\end{algorithm}
The Dolinar receiver gives a prediction on the unknown state and we perform Monte-Carlo simulation to evaluate the error probability.

For noiseless case $\ket{0}$ versus $\ket{\alpha}$, the measured photon distribution on $k$th slice follows Poisson distribution
\begin{equation}
    N^{(k)}\sim p_N(n,g|h)=\begin{cases}
    {\rm Pois}(n;(\gamma-u^{(k)})^2), & \textit{if $g=h$}\\
    {\rm Pois}(n;(\gamma+u^{(k)})^2), & \textit{otherwise}
    \end{cases}
\end{equation}
where ${\rm Pois}(\mu;\lambda)$ is the Poisson probability mass function. The conditional Bayesian probability of getting $N^{(k)}$ photons is $p(N^{(k)},g|h) = p_N(N^{(k)},g|h)$.

\subsection{Known sub-optimal receivers for quantum illumination target detection, phase sensing, and communication}
\label{app:OPA_PCR}

\subsubsection{Optical parametric amplifier receiver (OPAR)}

\begin{figure}[b]
    \centering
    \includegraphics[width=0.45\textwidth]{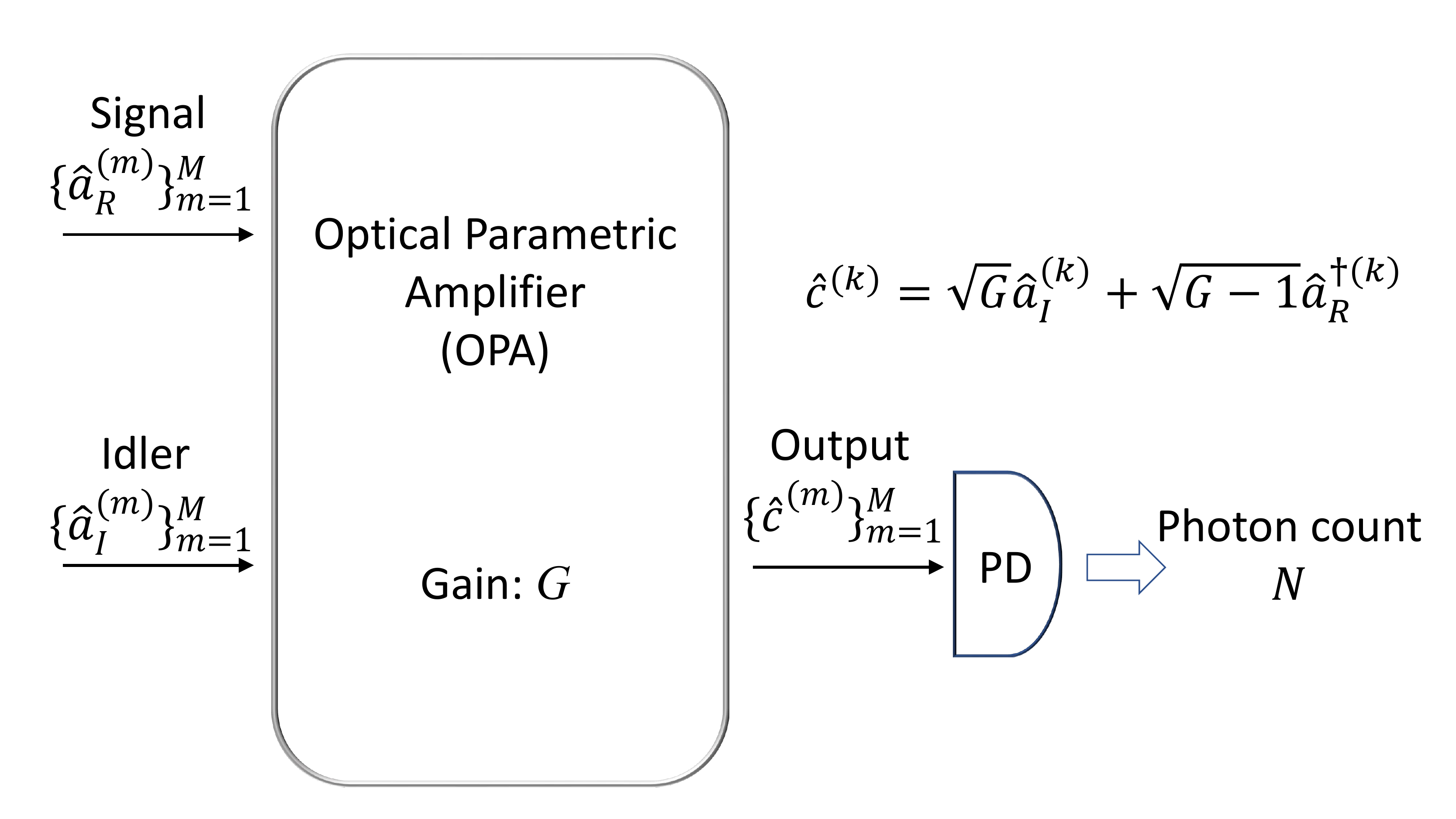}
    \caption{Schematic of the OPAR. $M$ i.i.d. return-idler pairs $\{\hat a_R^{(m)},\hat a_I^{(m)}\}_{m=1}^M$ are input to the receiver. An OPA component combines the idler and the conjugate of the signal, while
    only one of the output port is detected with photodetection. The total photon number over the $M$ pairs is collected. }
    \label{fig:schematicOPA}
\end{figure}

Fig.~\ref{fig:schematicOPA} shows the protocol of the OPAR. The OPAR applies parametric amplification across all the $M$ return-idler mode pairs $\{\hat a_{R}^{(m)},\hat a_{I}^{(m)}\}$ to recast the cross-correlations between the input modes into photon-number differences. The amplification produces $M$ output modes $\hat c^{(m)}=\sqrt{G} \hat a_{I}^{(m)}+\sqrt{G-1}\hat a_{R}^{\dagger(m)}, 1\le m\le M$. For two-mode Gaussian states with zero mean and covariance matrix specified by Eq.~\eqref{eq:CM_RI}, each output mode is in a thermal state with mean photon number 
\bal
\overline N(\theta,\kappa)\equiv & \expval{\hat c^{\dagger(m)} \hat c^{(m)}}\\
=&G N_S\!+\!(G\!-\!1)(\kappa N_S\!+\!N_B\!+\!1)\\
&+2\sqrt{G(G-1)\kappa N_S(1+N_S)}\cos\theta \,.
\label{eq:OPA_mean}
\eal
We collect the total photon number $\hat N=\sum_{m=1}^M \hat c^{\dagger(m)} \hat c^{(m)} $ across the $M$ modes. The probability mass function of the random-variable readout $N$ is~\cite{shi2020practical}
\begin{align}
&P_{N|\theta,\kappa}^{(M)}(n|\theta,\kappa)=
\nonumber
\\
&\!\binom {n\!\!+\!\!M\!\!-\!\!1}{n}\!\left(\frac{\overline N(\theta,\kappa)}{1\!+\!\overline N(\theta,\kappa)}\right)^{\!n}\!\!\left(\frac{1}{1\!+\!\overline N(\theta,\kappa)}\right)^{\!M},
\label{eq:Pn_OPA}
\end{align}
where $\binom{a}{b}$ is the binomial coefficient $a$ choose $b$.
Below we utilize the above measurement statistics to evaluate the performance of quantum illumination, phase sensing, and communication.

In quantum illumination scenario, the task is to discriminate between two channel hypotheses, $H_0:\Phi_{0,0}$ and $H_1:\Phi_{\kappa,0}$. When $M \gg 1$, due to the central limit theorem, Eq.~\eqref{eq:Pn_OPA} approximates to a Gaussian distribution, with mean and variance $\mu_0=G N_S+(G-1)(1+N_B)$, $\sigma_0^2 = \mu_0(\mu_0+1)$ for $H_0$, and $\mu_1 = G N_S+(G-1)(1+N_B+\kappa N_S)+2\sqrt{G(G-1)\kappa N_S(N_S+1)}$, $\sigma_1^2 = \mu_1(\mu_1+1)$ for $H_1$. One can make a near-optimum decision using a threshold detector that decides in favor of hypothesis $H_0$ if $N<N_{\rm th}$, and $H_1$ otherwise, with $N_{\rm th}\equiv\lceil M(\sigma_1\mu_0+\sigma_0\mu_1)/(\sigma_0+\sigma_1)\rceil$~\cite{Guha2009}. The error probability for target detection is
\be
P_{E,{\rm OPAR}} = \frac{1}{2}{\rm Erfc}\left(\sqrt{R_{\rm OPAR}^{\rm QI}M}\right)
\label{eq:pe_OPA}
\ee
where $R_{\rm OPAR}^{\rm QI} = (\mu_1 - \mu_0)^2/2(\sigma_0+\sigma_1)^2$. At the limit of $N_S\ll 1, \kappa\ll 1, N_B\gg 1$, $R_{\rm OPAR}^{\rm QI}\simeq \kappa N_S/(2N_B)$. Note that the exact optimal decision threshold is lengthy and only change the results slighly.

In the phase estimation scenario, the task is to estimate the parameter $\theta$ of quantum channel $\Phi_{\kappa,\theta}$. The Fisher information of OPAR is
\be 
\calF_{\rm OPAR} \equiv \sum_{n=0}^\infty\left(\partial_\theta\mathrm{log}P_{N|\theta,\kappa}^{(M)}(n|\theta,\kappa)\right)^2P_{N|\theta,\kappa}^{(M)}(n|\theta,\kappa)\,.
\ee 
Plugging in Eq.~\eqref{eq:Pn_OPA}, we find that the Fisher information depends on amplification gain $G$ as
\bal 
\calF_{\rm OPAR}(G)=\frac{4M(G-1)G\kappa N_S(1+N_S)\mathrm{sin}^2\theta}{\overline N(1+\overline N)}\,.
\label{eq:optG_OPAR}
\eal
We derive the optimal gain as
\be 
G_{\rm opt}^{\rm OPAR}\equiv {\rm argmax}_G \calF_{\rm OPAR}(G)=\max\{G^*,1\}
\ee 
where
\be 
G^*=1+\frac{\sqrt{N_S \left(N_S+1\right) \left(N_B'-1\right) N_B'}+N_S \left(N_S+1\right)}{ \left(N_B'-N_S-1\right) \left(N_B'+N_S\right)}\,,
\ee
and $N_B'\equiv N_B+\kappa  N_S+1$.
Here it is necessary to take the maximum between $G^*$ and $1$, because when $N_S>N_B/(1-\kappa)$, the optimum $G^*$ falls below 1, which is not physical. As a result, at the limit $N_S\ll 1$ we have $G^*=1+{\sqrt{N_S}}/{\sqrt{N_B(1+N_B)}}+O(N_S)$. In this regime, the optimum Fisher information is
$
\calF_{\rm OPAR} ={4M\kappa N_S\sin^2\theta}/{(1+N_B)} +O(N_S^{3/2})\,.
$



In the communication scenario, let us consider the BPSK modulation where $\theta\in\{0,\pi\}$ with equal probability $1/2$. Then the conditional statistics of Eq.~\eqref{eq:Pn_OPA} leads to the unconditional statistics $P_N^{(M)}(n)\equiv \sum_{\theta\in\{0,\pi\}} P_{N|\theta,\kappa}^{(M)}(n|\theta,\kappa)/2$. Using these two distributions, we obtain the Shannon information
\be 
I(N;\theta)=H(N)-H(N|\theta)\,,
\label{eq:Shannon_supp}
\ee
where 
\bal 
H(N|\theta)&=\!-\!\!\sum_{\theta\in\{0,\pi\}}\frac{1}{2}\sum_{n=0}^\infty P_{N|\theta,\kappa}^{(M)}(n|\theta,\kappa)\log_2 P_{N|\theta,\kappa}^{(M)}(n|\theta,\kappa)\,,\\
H(N)&=-\sum_{n=0}^\infty P_N^{(M)}(n)\log_2 P_N^{(M)}(n)\,.
\eal 
For simplicity of the description, in our simulation, we choose $M=1000$ to match the choice of PCR in quantum illumination where Gaussian approximation requires large $M$. Indeed, we find that it achieves a performance almost identical to the optimum choice of $M=1$. The optimality is due to the fact that data processing, e.g. summing over $M$ photon counts here, never increases Shannon information.

\subsubsection{Phase conjugate receiver (PCR)}

\begin{figure}[t]
    \centering
    \includegraphics[width=0.3\textwidth]{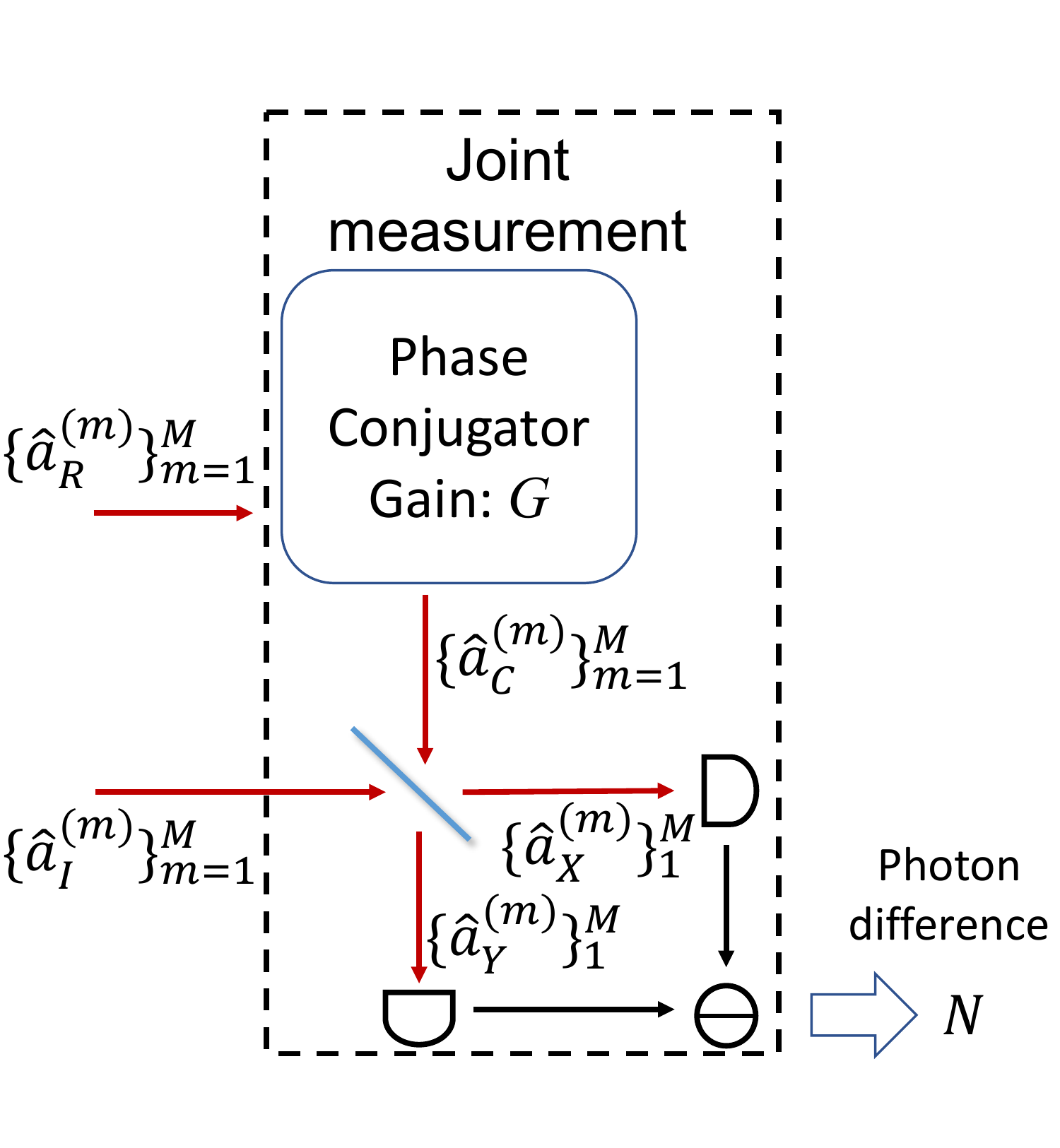}
    \caption{The schematic of PCR. $M$ i.i.d. correlated return-idler pairs $\{\hat a_R^{(m)},\hat a_I^{(m)}\}_{m=1}^M$ are input to the receiver. The receiver applies a joint measurement on the returned signals $\{\hat a_R^{(m)}\}$ and the idlers $\{\hat a_I^{(m)}\}$: first $\{\hat a_R^{(m)}\}$ are phase conjugated to produce $\{\hat a_C^{(m)}\}$, and then an interferometry is applied to $\{\hat a_C^{(m)}\}$ and $\{\hat a_I^{(m)}\}$ by a balanced beamsplitter, finally the total photon number difference over the $M$ pairs is collected. }
    \label{fig:schematic_PCR}
\end{figure}

Fig.~\ref{fig:schematic_PCR} shows the protocol of PCR. 
The inputs are $M$ i.i.d. return-idler pairs $\{\hat a_{R}^{(m)},\hat a_{I}^{(m)}\}$. A phase conjugator recasts $\{\hat a_{R}^{(m)}\}$ to their phase conjugates $\{\hat a_C^{(m)}\}$. Then the receiver recombines each phase conjugate $\hat a_C^{(m)}$ with the paired idler $\hat a_I^{(m)}$ on a $50:50$ beamsplitter. Finally, the total photon number difference $\hat N= \hat N_X-\hat N_Y$ is detected between the two arms $X,Y$ over the $M$ modes, where $\hat N_X=\sum_{m=1}^M\hat a_X^{(m)\dagger} \hat a_X^{(m)} $, $\hat N_Y=\sum_{m=1}^M\hat a_Y^{(m)\dagger} \hat a_Y^{(m)}$. The random readout $N$ is approximately a Gaussian random variable at the limit of $M\gg 1$ due to the central limit theorem, subject to the Gaussian probability density function
\be 
P_{N|\theta,\kappa}^{(M)}(n|\theta,\kappa)=\frac{1}{\sqrt{2\pi\sigma^2(\theta,\kappa)}}\exp{-\frac{\left(n-\mu\left(\theta,\kappa\right)\right)^2}{2\sigma^2\left(\theta,\kappa\right)}}
\label{eq:Pn_PCR}
\ee
with mean and variance
\bal 
\mu(\theta,\kappa)&=M\cdot 2C_{CI}\cos(\theta)\,,\\
\sigma^2(\theta,\kappa)&=M\cdot\left(N_I + 2 N_C N_I+  N_C+2 C_{CI}^2\cos(2\theta)\right)\,,
\label{eq:Pn_PCR_mean_var}
\eal 
where $N_C= ({G}-1)  ( \kappa N_S+{N_B}+1) , N_I= N_S$ and $C_{CI}=\sqrt{(G-1)\kappa N_S(1+N_S)} $. 


In the quantum illumination scenario, as a reminder, the task is to discriminate between two channel hypotheses, $H_0:\Phi_{0,0}$ and $H_1:\Phi_{\kappa,0}$. The mean and variance of the Gaussian statistics are different in the two hypotheses: $\mu_0=\mu(0,0)$, $\sigma_0^2=\sigma^2(0,0)$ for $H_0$, and $\mu_1=\mu(0,\kappa)$, $\sigma_1^2=\sigma^2(0,\kappa)$ for $H_1$. Using the near-optimum threshold detector with threshold $N_{\rm th}\equiv\lceil M(\sigma_1\mu_0+\sigma_0\mu_1)/(\sigma_0+\sigma_1)\rceil$~\cite{Guha2009}, the error probability for target detection is
\be
P_{E,{\rm PCR}} = \frac{1}{2}{\rm Erfc}\left(\sqrt{R_{\rm PCR}^{\rm QI}M}\right),
\label{eq:pe_PCR}
\ee
where $R_{\rm PCR}^{\rm QI} = \kappa N_S(N_S+1)/(2N_B+4N_SN_B+6N_S+4\kappa N_S^2+3\kappa N_S+2)$. At the $N_S\ll 1, \kappa\ll 1, N_B\gg 1$ limit, its performance $R_{\rm PCR}^{\rm QI}\simeq \kappa N_S/(2N_B)$ becomes identical to the OPAR. Away from the asymptotic parameter region, PCR typically has a slightly better performance than OPAR.

In the phase estimation scenario, one estimates the parameter $\theta$ of the channel $\Phi_{\theta,\kappa}$, the PCR yields Fisher information 
\begin{align}
\calF_{\rm PCR}(G)&=\!\int_{-\infty}^\infty \! {\rm d}n \left[\partial_\theta \ln\left(P_{N|\theta,\kappa}^{(M)}(n|\theta,\kappa)\right)\right]^2 \! P_{N|\theta,\kappa}^{(M)}(n|\theta,\kappa)
\nonumber
\\
&=\left[\partial_\theta \mu(\theta,\kappa) \right]^2/[\sigma^2(\theta,\kappa)/M].
\end{align} 
Substituting with Eq.~\eqref{eq:Pn_PCR_mean_var}, the Fisher information is dependent on the conjugator gain $G$ as
\be
\calF_{{\rm PCR}}(G)= M\cdot \frac{4 ({G}-1) \kappa {}  {} {N_S} ({N_S}+1)\sin^2\theta}{(N_I+N_C)+\left(2N_CN_I+2C_{CI}^2\cos(2\theta)\right)}.
\ee
It is easy to check that $\calF_{\rm PCR}(G)$ monotonically increases with $G$, while the gradient decays rapidly. As a result, one may regard the case of $G=2$ as almost saturating the large gain limit, and obtain a performance sufficiently close to the optimum
\begin{widetext}
\be 
\calF_{\rm PCR}(G=2)=\frac{4M \kappa  N_S \left(N_S+1\right)  \sin^2\theta }{ N_B(1+2N_S)+N_S \left(2\kappa  N_S+\kappa+3\right)+2 \kappa  \cos (2 \theta ) N_S \left(N_S+1\right)+1}.
\ee
\end{widetext}
In practice, the gain can be limited because the photon-photon interaction is intrinsically weak. When $N_S$ is sufficiently small, we can obtain a less stringent condition for $G$ to saturate the quantum advantage. Consider the weak gain limit $G-1\ll 1$, we have
\bal 
&\calF_{{\rm PCR}}(G)=\frac{4M   \kappa  {N_S} ({N_S}+1)\sin^2\theta} {1+N_B+N_S/(G-1)+O(N_S)}.
\label{eq:FI_PCR_asym}
\eal 
The term $N_S/(G-1)$ in the denominator will be negligible when
\be
 (G-1)(1+N_B) \gg N_S
\label{eq:gain_cond}
\,.\ee
Indeed, as long as this condition holds, the Fisher information of PCR reduces to the zero-order asymptotic formula $\calF_{{\rm PCR}}\simeq 4M\kappa N_S\sin^2\theta/(1+N_B)$, which saturates the optimum 3dB entanglement-assisted advantage over the classical coherent-state approach locally at $\theta=\pi/2$.

In the communication scenario, we consider the BPSK modulation such that $\theta\in\{0,\pi\}$ with equal probability $1/2$. Then the conditional statistics of Eq.~\eqref{eq:Pn_PCR} leads to the unconditional statistics $P_{N}^{(M)}(n)=\sum_{\theta\in\{0,\pi\}} P_{N|\theta,\kappa}^{(M)}(n|\theta,\kappa)/2$, and thereby the Shannon information is also given by  Eq.~\eqref{eq:Shannon_supp}, substituting $P_{N|\theta,\kappa}^{(M)}(n|\theta,\kappa)$ with Eq.~\eqref{eq:Pn_PCR}.
In the simulation, we choose $M=1000$ to validate the Gaussian approximation of Eq.~\eqref{eq:Pn_PCR}. Similar to the OPAR, we numerically find that the information rate does not decay significantly as $M$ increases up to 1000 in the parameter region of interest.

\section{Detailed analyses for correlation-to-displacement conversion}
\label{sec:details_C2D}

In this section, we provide detailed analyses for our results presented in the main text.

\subsection{Measurement statistics}
In the main text, we consider a pair of modes (denoted as `signal' and `idler' mode, corresponding to $K_A=K_B=1$ in Appendix~\ref{conditional_distribution}) in a TMSV state with mean photon number $N_S$, described by the covariance matrix Eq.~\eqref{eq:CM_SI}, resulting channel output with covariance matrix Eq.~\eqref{eq:CM_RI}.
By performing the heterodyne measurement on the returned mode, it is mapped to a coherent state with mean $\overline{\bm x}_\Pi\equiv (q_\Pi, p_\Pi)$ and identity covariance matrix, $V_{\Pi}=\mathbb{I}$. 
In general, for input state $\hat \rho_{RI}$ of a single pair of return and idler, conditioned on the heterodyne measurement result $(q_\Pi, p_\Pi)$, we produce an output state
\be 
\hat{\rho}_I= \frac{\expval{q_\Pi+i p_\Pi|\hat \rho_{RI}| q_\Pi+i p_\Pi}}{\tr \left[\expval{q_\Pi+i p_\Pi|\hat \rho_{RI}| q_\Pi+i p_\Pi}\right]},
\ee 
where $\ket{q_\Pi+i p_\Pi}$ is a coherent state with amplitude $q_\Pi+i p_\Pi$. Similar to Eq.~\eqref{projection_state}, we can express the input-output relation in characteristic function. Starting from the general two-mode input characteristic function
\begin{equation}
    \hat{\rho}_{RI} = \frac{1}{(2\pi)^2}\int_{\mathbb{R}^{4}}d\bm{z} \chi(\bm{z})\hat{D}(-\bm{z}),
\end{equation} 
we have the conditional state
\begin{align}
    \hat{\rho}_I &\propto \int {\rm d}\bm{z} \chi(\bm z)\bra{q_\Pi+i p_\Pi}\hat{D}(-\bm z)\ket{q_\Pi+i p_\Pi}
    \\
    &= \int {\rm d}\bm{z} \chi(\bm z_R,\bm z_I) \times e^{-\frac{1}{2}{\bm z}_R^T {\bm z}_R+i(\Omega \overline{\bm x}_\Pi)^T \bm z_R}\hat{D}\left(-{\bm z_I}\right).
\end{align}
Therefore the output has the characteristic function
\begin{align}
    &\chi_I(\bm \xi) = \int {\rm d}\bm{z}_R \chi(\bm z_R,\bm \xi) e^{-\frac{1}{2}{\bm z}_R^T {\bm z}_R-i(\Omega \overline{\bm x}_\Pi)^T \bm z_R}
\end{align}
For non-Gaussian input states, one needs to perform the integral to obtain the output characteristic function. For Gaussian states, we can make use of results in Appendix~\ref{conditional_distribution} to obtain analytical solution.

From Eqs.~\eqref{conditional_output}, the idler mode's covariance matrix and mean, and the distribution of measurement outcome are
\begin{subequations}
\begin{align}
V_I^\prime &= \left(2\frac{(1-\kappa+N_B)N_S}{\kappa N_S+N_B+1}+1\right)\mathbb{I}, \label{quadrature_conditional_output_CM}\\
\overline{\bm x}_I^\prime &= \frac{\sqrt{\kappa N_S(N_S+1)}}{\kappa N_S+N_B+1}\begin{pmatrix}
 \cos\theta q_\Pi+\sin\theta p_\Pi \\
 +\sin\theta q_\Pi-\cos\theta p_\Pi
\end{pmatrix}, \label{quadrature_conditional_output_mean} \\
p(\overline{\bm x}_\Pi) &= \frac{e^{-\frac{|\overline{\bm x}_\Pi|^2}{4(\kappa N_S+N_B+1)}}}{4\pi(\kappa N_S+N_B+1)}. \label{quadrature_conditional_output_p}
\end{align}
\end{subequations}
One can directly realize that the idler mode is in a displaced thermal state with mean $\overline{\bm x}^\prime_I$ and thermal photon number $E\equiv (1-\kappa+N_B)N_S/(\kappa N_S+N_B+1)$, as stated in the main text.
Formally, a displaced thermal state with complex mean $\alpha$ and mean thermal photon number $E$ is defined as
\be 
\hat\rho_{\alpha,E}\equiv \sum_{n=0}^\infty \hat D(\alpha)\frac{E^n}{(1+E)^{n+1}}\ketbra{n}{n} \hat D^\dagger(\alpha),
\label{eq:dts_app}
\ee 
where $\hat D(\alpha)=\exp\left(\alpha \hat{a}^\dagger-\alpha^\star \hat{a}\right)$ is the complex displacement operator acting on a mode $\hat{a}$ and $\ket{n}$ is a number state. Note that the comeplex displacement $\alpha=\expval{\hat a}=\expval{\hat q+i\hat p}/2$.

Using the distribution of quadratures Eq.~\eqref{quadrature_conditional_output_p}, we can obtain the distribution of the complex heterodyne readout on the $m$th returned mode
${\calM_m}=(q_{R_m}+ip_{R_m})/2$ as
\be 
p({\calM_m}) = \frac{e^{-\frac{|{\calM_m}|^2}{(\kappa N_S+N_B+1)}}}{\pi(\kappa N_S+N_B+1)}.
\label{eq:p_M}
\ee 
At the same time, the complex displacement of idler conditioned on the measurement result is 
\begin{align}
d_{m} &= \frac{\sqrt{\kappa N_S(N_S+1)}}{\kappa N_S+N_B+1}e^{i\theta}{\calM_m}^*
\end{align}
where $\calM_m^*$ denotes the complex conjugate.

Through the change of variables, one can write the total displacement amplitude square 
\be 
|d_T|^2\equiv \sum_{m=1}^M |d_m|^2 = \xi \sum_{i=1}^{2M} z_i^2,
\ee 
with 
\be 
\xi=\kappa N_S(N_S+1)/2(N_B+\kappa N_S+1)
\label{xi_definition}
\ee  
and $z_i \sim \calN(0,1)$ being a standard normal random variable, and thus we obtain the $\chi^2$ distribution of $|d_T|^2$
\be 
P_{\kappa}^{(M)}(x)\sim \frac{1}{\xi}\left(\frac{x}{\xi}\right)^{M-1}e^{-x/(2\xi)},
\label{p_overall}
\ee 
with mean $2M\xi$ and variance $4M\xi^2$, where 
$ 
\xi\equiv C_p^2/4v_\calM\,.
$

At the end of the section, we discuss the Gaussian approximation to the distribution Eq.~\eqref{p_overall} to enable a more efficient numerical simulation when $M \ge 10^7$. Note that the following approximation is not utilized in any proof in the paper. First, we define $Z \equiv \sum_{i=1}^{2M} z_i^2 \sim \chi^2(2M)$. By central limit theorem, at the limit of $M\gg 1$, $(Z-2M)/\sqrt{4M} \sim \calN(0,1)$ follows standard normal distribution. Therefore, when $M\gg1$, we can approximate the distribution 
\begin{equation}
    |d_T|^2 \sim \calN(2M\xi, 4M\xi^2)
\label{eq:dT_gaussian}
\end{equation}
as a Gaussian distribution with mean $2M\xi$ and variance $4M\xi^2$.

\subsection{Details on error probability analyses}
\label{app:LB_exp}

In this section, analyze the error probability in quantum illumination target detection, enabled by the conversion module. In particular, we obtain upper bound on the error probability limit $P_{{\rm C}\veryshortrightarrow{\rm D}}$ in Eq.~\eqref{eq:average_pe} of the main paper, then utilize the upper bound to obtain lower bound on the error exponent $r_{{\rm C}\veryshortrightarrow{\rm D}}$.

\begin{lemma}
\label{lemma:PCD_UB}
The error probability performance limit enabled by ${\rm C}\veryshortrightarrow{\rm D}$ conversion module in quantum illumination target detection 
\begin{align}
P_{{\rm C}\veryshortrightarrow{\rm D}}\le \frac{1}{2}\min_{s\in[0,1]} \left(1+\frac{4\xi}{\Lambda_{s}(1+2N_S)+\Lambda_{1-s}(1+2E)}\right)^{-M},
\label{PCD_UB}
\end{align}
where $\xi$ is defined in Eq.~\eqref{xi_definition} and we have defined the function
\be 
\Lambda_p(\nu) \equiv \frac{(\nu+1)^p+(\nu-1)^p}{(\nu+1)^p-(\nu-1)^p}.
\ee 
\end{lemma}

Before proving the lemma, we provide some discussions.
By choosing $s=1/2$, we can also obtain a slightly looser upper bound
\be 
P_{{\rm C}\veryshortrightarrow{\rm D}}\le \left(1+\frac{4\xi}{h(N_S)+h(E)}\right)^{-M},
\ee 
where $h(y)\equiv \Lambda_{1/2}(1+2y)=\left(\sqrt{y+1}+\sqrt{y}\right)^2$.

\begin{figure}[tbp]
    \centering
    \includegraphics[width=0.5\textwidth]{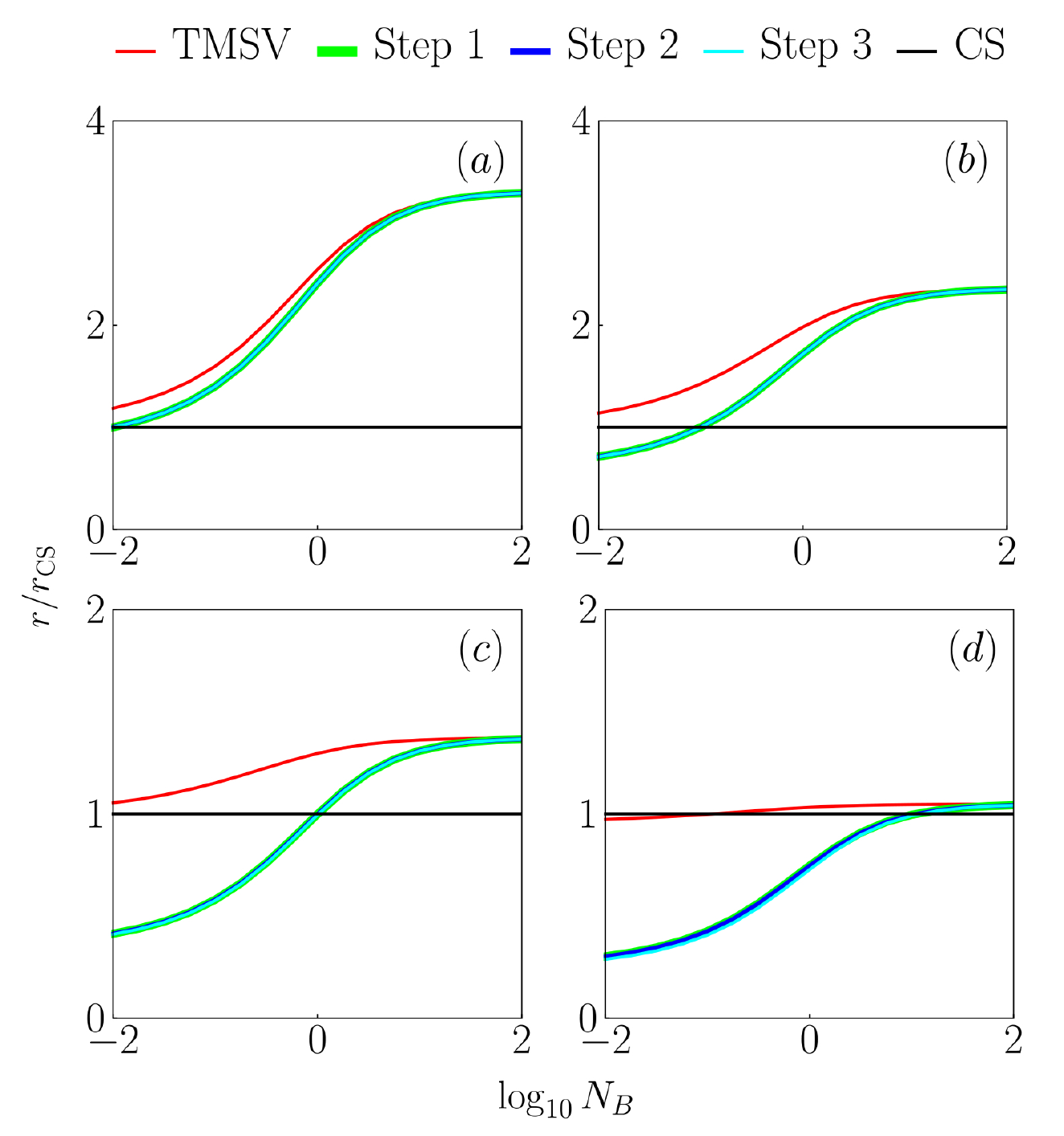}
    \caption{Comparison of the three lower bounds. Step 1, 2 and 3 corresponds to Eqs.~\eqref{eq:C11}, \eqref{eq:C12}, and \eqref{eq:C13} of the error exponent of correlation-to-displacement module and the QCB error exponent between the channel outputs using the TMSV source, under various $N_B$. (a-d) $N_S=0.01, 0.1,1,10$. Normalized by the coherent-state error exponent $r_{\rm CS}$. Channel transmissivity $\kappa=0.01$. 
    }
    \label{fig:QCB_Nb}
\end{figure}

When $\xi\ll1$, we can approximate the above upper bounds as exponential functions and obtain lower bounds on the error exponent.
So we obtain a lower bound of the conversion module
\begin{align}
&r_{{\rm C}\veryshortrightarrow{\rm D}}\ge
\max_{s\in[0,1]}\frac{4\xi}{\Lambda_{s}(1+2N_S)+\Lambda_{1-s}(1+2E)}
\label{eq:C11}
\\
&\ge 
\frac{2}{h(N_S)+h(E)}\frac{ {\kappa  N_S \left(N_S+1\right)}}{N_B+\kappa  N_S+1}
\label{eq:C12}
\\
&\ge 
(\sqrt{N_S+1}-\sqrt{N_S})^2\frac{ {\kappa  N_S \left(N_S+1\right)}}{N_B+\kappa  N_S+1},
\label{eq:C13}
\end{align} 
where in the last step we used the fact that the noise $E\le N_S$. A comparison of the above three lower bounds, normalized by the coherent-state Chernoff exponent
$ 
r_{\rm CS}=\kappa N_S(\sqrt{N_B+1}-\sqrt{N_B})^2
$~\cite{tan2008quantum}, is shown in Fig.~\ref{fig:QCB_Nb}. The lower bounds are shown to be always close to each other. 

From the above lower bound, we have
\be 
\frac{r_{{\rm C}\veryshortrightarrow{\rm D}} }{r_{\rm CS}}\ge 
\frac{(\sqrt{N_S+1}-\sqrt{N_S})^2}{(\sqrt{N_B+1}-\sqrt{N_B})^2} 
\frac{  \left(N_S+1\right)}{N_B+\kappa  N_S+1}.
\ee 
It is easy to see that when $\kappa\ll1$, the condition for advantage (${r_{{\rm C}\veryshortrightarrow{\rm D}} }/{r_{\rm CS}}>1$) is true as long as $N_S< N_B$.

Now we prove Lemma~\ref{lemma:PCD_UB}

\begin{proof}
To begin with, the Helstrom limit is upper bounded by the QCB for any number of copies of states~\cite{Audenaert2007,Pirandola2008}, therefore
\begin{align} 
&P_{\rm H}(\hat{\rho}_{0,N_S},\hat{\rho}_{\sqrt{x},E})
\\
&\le \frac{1}{2}\inf_{s\in[0,1]}Q_s\left(\hat{\rho}_{0,N_S},\hat{\rho}_{\sqrt{x},E}\right)
\\
&=\frac{1}{2}\inf_{s\in[0,1]}\overline{Q}_{s}\exp{-\frac{1}{2}{\bm d}^T\left(\tilde{V}_1(s)+\tilde{V}_2(1-s)\right)^{-1}{\bm d}}
\\
&=\frac{1}{2}\inf_{s\in[0,1]}\overline{Q}_{s}\exp{-\frac{2x}{\Lambda_{s}(1+2N_S)+\Lambda_{1-s}(1+2E)}},
\end{align}
where we utilize the definition of QCB in Appendix~\ref{app:QCB}, with ${\bm d}=(2\sqrt{x},0)$ and 
$
\tilde{V}_1(s)= \Lambda_{s}(1+2N_S)\mathbb{I}, \tilde{V}_2(s)= \Lambda_{s}(1+2E)\mathbb{I}.
$ 

Therefore,
\begin{align}
&P_{{\rm C}\veryshortrightarrow{\rm D}}=\int {\rm d}x P_{\kappa}^{(M)}(x) P_{\rm H}(\hat{\rho}_{0,N_S},\hat{\rho}_{\sqrt{x},E}) 
\\
&\le 
\int {\rm d}x P_{\kappa}^{(M)}(x)
\frac{1}{2}\inf_{s\in[0,1]}\left[\overline{Q}_{s}\right. \nonumber \\
&\left.\exp{-\frac{2x}{\Lambda_{s}(1+2N_S)+\Lambda_{1-s}(1+2E)}}\right]
\\
&\le \frac{1}{2}\inf_{s\in[0,1]} \left[\overline{Q}_{s}
\int {\rm d}x P_{\kappa}^{(M)}(x) \right. \nonumber\\
&\left.\exp{-\frac{2x}{\Lambda_{s}(1+2N_S)+\Lambda_{1-s}(1+2E)}}\right]
\\
&= \frac{1}{2}\inf_{s\in[0,1]} \overline{Q}_{s}\left(1+\frac{4\xi}{\Lambda_{s}(1+2N_S)+\Lambda_{1-s}(1+2E)}\right)^{-M}
\\
&\le 
\frac{1}{2}\min_{s\in[0,1]} \left(1+\frac{4\xi}{\Lambda_{s}(1+2N_S)+\Lambda_{1-s}(1+2E)}\right)^{-M}.
\end{align} 

\end{proof}

\begin{figure*}
    \centering
    \includegraphics[width=0.75\textwidth]{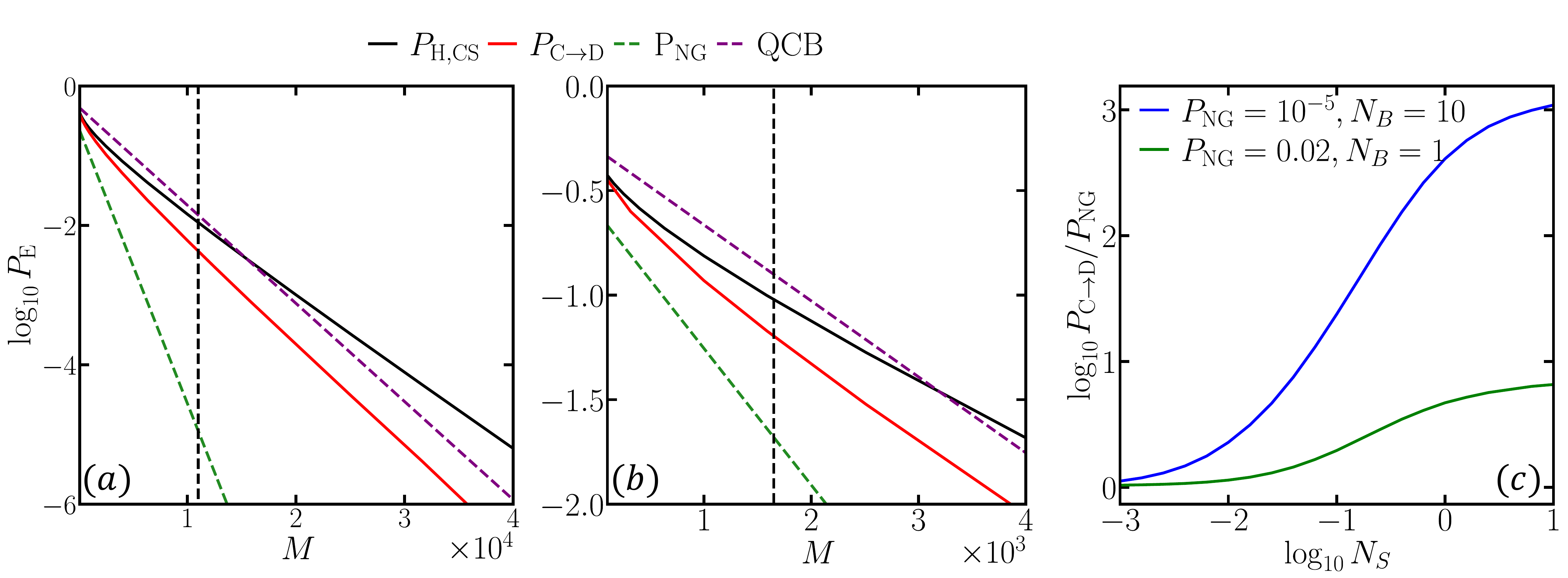}
    \caption{Quantum illumination error probability versus number of identical copies $M$ with (a) $N_S=1$, $N_B=10$ and (b) $N_S=0.3, N_B=1$. Vertical dashed lines indicate corresponding error probability with $P_{\rm NG}=10^{-5}$ and $P_{\rm NG}=0.02$ separately. (c) Error probability ratio $P_{{\rm C}\veryshortrightarrow {\rm D}}/P_{\rm NG}$ versus $N_S$ for $P_{\rm NG}$ and $N_B$ chosen from (a) and (b). In all cases $\kappa=0.01$.}
    \label{fig:illumination_largeN}
    \centering
    \includegraphics[width=0.75\textwidth]{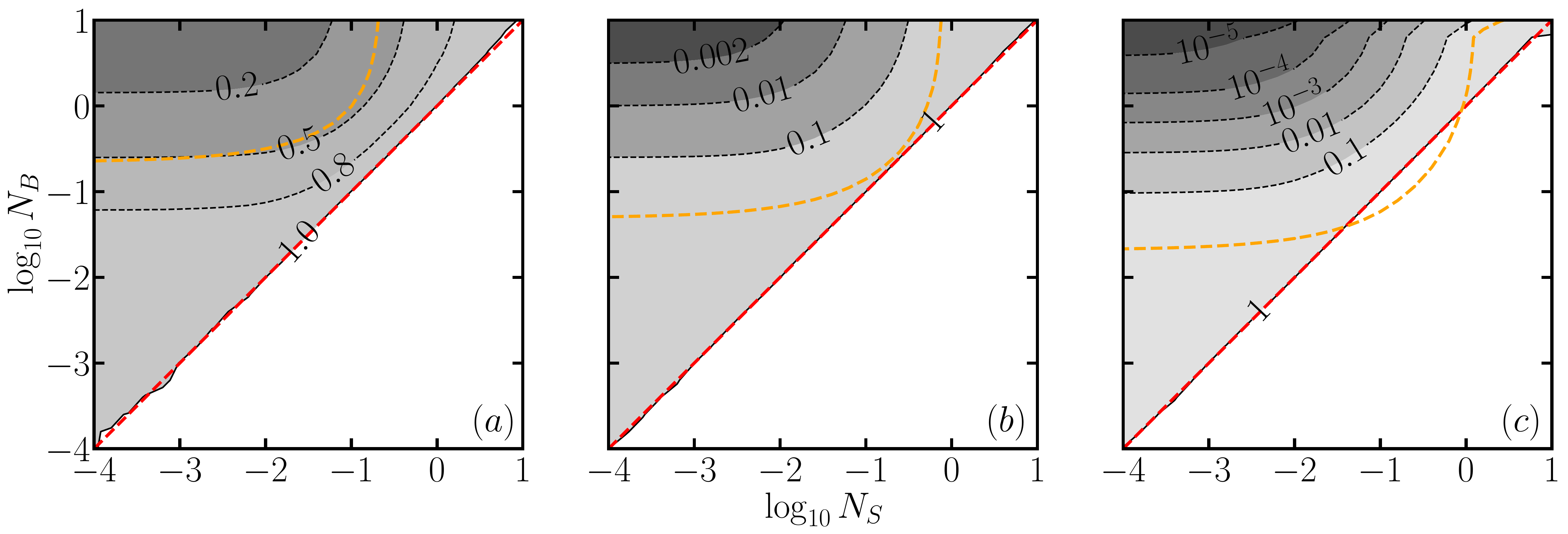}
    \caption{Quantum illumination error probability ratio $P_{{\rm C}\veryshortrightarrow {\rm D}}/P_{\rm H,CS}$ (plot in logarithmic scale) versus $N_S$, $N_B$ with $P_{\rm H,CS}=0.1, 0.01, 0.001$ (from left to right). Red dashed lines indicate the boundary of quantum advantage by the conversion module $N_S\le N_B$ and orange dashed curves represent the quantum advantage boundary by QCB. The discontinuity of curve in the upper right of (c) is due to the integer requirement of optimal $M$ to hold $P_{\rm H,CS}=0.001$. In all cases $\kappa=0.01$.}
    \label{fig:illumination_contours}
\end{figure*}

To complete the error probability analysis of our conversion module in the entire parameter region, in addition to the figures presented in the main text, we study the case where the parameters $N_S, N_B$ are both large but $N_S\le N_B$. We see that quantum advantage still exists in Fig.~\ref{fig:illumination_largeN}(a)-(b). In the non-asymptotic region, we also see that quantum advantage can only be revealed by our module, not the known QCB. Note when the brightness is large, there is a relatively large gap to the Nair-Gu bound in Eq.~\eqref{eq:nair_gu} of the main text, which is further confirmed in Fig.~\ref{fig:illumination_largeN}(c). 

In the main text, we show the error probability ratio with fixed $P_{\rm H,CS}=0.05$; here, we extend to $P_{\rm H, CS}=0.1, 0.01, 0.001$ in Fig.~\ref{fig:illumination_contours}(a)-(c) to cover both the non-asymptotic and asymptotic regions. With the conversion module, the error probability decreases with a smaller fixed $P_{\rm H,CS}$, indicating a larger quantum advantage. At the same time, the parameter region where quantum advantage can be predicted by QCB also increases. Note that there exist a region where $P_{\rm QCB}< P_{\rm H,CS}<P_{{\rm C}\veryshortrightarrow {\rm D}}$ and quantum advantage can only be revealed by QCB, as shown in Fig.~\ref{fig:illumination_contours}(c). The advantage can also be seen from Fig.~\ref{fig:QCB_Nb}(b),(c), when $N_B\lessapprox N_S$ , the error exponent lower bound of conversion module is smaller than the coherent state one, while the QCB error exponent is larger than it.

\subsection{Quantum Fisher information for phase sensing}
\label{sec:Fisher}

In this section, we evaluate the QFI for phase sensing, utilizing Eq.~\eqref{eq:QFI_Gaussian_supp}, where
the parameter is the signal phase shift $\theta$.

A displaced thermal state $\rho_{\sqrt{x}e^{i\theta},y}$ defined in Eq.~\eqref{eq:dts_app} has the mean and covariance matrix
\bal 
\bm d&=[\sqrt{x} e^{i\theta},\sqrt{x} e^{-i\theta}]^T,
\\
\Sigma&=\left(
\begin{array}{cc}
 0 &  y+1/2 \\
  y+1/2 & 0 \\
\end{array}
\right)\,.
\eal
Thus the QFI for phase sensing
\be 
\calF_{\rm DTS}=\frac{4x}{1+2y}\,.
\label{QFI_DTS}
\ee
Consider $M$ independent and identically distributed (i.i.d.) probes estimating the lossy channel $\Phi_{\kappa,\theta}$ (defined in Eq.~\eqref{input_output_main} of the main text) with thermal noise $N_B$, each with mean photon number $N_S$. For a classical protocol using coherent-state probes $\ket {\sqrt{N_S}}^{\otimes M}$, observe that the channel output $[\Phi_{\kappa,\theta}(\ketbra{\sqrt{N_S}}{\sqrt{N_S}})]^{\otimes M}$ is a product of displaced thermal states. Then the $M$ channel outputs can be combined into a single mode in a displaced thermal state by a balanced $M$-port beamsplitter. This processing does not change the QFI, because the beamsplitter transform is a unitary and the output is again a product state, where the additional noise modes can be discarded. The output state has $x=M\kappa N_S$, $y=N_B$, thus
\be 
\calF_{\rm CS}=\frac{4M\kappa N_S}{1+2N_B}\,.
\ee
Similarly, for an entanglement-assisted protocol using correlation-to-displacement ($\rm C\veryshortrightarrow D$) module with TMSV probes, the outputs at the idler ports are combined into a displaced thermal state. The random readouts of heterodyne detection at the signal ports determines the squared mean $x$ of the displaced thermal state to be in the $\chi^2$ distribution $P_\kappa^{(M)}(x)$ defined as Eq.~\eqref{p_overall}. Thus
\be 
\calF_{{\rm C}\veryshortrightarrow{\rm D}}\equiv\int {\rm d}x P_{\kappa}^{(M)}(x) \calF_\theta(\hat{\rho}_{e^{i\theta}\sqrt{x},E})=\frac{8M\xi}{1+2E}\,.
\ee 
where $\calF_\theta(\hat{\rho}_{e^{i\theta}\sqrt{x},E})=4x/(1+2E)$ is the QFI of the displaced thermal state conditioned on a specific $x$. Plugging the definitions of $\xi$, $E$ in the main text, we obtain
\be 
\calF_{{\rm C}\veryshortrightarrow{\rm D}}=\frac{4 M\kappa  N_S \left(N_S+1\right)}{1+N_B+N_S \left(2 N_B+2-\kappa \right)}.
\ee

At the neighborhood of true value, the QFI of a displaced thermal state is achieved by homodyne measurement. This can be seen as follows. For $\hat\rho_{\sqrt{x}e^{i\theta},y}$, suppose one first apply a phase rotation of angle $\theta_c$, the state becomes $\hat\rho_{\sqrt{x}e^{i(\theta+\theta_c)},y}$. Then we apply homodyne detection, giving the random readout $Q$ subject to the distribution
\be 
p_Q(q)=\frac{1}{\sqrt{2\pi\sigma^2}}\exp{-\frac{\left(Q-\sqrt{2}\Re\left(d\right)\right)^2}{2\sigma^2}}
\ee
where $\sigma^2=1/2+y$, $d=\sqrt{x}e^{i\theta}$.
Thus the Fisher information of homodyne measurement, depending on a phase compensation $\theta_c$, can be calculated from the distribution as
\be 
\calF_{\rm hom}(x,E,\theta)=\frac{4x}{1+2y}\sin^2(\theta+\theta_c)\,.
\ee
Now we see that homodyne measurement achieves the QFI in Eq.~\eqref{QFI_DTS} locally, which is true only when the prior knowledge is sufficient such that $\theta+\theta_c$ is close to $\pi/2$, while its performance decays rapidly when $\theta_c$ deviates from the ideal compensation $\pi/2-\theta$. When prior knowledge is insufficient, an adaptive policy can be designed to approach the ideal compensation, as the number of available probes is sufficiently large. 
\begin{figure}[t]
    \centering
    \includegraphics[width=0.25\textwidth]{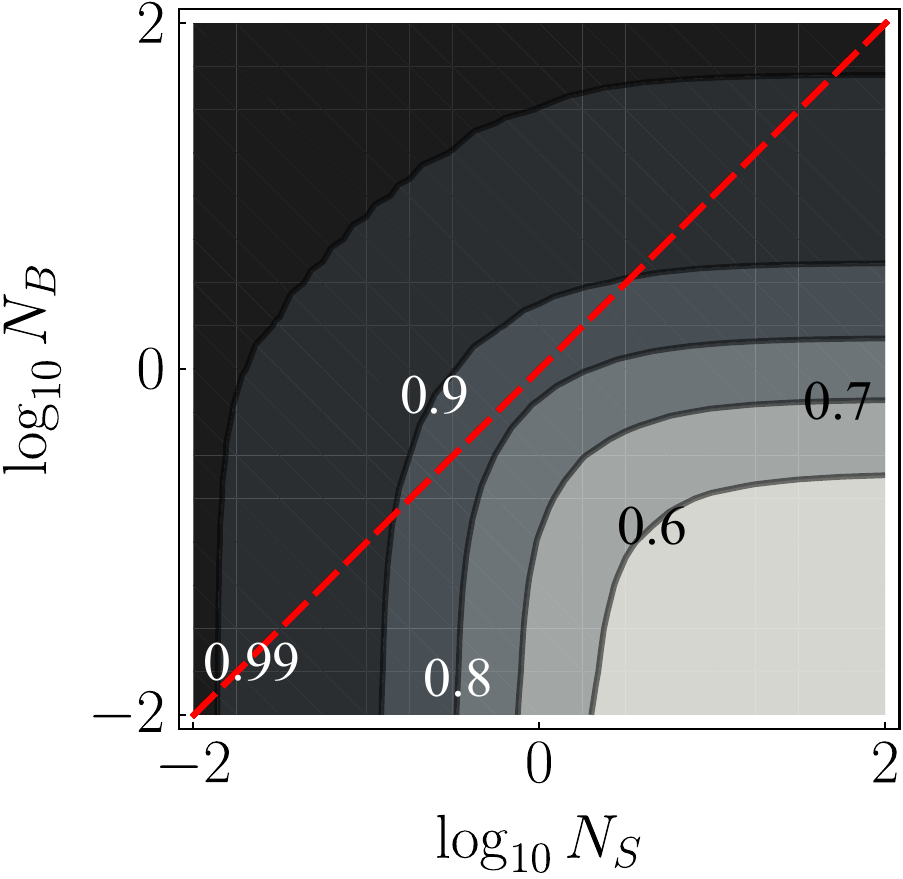}
     \caption{The ratio in quantum Fisher information of the correlation-to-displacement conversion module over the TMSV limit under various $N_S,N_B$. The red dashed diagonal line indicates the boundary of quantum-enhanced region $N_S=N_B/(1-\kappa)$. $\kappa=0.01$. }
    \label{fig:qfi_limits_NbNs_TMSV}
\end{figure}

It is worthwhile to note that the $\rm C\veryshortrightarrow D$ module is optimal for TMSV-based phase estimation: it achieves the QFI of the channel output of TMSV sources \cite{shi2020practical}
\be 
\calF_{\rm TMSV}=\frac{4 M\kappa  N_S \left(N_S+1\right)}{1+N_B(1+ 2N_S)+N_S(1-\kappa)} 
\label{eq:QFI_TMSV_supp}
\ee
at the limit of $N_B\gg 1$. It is verified in the numerical evaluation as shown in Fig.~\ref{fig:qfi_limits_NbNs_TMSV}.

\subsection{Entanglement-assisted communication rate analyses}
\label{sec:holevo}

We evaluate the Holevo information~\cite{holevo1973bounds,wilde2013quantum} of the output ensemble of our $\rm C\veryshortrightarrow D$ module, using phase-encoded TMSV source. The Holevo information is a tight upper bound on the information rate of a channel given a specific encoding ensemble $\{p_\theta,\hat\rho_\theta\}$, which is achieved by the optimum receiver. The ultimate capacity can be obtained by optimizing the Holevo information over $\{p_\theta,\hat \rho_\theta\}$. In general $\theta$ can be an arbitrary parameter, while we specify it to be the phase shift in this paper. We consider repetition coding that yields $M$ i.i.d. copies of the output ensemble. Given $\theta$, the output state is $\hat\rho_{\sqrt{X}e^{i\theta},E}$ defined by Eq.~\eqref{eq:dts_app}, where $X$ is a random readout under $\chi^2$ distribution $P_{\kappa}^{(M)}(x)$ defined in Eq.~\eqref{p_overall}. Let the encoding phase be a random variable $\Theta$ subject to probability distribution $P_\Theta$. Denote the output quantum system as $O$. In the communication protocol, the readouts $X$, the output quantum system $O$ along with the input symbol $\Theta$ is in a classical-quantum state
\bal 
&\hat{\sigma}_{XO\Theta}=\\
&\int \! {\rm d}\theta p_\Theta(\theta) \! \int  \! {\rm d}x P_{\kappa}^{(M)}(x) \state{x}_X \otimes \left(\hat{\rho}_{\sqrt{x}e^{i\theta},E}\right)_O\otimes \ketbra{\theta}{\theta}_\Theta\,.
\eal 
The overall Holevo information about the input symbol is 
\bal 
\chi_{\rm C\veryshortrightarrow D}&\equiv \frac{1}{M} [S(XO)_{\hat \sigma}-S(XO|\Theta)_{\hat \sigma}]\\
&= \frac{1}{M}\!\int \! {\rm d}x p_X(x) \left[S(O|X=x)_{\hat \sigma} \! -\! S(O|\Theta,X=x)_{\hat \sigma}\right]\\
&= \frac{1}{M}\int {\rm d}x P_{\kappa}^{(M)}(x) \chi\left(\{p_\Theta,\hat{\rho}_{\sqrt{x}e^{i\Theta},E}\}\right)\,.
\label{eq:overallHolevo_supp}
\eal
The second equality is due to the joint entropy theorem \cite{nielsen2002quantum} given the orthogonality of $\{\ket{x}_X\}$. Here $\chi\left(\{p_\Theta,\hat{\rho}_{\sqrt{x}e^{i\Theta},E}\}\right)$ can be efficiently evaluated in the following example.

Consider continuous PSK (CPSK) modulation on TMSV sources with $P_\Theta(\theta)=1/2\pi, \theta\in [0,2\pi)$. The output ensemble of the $\rm C\veryshortrightarrow D$ module yields the Holevo information
\bal
&\chi\left(\{P_\Theta,\hat{\rho}_{\sqrt{x}e^{i\Theta},E}\}\right)
\\
&=S(\int {\rm d}\theta P_\Theta(\theta)\hat{\rho}_{\sqrt{x}e^{i\theta},E})-\int {\rm d}\theta P_\Theta(\theta)S(\hat{\rho}_{\sqrt{x}e^{i\theta},E})
\\
&=H\left[\{P(n|X=x)\}\right]-g(E).
\label{eq:almost_closed_form_supp}
\eal 
The third line follows from the following. The conditional states $\{\hat{\rho}_{\sqrt{x}e^{i\theta},E}\}$ are Gaussian states with identical entropy,
$
S(\hat{\rho}_{\sqrt{x}e^{i\theta},E})=g(E),
$ 
where $g(n)=(n+1)\log_2(n+1)-n \log_2 n$ is the entropy of a thermal state with mean photon number $n$. Thus
$
\int {\rm d}\theta P_\Theta(\theta)S(\hat{\rho}_{\sqrt{x}e^{i\theta},E})=g(E).
$
Meanwhile, the unconditional state $\int {\rm d}\theta P_\Theta(\theta)\hat{\rho}_{\sqrt{x}e^{i\theta},E}$ is completely dephased due to $[0,2\pi)$ uniform phase encoding thus its eigenbasis is the photon number Fock basis. Its distribution on the Fock basis is
\be 
P(n|X=x)=E^n (E+1)^{-n-1} e^{-\frac{x}{E+1}} L_n\left(-\frac{x}{E (E+1)}\right),
\ee 
where $L_n(x)$ is the $n$th Laguerre polynomial. 
Thus the unconditional entropy reduces to the Shannon entropy of the photon number distribution
\be 
H\left[\{P(n|X=x)\}\right]\equiv-\sum_{n=0}^\infty P(n|X=x)\log \left(P(n|X=x)\right).
\label{eq:S_app}
\ee
Combining Eqs.~\eqref{eq:overallHolevo_supp} and~\eqref{eq:almost_closed_form_supp}, we have the Holevo information for CPSK
\be 
\chi_{\rm C\veryshortrightarrow D}^{\rm CPSK}=\frac{1}{M}\left[\int {\rm d}x P_{\kappa}^{(M)}(x) H\left[\{P(n|X=x)\}\right]-g(E)\right].
\label{chi_CD_exact}
\ee 
We can adopt Eq.~\eqref{chi_CD_exact} for efficient numerical evaluation. Below, we further obtain some asymptotic results.

At the limit $M\to \infty$, $x$ converges to $2M\xi$ with probability by law of large numbers. Then we have a closed-form formula
\bal 
&\chi_{{\rm C}\veryshortrightarrow {\rm D}}^{\rm CPSK}=\frac{1}{M}\left[H\left[\{P(n|X=2M\xi)\}\right]-g(E)\right]\\
&=
\frac{\kappa N_S \left[ \ln \left(\frac{1}{N_S }\right)+\calR_{\rm C\veryshortrightarrow D}\right]}{ \left(N_B+1\right)\ln 2}+O(N_S^2)\\
   &=\frac{\kappa  N_S \ln(1/N_S)}{(N_B+1)\ln 2}+O(N_S)\,,
\label{eq:chi_CPSK_supp}
\eal 
where $\calR_{\rm C\veryshortrightarrow D}= \frac{2 \left(-N_B+\kappa -1\right) }{\kappa M}\tanh ^{-1}\left(\frac{\kappa  M}{2 N_B-2 \kappa +\kappa  M+2}\right)+\left(\ln
   \left(N_B+1\right)+1\right)-\ln\left(N_B+\kappa  (M-1)+1\right)$ is independent on $N_S$. In the last two lines, we are expanding at the $N_S\to 0$ limit. Here $|O(x)|/x<\infty $ as $x\to 0$.

Note that the above scaling at $N_S\to 0$ saturates the EA classical capacity in Eq.~\eqref{CE_expansion}, therefore is asymptotically optimal. At the same time, the information per symbol is strictly higher than the case of $M\gg1$, therefore the optimal scaling applies to any finite $M$.\\

One may follow a similar route to solve the binary PSK (BPSK) case, where $P_\Theta(0)=P_\Theta(\pi)=1/2$. The conditional entropy is the same as that in the CPSK, 
$
S(\hat{\rho}_{\sqrt{x}e^{i\Theta},E})=g(E)
$. 
The evaluation of the unconditional entropy is more challenging: it is now a Von Neumann entropy where eigenvalues of the density operator $\hat{\overline{ \rho}}=\int {\rm d}\theta P_\Theta(\theta)\hat{\rho}_{\sqrt{x}e^{i\theta},E}$ are to be solved. Nevertheless, we find that a closed-form formula is still available at the limit of $M\gg1$ (such that $x\to 2M\xi$) and $N_S\to 0$. Indeed, the performance of BPSK is almost identical to the CPSK case in the parameter region of Fig.~\ref{fig:EACOMM_c2d_limits} in the main text. Below, we approximate the eigenvalues via matrix perturbation theory. We consider the representation in Fock basis 
\begin{widetext}
\bal 
&\rho_{mn}=\expval{m|\hat{\overline{ \rho}}|n}
=\frac{\sqrt{m!}  E^n \left| x\right|^{(m-n)/2} 
      ~_1\tilde{F}_1\left(m+1;m-n+1;\frac{x}{E^2+E}\right)
   e^{-\frac{x}{E}+i \theta (m-n)}}{(E+1)^{m+1}\sqrt{n!}},
\eal
\end{widetext}
where $_1\tilde{F}_1(a;b;z)$ is the regularized confluent hypergeometric function. In the numerical evaluation, we truncate $\rho$ in finite dimension $d\times d$.

In the final approximation of eigenvalues, we expect to keep the infinitesimal terms up to $O(N_S)$. We find that $d=3$ is sufficient, the error analysis is deferred to the end of this section. Now, we apply Taylor expansion to each matrix entry $\rho_{mn}$ as 
\be 
\rho_{mn}=\tilde\rho_{mn}+\delta\rho_{mn}\,,
\ee
where the approximation $\tilde \rho_{mn}\sim O(N_S^{d})$ omits higher order term $\delta\rho_{mn} \sim O(N_S^{d+1})$. With $d=3$, the eigenvalues of $\tilde \rho$ can be solved analytically, and thus we obtain the Holevo information
\bal 
&\chi_{{\rm C}\veryshortrightarrow {\rm D}}^{\rm BPSK}=\frac{1}{M}\left[S( \rho)-g(E)\right]+O(\delta_d)\\
=&
\frac{\kappa N_S \left[ \ln \left(\frac{1}{N_S }\right)+\calR_{\rm C\veryshortrightarrow D}\right]}{ \left(N_B+1\right)\ln 2}+O(\delta_{N_S})+O(\delta_d)\\
=&\frac{\kappa  N_S \ln(1/N_S)}{(N_B+1)\ln 2}+O(N_S)+O(\delta_{N_S})+O(\delta_d)\,,
\label{eq:chi_BPSK_supp}
\eal 
where $\delta_d$ is the maximal error in eigenvalues from the matrix truncation and $\delta_{N_S}$ is from the matrix Taylor expansion, the residue $\calR_{\rm C\veryshortrightarrow D}$ is the same as that defined below the CPSK case Eq.~\eqref{eq:chi_CPSK_supp}.
The leading term of the matrix Taylor expansion at the second equality coincides with Eq.~\eqref{eq:chi_CPSK_supp}.

At last, we analyze the errors $\delta_d, \delta_{N_S}$ in Eq.~\eqref{eq:chi_BPSK_supp}. First, let us consider $\delta_d$. Define the true eigenvalues of the operator $\hat {\overline{ \rho}}$ as $\mu_1\ge \mu_2\ge \ldots $ and the eigenvalues of the truncated representation $\rho$ as $\lambda_1\ge \lambda_2\ge \ldots\ge \lambda_d$. Then $\delta_d \equiv \max_{i\in [d]}|\mu_i-\lambda_i|$. Using Theorem 4.14 in Ref.~\cite{stewart1990matrix}, we have
\be 
|\delta_d| =||{\rm diag} \left(\mu_i-\lambda_i\right)||_2\le  || X||_2\,,
\ee 
where $ X= \rho^{(\infty)}  I_d- I_d  \rho$, $ I_d$ is a $d\times d$ matrix representation of projector that implements the cutoff, $ \rho^{(\infty)}$ is the exact infinite-dimensional matrix representation of the operator $\hat {\overline{\rho}}$. Here the matrix 2-norm is defined using vector 2-norm: for $d\times d$ matrix $A\in \mathbb R^{d\times d}$, $||A||_2\equiv \sup_{x\neq 0} ||Ax||_2/||x||_2, \forall x\in \mathbb R^{ d}$. 
Observe that the 2-norm $|| X||_2=O(N_S^{d/2})$. Thus $d=3$ is sufficient to suppress the error to $O(N_S^{3/2})$.
Next, we consider $\delta_{N_S}$. Define the eigenvalues of ${\rho}$ as $\{\lambda_i\}_{i=1}^{d}$, and the eigenvalues of ${\tilde{\rho}}$ as $\{\tilde \lambda_i\}_{i=1}^{d}$. The error in eigenvalues is equal to the Hausdorff distance $|\delta_{N_S}|\equiv\max_i |\tilde\lambda_i-\lambda_i|= hd(\rho,{\tilde{\rho}})$, when the perturbation is small such that the eigenvalues are still pairwise matched: $j=\argmin_{j'}|\tilde \lambda_{j'}-\lambda_j|$.
Note that $||\delta{\rho}||_2=O(N_S^{d+1})$. According to Elsner's theorem\cite{stewart1990matrix}, the error is upper bounded by
\be 
|\delta_{N_S}|\le (||{\rho}||_2+||\tilde{{\rho}}||_2)^{1-1/d}||\delta {\rho}||_2^{1/d}=O(N_S^{1+1/d})\,,
\ee
which is much smaller than $O(N_S)$.
Finally, we see that the overall error $|\delta_d|+|\delta_{N_S}|\ll O(N_S)$ when $d=3$, $N_S\to 0$. 


\subsection{Proof of Theorem 1 of channel pattern classification in the main text on error exponent}
\label{general_pattern}

The error exponent of multiple-hypothesis testing is given by the worst case of binary hypothesis testing between any of the two hypotheses involved~\cite{li2016discriminating,nussbaum2011asymptotic,Audenaert2007}.
More precisely, this comes from two inequalities that are true for any number $n\ge 1$ of iid states~\cite{li2016discriminating}
\begin{align}
&P_H(\{p_1\hat \rho_1^{\otimes n},\cdots, p_r \hat\rho_r^{\otimes n}\})
\le 10(r-1)^2 C_r^2 (n+1)^{2d}
\nonumber
\\
& \quad \times \max\{p_1,\cdots,p_r\}
\max_{i,j}\left({\rm inf}_{s\in[0,1]}Q_s\left(\hat{\rho}_i,\hat{\rho}_j\right)\right)^n,
\label{PH_precise_multi}
\end{align}
which is asymptotically tight in the error exponent.

\subsubsection{Classical performance}
Due to the convexity
of the Helstrom limit and the quantum Chernoff bound (see ref~\cite{zhuang2021quantum} supplemental materials), the optimal classical strategy is to utilize a product of coherent state $\hat{\rho}_{\rm C}=\otimes_{m=1}^M\ket{\alpha_m}$ as the probe input, which leads to the output of a product of displaced thermal states
$
\otimes_{m=1}^M \hat{\rho}_{\exp\left(i\theta^{(h)}_m\right)\sqrt{\kappa^{(h)}_m}\alpha_m, N_B}.
$ 
When one has $n\gg1$ copies of the coherent states, the iid nature of the output states allows us to focus on the quantum channel discrimination (QCD) between the worst pair of two channels $\Phi_{\bm \kappa^{(h)}, \bm \theta^{(h)}}$, with $h=1,2$. For the two displaced thermal state, we have the mean
\begin{align}
    \overline{\bm x}_h &= \left(q_1^{(h)},p_1^{(h)},\dots,q_M^{(h)},p_M^{(h)}\right)^T,
\end{align}
where $q_m^{(h)}=2\Re{e^{i\theta_m^{(h)}}\sqrt{\kappa_m^{(h)}}\alpha_m}$ and $p_m^{(h)}=2\Im{e^{i\theta_m^{(h)}}\sqrt{\kappa_m^{(h)}}\alpha_m}$. The covariance matrix is diagonal $B\mathbb{I}$ where $B\equiv 2N_B+1 $. Now we evaluate the quantum Chernoff bound according to Appendix~\ref{app:QCB}. First, the quantity
\begin{equation}
\begin{split}
    Q_s &= \overline{Q}_se^{-\frac{1}{2}{\bm d}^T[\tilde{V}_1(s)+\tilde{V}_2(1-s)]^{-1}{\bm d}}\\
    &= \overline{Q}_s e^{-\frac{1}{2(\Lambda_s(B)+\Lambda_{1-s}(B))}{\bm d}^T{\bm d}}\\
    &= \overline{Q}_s e^{-\frac{1}{2(\Lambda_s(B)+\Lambda_{1-s}(B))}\sum_{m=1}^M (q_m^{(1)}-q_m^{(2)})^2+(p_m^{(1)}-p_m^{(2)})^2}\\
    &= \overline{Q}_s e^{-\frac{2}{\Lambda_s(B)+\Lambda_{1-s}(B)}\sum_{m=1}^M |e^{i\theta_m^{(1)}}\sqrt{\kappa_m^{(1)}}\alpha_m-e^{i\theta_m^{(2)}}\sqrt{\kappa_m^{(2)}}\alpha_m|^2}\\
    &= \overline{Q}_s e^{-\frac{2}{\Lambda_s(B)+\Lambda_{1-s}(B)}\sum_{m=1}^M \delta_m|\alpha_m|^2}
    \label{eq:Qs}
\end{split}
\end{equation}
where we define $\delta_m \equiv|e^{i\theta^{(1)}_m}\sqrt{\kappa^{(1)}_m}-e^{i\theta^{(2)}_m}\sqrt{\kappa^{(2)}_m}|^2$ and
\begin{equation}
    \overline{Q}_s = \left(\frac{ G_s(B)G_{1-s}(B)}{\Lambda_s(B)+\Lambda_{1-s}(B)}\right)^M = \frac{1}{2^M}.
    \label{eq:Qs_2}
\end{equation}
Therefore, the quantum Chernoff bound is
\begin{equation}
\begin{split}
    P_{\rm QCB}^{\rm QCD} &= \frac{1}{2}{\rm inf}_{s\in[0,1]}Q_s\\
    & = \frac{1}{2}Q_{1/2}\\
    &= \frac{\overline{Q}_{1/2}}{2}\exp\left[-\frac{1}{\Lambda_{1/2}(B)}\sum_{m=1}^M \delta_m|\alpha_m|^2\right]\\
    &= \frac{G_{1/2}(B)^{2M}}{2\Lambda_{1/2}(B)^M}\exp\left[-\frac{1}{\Lambda_{1/2}(B)}\sum_{m=1}^M \delta_m|\alpha_m|^2\right]\\
    &= \frac{1}{2}\exp\left[-\sum_{m=1}^M \delta_m|\alpha_m|^2(\sqrt{N_B+1}-\sqrt{N_B})^2\right], \label{eq:channel_error}
\end{split}
\end{equation}
where in the second line we utilize the fact that the minimum of $Q_s$ takes place at $s=1/2$. This is because $\overline{Q_s}$ in Eq.~\eqref{eq:Qs_2} is independent on $s$ and the exponent $1/(\Lambda_s(B)+\Lambda_{1-s}(B))$ in Eq.~\eqref{eq:Qs} is symmetric on $s$ and strictly concave as its second order derivative is negative
\begin{align}
    &\partial_s^2 \left(\frac{1}{\Lambda_s(B)+\Lambda_{1-s}(B)}\right)\nonumber \\ 
    &= -\frac{1}{4}\frac{(B+1)^{2s-1}+ (B-1)^{2s-1}}{(B^2-1)^{s-1}}\log ^2\left(\frac{B-1}{B+1}\right) < 0
\end{align}
due to $B > 1$. Therefore $\max_{s\in[0,1]}\{1/(\Lambda_s(B)+\Lambda_{1-s}(B))\} = 1/2\Lambda_{1/2}(B)$ where ${\rm inf}_{s\in[0,1]}Q_s = Q_{1/2}$.

To conclude, we solve the error exponent between the discrimination of any two channels via the coherent state input as
\begin{align}
    P_{\rm CS}^{\rm QCD} &\sim \exp\left[-\sum_m \delta_m |\alpha_m|^2\left(\sqrt{N_B+1}-\sqrt{N_B}\right)^2\right] \nonumber \\ 
    &\sim \exp\left[-\sum_m \delta_m |\alpha_m|^2/4N_B\right]
    \label{CS:exponent_pattern}
\end{align}
where we approximate $\left(\sqrt{N_B+1}-\sqrt{N_B}\right)^2 = N_B(\sqrt{1+1/N_B}-1)^2 = N_B(1/2N_B+O(N_B^2))^2\simeq 1/4N_B$ at the $N_B\gg1$ limit.

Note that the above exponent is tight asymptotically, when one consider the iid copies of the input coherent state $\hat{\rho}_{\rm C}^{\otimes n}$.

\subsubsection{Entangled performance}
For the entangled strategy, one inputs a product of TMSV, each mode pair with mean photon number $N_{S,m}=|\alpha_m|^2$ matching that of the classical input. For each channel, we have $n$ identical copies of TMSV. Combining the $M$ channels, the overall output state has the identical copy form of $\{\rho_h^{\otimes n}\}_h$, where each $\rho_h$ has $M$ modes. Once we apply the conversion module to the output state, one caveat is that the iid structure of output states is no longer preserved as the measurement outcome is random in different copies. The way out of such a dilemma is as the following: 
when provided with $n$ copies of output states $\{\rho_h^{\otimes n}\}_h$ from the quantum channel, one applies the conversion module on each copy, and produce product of displaced thermal states $\{\hat\sigma_h\}_h$, where each state is $nM$ modes and
\be 
\hat \sigma_h=\otimes_{m=1}^M \left[\otimes_{\ell=1}^n \hat{\rho}_{\zeta_m^{(h)} e^{ i \theta_m^{(h)} }{\calM_{m,\ell}}^{*},E_m^{(h)}}\right]
\label{sigma_h}
\ee   
is conditioned on the measurement results $\{\calM_{m,\ell}, 1\le \ell\le n, 1\le m \le M\}$. The constant
\begin{align}
\zeta_m^{(h)}\equiv   \frac{\sqrt{\kappa_m^{(h)}  N_{S,m} \left(N_{S,m}+1\right)}}{N_B+\kappa_m^{(h)}  N_{S,m}+1},
\end{align} 
and mean thermal photon number
\be 
E_m^{(h)}=\frac{N_{S,m} \left(N_B-\kappa_m^{(h)} +1\right)}{ N_B+\kappa_m^{(h)}  N_{S,m}+1}.
\ee
The measurement result ${\calM_{m,\ell}}=(q_{\calM_{m,\ell}}+ip_{\calM_{m,\ell}})/2$, with each quadrature output obeying a zero-mean Gaussian distribution with variance 
$
(N_B + \kappa_m^{(h)} N_S+1)/2\simeq N_B/2
$.
We consider the $N_B\gg1$ limit, then $E_m^{(h)}\simeq N_{S,m}\simeq N_S$ is a constant noise background.

Because the conversion module is a quantum process, we have the Helstrom limit of the entangled input case
\begin{align}
&P_H(\{p_h\rho_h^{\otimes n}\})
\le 
\mathbb{E}[P_H(\{p_h\sigma_h\})]
\le 10(r-1)^2 C_r^2 2^{2d}
\nonumber
\\
& \quad \times \max\{p_1,\cdots,p_r\}
\mathbb{E}\max_{i,j}
\left[{\rm inf}_{s\in[0,1]}Q_s\left(\hat{\sigma}_i,\hat{\sigma}_j\right)\right],
\label{PH_UB}
\end{align}
where the expectation value is over the measurement statistics. In the last step, we applied Ineq.~\eqref{PH_precise_multi} with the single-copy case and provide an upper bound on the error probability directly. Then we can still reduce the calculation to the Chernoff exponents for the displaced thermal states. However, now each state $\hat\sigma_h$ has $nM$ modes and are dependent on the measurement result.

Now we consider the binary exponent ${\rm inf}_{s\in[0,1]}Q_s\left(\hat{\sigma}_i,\hat{\sigma}_j\right)$ in Ineq.~\eqref{PH_UB}.
Conditioned on the measurement result, following Eq.~\eqref{eq:channel_error}, we have
\small
\begin{equation}
\begin{split}
    &{\rm inf}_{s\in[0,1]}Q_s\left(\hat{\sigma}_i,\hat{\sigma}_j\right)
   \simeq 
   \\
   & \exp\left[- \frac{ \sum_mN_{S,m}(N_{S,m}+1)\delta_m\sum_{\ell=1}^n|{\calM_{m,\ell}}^*|^2}{(N_B+1)^2(\sqrt{N_{S,m}+1}+\sqrt{N_{S,m}})^2}\right],
   \\
   &\simeq \exp\left[- n\frac{ \sum_mN_{S,m}\delta_m^{(i,j)}}{N_B}\right],
\end{split}
\end{equation}
\normalsize
where $\delta_m^{(i,j)} \equiv|e^{i\theta^{(i)}_m}\sqrt{\kappa^{(i)}_m}-e^{i\theta^{(j)}_m}\sqrt{\kappa^{(j)}_m}|^2$. In the asymptotic limit of $n\gg1$, $\sum_{\ell=1}^n|{\calM_{m,\ell}}^*|^2$ converges to its mean $ nN_B$, and therefore we have the last step of approximation. We have also applied the asymptotic limit $N_B\gg1$ and $N_S\ll1$.

Then, we have from Eq.~\eqref{PH_UB},
\begin{align}
&P_H(\{p_h\rho_h^{\otimes n}\})
\le 
10(r-1)^2 C_r^2 2^{2d}
\nonumber
\\
& \quad \times \max\{p_1,\cdots,p_r\}
\max_{i,j}
\exp\left[- n\frac{ \sum_mN_{S,m}\delta_m^{(i,j)}}{N_B}\right],
\end{align}

Comparing Eq.~\eqref{CS:exponent_pattern} and above, in terms of general pattern classification, entanglement combined with our conversion module enables a 6dB advantage in the error exponent.


\subsection{Simulation of noisy Dolinar receiver}

In this section, we generalize the Dolinar receiver described in Appendix~\ref{app:receivers_summary} to the noisy coherent state case and describe the numerical evaluation of the performance. 
When there is noise $N_{B,h}$ for both coherent states in the two hypotheses, original candidate states become displaced thermal state, $\hat{\rho}_0=\hat{\rho}_{0,N_{B,0}}$ and $\hat{\rho}_1=\hat{\rho}_{\alpha,N_{B,1}}$. For simplicity, we only consider the case with equal thermal noise $N_{B,0}\simeq N_{B,1} = N_B$. 

When one slices the input state into many slices, as in the Dolinar receiver presented in Appendix~\ref{app:receivers_summary}, the thermal noise part is no longer independent from each other between the different slices, which creates a challenge in numerical evaluating the performance. To solve this problem, we make use of the fact that $\hat{\rho}_h$ has positive P-function, and therefore can be realized by generating random coherent state $\ket{\alpha_h}$ where $\alpha_0 = r_0$ and $\alpha_1 = \alpha + r_1$ with $\{r_h\}_{h=0}^1$ are random complex number. The module $|r_h|$ follows exponential distribution $|r_h|^2 \sim {\rm Exp}(1/N_B)$ and the argument is uniformly random as ${\rm arg}(r_h) \sim U(0,2\pi)$~\cite{Lachs_1965}. As the measured states are still coherent states, the probability distribution of measured photons also follows Poisson distribution $N^{(k)}\sim p_N(n,g|h)$,
\begin{equation}
     p_N(n,g|h)= \begin{cases}
    {\rm Pois}(n;|\frac{\alpha_h}{\sqrt{S}}-\gamma+u^{(k)}|^2), & \textit{if $g=0$}\\
    {\rm Pois}(n;|\frac{\alpha_h}{\sqrt{S}}-\gamma-u^{(k)}|^2), & \textit{otherwise}
    \end{cases}
\end{equation}
Recall that photon number probability distribution for a displaced thermal state $\hat{\rho}_{\alpha,N_B}$ is~\cite{Lachs_1965}
\begin{align}
    P_{\alpha,N_B}(n)
    =& e^{-\frac{|\alpha|^2}{N_B+1}}\frac{N_B^n}{(1+N_B)^{n+1}}
    \nonumber
    \\
    &\quad \times {}_1F_1\left(-n,1,-\frac{|\alpha|^2}{N_B(N_B+1)}\right),
\end{align}
where ${}_1F_1(a,b;z)$ is the confluent hypergeometric function of the first kind. The conditional Bayesian probability of getting $N^{(k)}$ photon is thus
\begin{equation}
    p(N^{(k)},g|h) = \begin{cases}
    P_{\gamma-u^{(k)},N_B/S}(N^{(k)}), & \textit{if $g=h$}\\
    P_{\gamma+u^{(k)},N_B/S}(N^{(k)}), & \textit{otherwise}
    \end{cases}.
\end{equation}


\begin{thebibliography}{74}%
\makeatletter
\providecommand \@ifxundefined [1]{%
 \@ifx{#1\undefined}
}%
\providecommand \@ifnum [1]{%
 \ifnum #1\expandafter \@firstoftwo
 \else \expandafter \@secondoftwo
 \fi
}%
\providecommand \@ifx [1]{%
 \ifx #1\expandafter \@firstoftwo
 \else \expandafter \@secondoftwo
 \fi
}%
\providecommand \natexlab [1]{#1}%
\providecommand \enquote  [1]{``#1''}%
\providecommand \bibnamefont  [1]{#1}%
\providecommand \bibfnamefont [1]{#1}%
\providecommand \citenamefont [1]{#1}%
\providecommand \href@noop [0]{\@secondoftwo}%
\providecommand \href [0]{\begingroup \@sanitize@url \@href}%
\providecommand \@href[1]{\@@startlink{#1}\@@href}%
\providecommand \@@href[1]{\endgroup#1\@@endlink}%
\providecommand \@sanitize@url [0]{\catcode `\\12\catcode `\$12\catcode
  `\&12\catcode `\#12\catcode `\^12\catcode `\_12\catcode `\%12\relax}%
\providecommand \@@startlink[1]{}%
\providecommand \@@endlink[0]{}%
\providecommand \url  [0]{\begingroup\@sanitize@url \@url }%
\providecommand \@url [1]{\endgroup\@href {#1}{\urlprefix }}%
\providecommand \urlprefix  [0]{URL }%
\providecommand \Eprint [0]{\href }%
\providecommand \doibase [0]{https://doi.org/}%
\providecommand \selectlanguage [0]{\@gobble}%
\providecommand \bibinfo  [0]{\@secondoftwo}%
\providecommand \bibfield  [0]{\@secondoftwo}%
\providecommand \translation [1]{[#1]}%
\providecommand \BibitemOpen [0]{}%
\providecommand \bibitemStop [0]{}%
\providecommand \bibitemNoStop [0]{.\EOS\space}%
\providecommand \EOS [0]{\spacefactor3000\relax}%
\providecommand \BibitemShut  [1]{\csname bibitem#1\endcsname}%
\let\auto@bib@innerbib\@empty
\bibitem [{\citenamefont {Giovannetti}\ \emph {et~al.}(2006)\citenamefont
  {Giovannetti}, \citenamefont {Lloyd},\ and\ \citenamefont
  {Maccone}}]{giovannetti2006}%
  \BibitemOpen
  \bibfield  {author} {\bibinfo {author} {\bibfnamefont {V.}~\bibnamefont
  {Giovannetti}}, \bibinfo {author} {\bibfnamefont {S.}~\bibnamefont {Lloyd}},\
  and\ \bibinfo {author} {\bibfnamefont {L.}~\bibnamefont {Maccone}},\
  }\bibfield  {title} {\bibinfo {title} {Quantum metrology},\ }\href@noop {}
  {\bibfield  {journal} {\bibinfo  {journal} {Phys. Rev. Lett.}\ }\textbf
  {\bibinfo {volume} {96}},\ \bibinfo {pages} {010401} (\bibinfo {year}
  {2006})}\BibitemShut {NoStop}%
\bibitem [{\citenamefont {Giovannetti}\ \emph {et~al.}(2011)\citenamefont
  {Giovannetti}, \citenamefont {Lloyd},\ and\ \citenamefont
  {Maccone}}]{giovannetti2011advances}%
  \BibitemOpen
  \bibfield  {author} {\bibinfo {author} {\bibfnamefont {V.}~\bibnamefont
  {Giovannetti}}, \bibinfo {author} {\bibfnamefont {S.}~\bibnamefont {Lloyd}},\
  and\ \bibinfo {author} {\bibfnamefont {L.}~\bibnamefont {Maccone}},\
  }\bibfield  {title} {\bibinfo {title} {Advances in quantum metrology},\
  }\href@noop {} {\bibfield  {journal} {\bibinfo  {journal} {Nat. Photonics}\
  }\textbf {\bibinfo {volume} {5}},\ \bibinfo {pages} {222} (\bibinfo {year}
  {2011})}\BibitemShut {NoStop}%
\bibitem [{\citenamefont {Sidhu}\ and\ \citenamefont
  {Kok}(2020)}]{sidhu2020geometric}%
  \BibitemOpen
  \bibfield  {author} {\bibinfo {author} {\bibfnamefont {J.~S.}\ \bibnamefont
  {Sidhu}}\ and\ \bibinfo {author} {\bibfnamefont {P.}~\bibnamefont {Kok}},\
  }\bibfield  {title} {\bibinfo {title} {Geometric perspective on quantum
  parameter estimation},\ }\href@noop {} {\bibfield  {journal} {\bibinfo
  {journal} {AVS Quantum Science}\ }\textbf {\bibinfo {volume} {2}},\ \bibinfo
  {pages} {014701} (\bibinfo {year} {2020})}\BibitemShut {NoStop}%
\bibitem [{\citenamefont {Lawrie}\ \emph {et~al.}(2019)\citenamefont {Lawrie},
  \citenamefont {Lett}, \citenamefont {Marino},\ and\ \citenamefont
  {Pooser}}]{lawrie2019quantum}%
  \BibitemOpen
  \bibfield  {author} {\bibinfo {author} {\bibfnamefont {B.~J.}\ \bibnamefont
  {Lawrie}}, \bibinfo {author} {\bibfnamefont {P.~D.}\ \bibnamefont {Lett}},
  \bibinfo {author} {\bibfnamefont {A.~M.}\ \bibnamefont {Marino}},\ and\
  \bibinfo {author} {\bibfnamefont {R.~C.}\ \bibnamefont {Pooser}},\ }\bibfield
   {title} {\bibinfo {title} {Quantum sensing with squeezed light},\
  }\href@noop {} {\bibfield  {journal} {\bibinfo  {journal} {ACS Photonics}\
  }\textbf {\bibinfo {volume} {6}},\ \bibinfo {pages} {1307} (\bibinfo {year}
  {2019})}\BibitemShut {NoStop}%
\bibitem [{\citenamefont {T{\'o}th}\ and\ \citenamefont
  {Apellaniz}(2014)}]{toth2014quantum}%
  \BibitemOpen
  \bibfield  {author} {\bibinfo {author} {\bibfnamefont {G.}~\bibnamefont
  {T{\'o}th}}\ and\ \bibinfo {author} {\bibfnamefont {I.}~\bibnamefont
  {Apellaniz}},\ }\bibfield  {title} {\bibinfo {title} {Quantum metrology from
  a quantum information science perspective},\ }\href@noop {} {\bibfield
  {journal} {\bibinfo  {journal} {J. Phys. A: Math. Theor.}\ }\textbf {\bibinfo
  {volume} {47}},\ \bibinfo {pages} {424006} (\bibinfo {year}
  {2014})}\BibitemShut {NoStop}%
\bibitem [{\citenamefont {Pirandola}\ \emph {et~al.}(2018)\citenamefont
  {Pirandola}, \citenamefont {Bardhan}, \citenamefont {Gehring}, \citenamefont
  {Weedbrook},\ and\ \citenamefont {Lloyd}}]{pirandola2018advances}%
  \BibitemOpen
  \bibfield  {author} {\bibinfo {author} {\bibfnamefont {S.}~\bibnamefont
  {Pirandola}}, \bibinfo {author} {\bibfnamefont {B.~R.}\ \bibnamefont
  {Bardhan}}, \bibinfo {author} {\bibfnamefont {T.}~\bibnamefont {Gehring}},
  \bibinfo {author} {\bibfnamefont {C.}~\bibnamefont {Weedbrook}},\ and\
  \bibinfo {author} {\bibfnamefont {S.}~\bibnamefont {Lloyd}},\ }\bibfield
  {title} {\bibinfo {title} {Advances in photonic quantum sensing},\
  }\href@noop {} {\bibfield  {journal} {\bibinfo  {journal} {Nat. Photonics}\
  }\textbf {\bibinfo {volume} {12}},\ \bibinfo {pages} {724} (\bibinfo {year}
  {2018})}\BibitemShut {NoStop}%
\bibitem [{\citenamefont {Degen}\ \emph {et~al.}(2017)\citenamefont {Degen},
  \citenamefont {Reinhard},\ and\ \citenamefont
  {Cappellaro}}]{degen2017quantum}%
  \BibitemOpen
  \bibfield  {author} {\bibinfo {author} {\bibfnamefont {C.~L.}\ \bibnamefont
  {Degen}}, \bibinfo {author} {\bibfnamefont {F.}~\bibnamefont {Reinhard}},\
  and\ \bibinfo {author} {\bibfnamefont {P.}~\bibnamefont {Cappellaro}},\
  }\bibfield  {title} {\bibinfo {title} {Quantum sensing},\ }\href
  {https://doi.org/10.1103/RevModPhys.89.035002} {\bibfield  {journal}
  {\bibinfo  {journal} {Rev. Mod. Phys.}\ }\textbf {\bibinfo {volume} {89}},\
  \bibinfo {pages} {035002} (\bibinfo {year} {2017})}\BibitemShut {NoStop}%
\bibitem [{\citenamefont {Zhang}\ and\ \citenamefont
  {Zhuang}(2021)}]{zhang2021dqs}%
  \BibitemOpen
  \bibfield  {author} {\bibinfo {author} {\bibfnamefont {Z.}~\bibnamefont
  {Zhang}}\ and\ \bibinfo {author} {\bibfnamefont {Q.}~\bibnamefont {Zhuang}},\
  }\bibfield  {title} {\bibinfo {title} {Distributed quantum sensing},\ }\href
  {https://doi.org/10.1088/2058-9565/abd4c3} {\bibfield  {journal} {\bibinfo
  {journal} {Quantum Sci. and Technol.}\ }\textbf {\bibinfo {volume} {6}},\
  \bibinfo {pages} {043001} (\bibinfo {year} {2021})}\BibitemShut {NoStop}%
\bibitem [{\citenamefont {Gisin}\ and\ \citenamefont
  {Thew}(2007)}]{gisin2007quantum}%
  \BibitemOpen
  \bibfield  {author} {\bibinfo {author} {\bibfnamefont {N.}~\bibnamefont
  {Gisin}}\ and\ \bibinfo {author} {\bibfnamefont {R.}~\bibnamefont {Thew}},\
  }\bibfield  {title} {\bibinfo {title} {Quantum communication},\ }\href@noop
  {} {\bibfield  {journal} {\bibinfo  {journal} {Nature photonics}\ }\textbf
  {\bibinfo {volume} {1}},\ \bibinfo {pages} {165} (\bibinfo {year}
  {2007})}\BibitemShut {NoStop}%
\bibitem [{\citenamefont {Kimble}(2008)}]{kimble2008quantum}%
  \BibitemOpen
  \bibfield  {author} {\bibinfo {author} {\bibfnamefont {H.~J.}\ \bibnamefont
  {Kimble}},\ }\bibfield  {title} {\bibinfo {title} {The quantum internet},\
  }\href@noop {} {\bibfield  {journal} {\bibinfo  {journal} {Nature}\ }\textbf
  {\bibinfo {volume} {453}},\ \bibinfo {pages} {1023} (\bibinfo {year}
  {2008})}\BibitemShut {NoStop}%
\bibitem [{\citenamefont {Wilde}(2013)}]{wilde2013quantum}%
  \BibitemOpen
  \bibfield  {author} {\bibinfo {author} {\bibfnamefont {M.~M.}\ \bibnamefont
  {Wilde}},\ }\href@noop {} {\emph {\bibinfo {title} {Quantum information
  theory}}}\ (\bibinfo  {publisher} {Cambridge University Press},\ \bibinfo
  {year} {2013})\BibitemShut {NoStop}%
\bibitem [{\citenamefont {Wehner}\ \emph {et~al.}(2018)\citenamefont {Wehner},
  \citenamefont {Elkouss},\ and\ \citenamefont {Hanson}}]{wehner2018quantum}%
  \BibitemOpen
  \bibfield  {author} {\bibinfo {author} {\bibfnamefont {S.}~\bibnamefont
  {Wehner}}, \bibinfo {author} {\bibfnamefont {D.}~\bibnamefont {Elkouss}},\
  and\ \bibinfo {author} {\bibfnamefont {R.}~\bibnamefont {Hanson}},\
  }\bibfield  {title} {\bibinfo {title} {Quantum internet: A vision for the
  road ahead},\ }\href@noop {} {\bibfield  {journal} {\bibinfo  {journal}
  {Science}\ }\textbf {\bibinfo {volume} {362}},\ \bibinfo {pages} {eaam9288}
  (\bibinfo {year} {2018})}\BibitemShut {NoStop}%
\bibitem [{\citenamefont {Tan}\ \emph {et~al.}(2008)\citenamefont {Tan},
  \citenamefont {Erkmen}, \citenamefont {Giovannetti}, \citenamefont {Guha},
  \citenamefont {Lloyd}, \citenamefont {Maccone}, \citenamefont {Pirandola},\
  and\ \citenamefont {Shapiro}}]{tan2008quantum}%
  \BibitemOpen
  \bibfield  {author} {\bibinfo {author} {\bibfnamefont {S.-H.}\ \bibnamefont
  {Tan}}, \bibinfo {author} {\bibfnamefont {B.~I.}\ \bibnamefont {Erkmen}},
  \bibinfo {author} {\bibfnamefont {V.}~\bibnamefont {Giovannetti}}, \bibinfo
  {author} {\bibfnamefont {S.}~\bibnamefont {Guha}}, \bibinfo {author}
  {\bibfnamefont {S.}~\bibnamefont {Lloyd}}, \bibinfo {author} {\bibfnamefont
  {L.}~\bibnamefont {Maccone}}, \bibinfo {author} {\bibfnamefont
  {S.}~\bibnamefont {Pirandola}},\ and\ \bibinfo {author} {\bibfnamefont
  {J.~H.}\ \bibnamefont {Shapiro}},\ }\bibfield  {title} {\bibinfo {title}
  {Quantum illumination with gaussian states},\ }\href@noop {} {\bibfield
  {journal} {\bibinfo  {journal} {Phys. Rev. Lett.}\ }\textbf {\bibinfo
  {volume} {101}},\ \bibinfo {pages} {253601} (\bibinfo {year}
  {2008})}\BibitemShut {NoStop}%
\bibitem [{\citenamefont {Zhuang}(2021{\natexlab{a}})}]{zhuang2021quantum}%
  \BibitemOpen
  \bibfield  {author} {\bibinfo {author} {\bibfnamefont {Q.}~\bibnamefont
  {Zhuang}},\ }\bibfield  {title} {\bibinfo {title} {Quantum ranging with
  gaussian entanglement},\ }\href@noop {} {\bibfield  {journal} {\bibinfo
  {journal} {Phys. Rev. Lett.}\ }\textbf {\bibinfo {volume} {126}},\ \bibinfo
  {pages} {240501} (\bibinfo {year} {2021}{\natexlab{a}})}\BibitemShut
  {NoStop}%
\bibitem [{\citenamefont {Zhuang}\ and\ \citenamefont
  {Shapiro}(2022)}]{zhuang2022ultimate}%
  \BibitemOpen
  \bibfield  {author} {\bibinfo {author} {\bibfnamefont {Q.}~\bibnamefont
  {Zhuang}}\ and\ \bibinfo {author} {\bibfnamefont {J.~H.}\ \bibnamefont
  {Shapiro}},\ }\bibfield  {title} {\bibinfo {title} {Ultimate accuracy limit
  of quantum pulse-compression ranging},\ }\href
  {https://doi.org/10.1103/PhysRevLett.128.010501} {\bibfield  {journal}
  {\bibinfo  {journal} {Phys. Rev. Lett.}\ }\textbf {\bibinfo {volume} {128}},\
  \bibinfo {pages} {010501} (\bibinfo {year} {2022})}\BibitemShut {NoStop}%
\bibitem [{\citenamefont {Bennett}\ \emph {et~al.}(2002)\citenamefont
  {Bennett}, \citenamefont {Shor}, \citenamefont {Smolin},\ and\ \citenamefont
  {Thapliyal}}]{Bennett2002}%
  \BibitemOpen
  \bibfield  {author} {\bibinfo {author} {\bibfnamefont {C.}~\bibnamefont
  {Bennett}}, \bibinfo {author} {\bibfnamefont {P.}~\bibnamefont {Shor}},
  \bibinfo {author} {\bibfnamefont {J.}~\bibnamefont {Smolin}},\ and\ \bibinfo
  {author} {\bibfnamefont {A.}~\bibnamefont {Thapliyal}},\ }\bibfield  {title}
  {\bibinfo {title} {Entanglement-assisted capacity of a quantum channel and
  the reverse shannon theorem},\ }\href
  {https://doi.org/10.1109/TIT.2002.802612} {\bibfield  {journal} {\bibinfo
  {journal} {IEEE Trans. Inf. Theory,}\ }\textbf {\bibinfo {volume} {48}},\
  \bibinfo {pages} {2637} (\bibinfo {year} {2002})}\BibitemShut {NoStop}%
\bibitem [{\citenamefont {Hao}\ \emph {et~al.}(2021)\citenamefont {Hao},
  \citenamefont {Shi}, \citenamefont {Li}, \citenamefont {Shapiro},
  \citenamefont {Zhuang},\ and\ \citenamefont {Zhang}}]{hao2021entanglement}%
  \BibitemOpen
  \bibfield  {author} {\bibinfo {author} {\bibfnamefont {S.}~\bibnamefont
  {Hao}}, \bibinfo {author} {\bibfnamefont {H.}~\bibnamefont {Shi}}, \bibinfo
  {author} {\bibfnamefont {W.}~\bibnamefont {Li}}, \bibinfo {author}
  {\bibfnamefont {J.~H.}\ \bibnamefont {Shapiro}}, \bibinfo {author}
  {\bibfnamefont {Q.}~\bibnamefont {Zhuang}},\ and\ \bibinfo {author}
  {\bibfnamefont {Z.}~\bibnamefont {Zhang}},\ }\bibfield  {title} {\bibinfo
  {title} {Entanglement-assisted communication surpassing the ultimate
  classical capacity},\ }\href {https://doi.org/10.1103/PhysRevLett.126.250501}
  {\bibfield  {journal} {\bibinfo  {journal} {Phys. Rev. Lett.}\ }\textbf
  {\bibinfo {volume} {126}},\ \bibinfo {pages} {250501} (\bibinfo {year}
  {2021})}\BibitemShut {NoStop}%
\bibitem [{\citenamefont {Zhang}\ \emph {et~al.}(2013)\citenamefont {Zhang},
  \citenamefont {Tengner}, \citenamefont {Zhong}, \citenamefont {Wong},\ and\
  \citenamefont {Shapiro}}]{zhang2013}%
  \BibitemOpen
  \bibfield  {author} {\bibinfo {author} {\bibfnamefont {Z.}~\bibnamefont
  {Zhang}}, \bibinfo {author} {\bibfnamefont {M.}~\bibnamefont {Tengner}},
  \bibinfo {author} {\bibfnamefont {T.}~\bibnamefont {Zhong}}, \bibinfo
  {author} {\bibfnamefont {F.~N.~C.}\ \bibnamefont {Wong}},\ and\ \bibinfo
  {author} {\bibfnamefont {J.~H.}\ \bibnamefont {Shapiro}},\ }\bibfield
  {title} {\bibinfo {title} {Entanglement's benefit survives an
  entanglement-breaking channel},\ }\href
  {https://doi.org/10.1103/PhysRevLett.111.010501} {\bibfield  {journal}
  {\bibinfo  {journal} {Phys. Rev. Lett.}\ }\textbf {\bibinfo {volume} {111}},\
  \bibinfo {pages} {010501} (\bibinfo {year} {2013})}\BibitemShut {NoStop}%
\bibitem [{\citenamefont {Assouly}\ \emph {et~al.}(2022)\citenamefont
  {Assouly}, \citenamefont {Dassonneville}, \citenamefont {Peronnin},
  \citenamefont {Bienfait},\ and\ \citenamefont
  {Huard}}]{assouly2022demonstration}%
  \BibitemOpen
  \bibfield  {author} {\bibinfo {author} {\bibfnamefont {R.}~\bibnamefont
  {Assouly}}, \bibinfo {author} {\bibfnamefont {R.}~\bibnamefont
  {Dassonneville}}, \bibinfo {author} {\bibfnamefont {T.}~\bibnamefont
  {Peronnin}}, \bibinfo {author} {\bibfnamefont {A.}~\bibnamefont {Bienfait}},\
  and\ \bibinfo {author} {\bibfnamefont {B.}~\bibnamefont {Huard}},\ }\bibfield
   {title} {\bibinfo {title} {Demonstration of quantum advantage in microwave
  quantum radar},\ }\href@noop {} {\bibfield  {journal} {\bibinfo  {journal}
  {arXiv:2211.05684}\ } (\bibinfo {year} {2022})}\BibitemShut {NoStop}%
\bibitem [{\citenamefont {Zhuang}\ \emph {et~al.}(2017)\citenamefont {Zhuang},
  \citenamefont {Zhang},\ and\ \citenamefont {Shapiro}}]{zhuang2017optimum}%
  \BibitemOpen
  \bibfield  {author} {\bibinfo {author} {\bibfnamefont {Q.}~\bibnamefont
  {Zhuang}}, \bibinfo {author} {\bibfnamefont {Z.}~\bibnamefont {Zhang}},\ and\
  \bibinfo {author} {\bibfnamefont {J.~H.}\ \bibnamefont {Shapiro}},\
  }\bibfield  {title} {\bibinfo {title} {Optimum mixed-state discrimination for
  noisy entanglement-enhanced sensing},\ }\href
  {https://doi.org/10.1103/PhysRevLett.118.040801} {\bibfield  {journal}
  {\bibinfo  {journal} {Phys. Rev. Lett.}\ }\textbf {\bibinfo {volume} {118}},\
  \bibinfo {pages} {040801} (\bibinfo {year} {2017})}\BibitemShut {NoStop}%
\bibitem [{\citenamefont {Audenaert}\ \emph {et~al.}(2007)\citenamefont
  {Audenaert}, \citenamefont {Calsamiglia}, \citenamefont {Mu\~noz Tapia},
  \citenamefont {Bagan}, \citenamefont {Masanes}, \citenamefont {Acin},\ and\
  \citenamefont {Verstraete}}]{Audenaert2007}%
  \BibitemOpen
  \bibfield  {author} {\bibinfo {author} {\bibfnamefont {K.~M.~R.}\
  \bibnamefont {Audenaert}}, \bibinfo {author} {\bibfnamefont {J.}~\bibnamefont
  {Calsamiglia}}, \bibinfo {author} {\bibfnamefont {R.}~\bibnamefont {Mu\~noz
  Tapia}}, \bibinfo {author} {\bibfnamefont {E.}~\bibnamefont {Bagan}},
  \bibinfo {author} {\bibfnamefont {L.}~\bibnamefont {Masanes}}, \bibinfo
  {author} {\bibfnamefont {A.}~\bibnamefont {Acin}},\ and\ \bibinfo {author}
  {\bibfnamefont {F.}~\bibnamefont {Verstraete}},\ }\bibfield  {title}
  {\bibinfo {title} {Discriminating states: The quantum chernoff bound},\
  }\href {https://doi.org/10.1103/PhysRevLett.98.160501} {\bibfield  {journal}
  {\bibinfo  {journal} {Phys. Rev. Lett.}\ }\textbf {\bibinfo {volume} {98}},\
  \bibinfo {pages} {160501} (\bibinfo {year} {2007})}\BibitemShut {NoStop}%
\bibitem [{\citenamefont {Pirandola}\ and\ \citenamefont
  {Lloyd}(2008)}]{Pirandola2008}%
  \BibitemOpen
  \bibfield  {author} {\bibinfo {author} {\bibfnamefont {S.}~\bibnamefont
  {Pirandola}}\ and\ \bibinfo {author} {\bibfnamefont {S.}~\bibnamefont
  {Lloyd}},\ }\bibfield  {title} {\bibinfo {title} {Computable bounds for the
  discrimination of gaussian states},\ }\href
  {https://doi.org/10.1103/PhysRevA.78.012331} {\bibfield  {journal} {\bibinfo
  {journal} {Phys. Rev. A}\ }\textbf {\bibinfo {volume} {78}},\ \bibinfo
  {pages} {012331} (\bibinfo {year} {2008})}\BibitemShut {NoStop}%
\bibitem [{\citenamefont {Nussbaum}\ \emph {et~al.}(2011)\citenamefont
  {Nussbaum}, \citenamefont {Szko{\l}a} \emph
  {et~al.}}]{nussbaum2011asymptotic}%
  \BibitemOpen
  \bibfield  {author} {\bibinfo {author} {\bibfnamefont {M.}~\bibnamefont
  {Nussbaum}}, \bibinfo {author} {\bibfnamefont {A.}~\bibnamefont {Szko{\l}a}},
  \emph {et~al.},\ }\bibfield  {title} {\bibinfo {title} {An asymptotic error
  bound for testing multiple quantum hypotheses},\ }\href@noop {} {\bibfield
  {journal} {\bibinfo  {journal} {Ann. Statist.}\ }\textbf {\bibinfo {volume}
  {39}},\ \bibinfo {pages} {3211} (\bibinfo {year} {2011})}\BibitemShut
  {NoStop}%
\bibitem [{\citenamefont {Li}(2016)}]{li2016discriminating}%
  \BibitemOpen
  \bibfield  {author} {\bibinfo {author} {\bibfnamefont {K.}~\bibnamefont
  {Li}},\ }\bibfield  {title} {\bibinfo {title} {Discriminating quantum states:
  The multiple chernoff distance},\ }\href@noop {} {\bibfield  {journal}
  {\bibinfo  {journal} {Ann. Statist.}\ }\textbf {\bibinfo {volume} {44}},\
  \bibinfo {pages} {1661} (\bibinfo {year} {2016})}\BibitemShut {NoStop}%
\bibitem [{\citenamefont {Dolinar}(1973)}]{dolinar_processing_1973}%
  \BibitemOpen
  \bibfield  {author} {\bibinfo {author} {\bibfnamefont {S.~J.}\ \bibnamefont
  {Dolinar}},\ }\href@noop {} {\emph {\bibinfo {title} {Processing and
  Transmission of Information}}},\ \bibinfo {type} {Technical Report}\
  (\bibinfo  {institution} {Research Laboratory of Electronics (RLE) at the
  Massachusetts Institute of Technology (MIT)},\ \bibinfo {year}
  {1973})\BibitemShut {NoStop}%
\bibitem [{\citenamefont {Guha}(2011)}]{guha2011structured}%
  \BibitemOpen
  \bibfield  {author} {\bibinfo {author} {\bibfnamefont {S.}~\bibnamefont
  {Guha}},\ }\bibfield  {title} {\bibinfo {title} {Structured optical receivers
  to attain superadditive capacity and the holevo limit},\ }\href@noop {}
  {\bibfield  {journal} {\bibinfo  {journal} {Phys. Rev. Lett.}\ }\textbf
  {\bibinfo {volume} {106}},\ \bibinfo {pages} {240502} (\bibinfo {year}
  {2011})}\BibitemShut {NoStop}%
\bibitem [{\citenamefont {Weedbrook}\ \emph {et~al.}(2012)\citenamefont
  {Weedbrook}, \citenamefont {Pirandola}, \citenamefont
  {Garc{\'\i}a-Patr{\'o}n}, \citenamefont {Cerf}, \citenamefont {Ralph},
  \citenamefont {Shapiro},\ and\ \citenamefont
  {Lloyd}}]{weedbrook2012gaussian}%
  \BibitemOpen
  \bibfield  {author} {\bibinfo {author} {\bibfnamefont {C.}~\bibnamefont
  {Weedbrook}}, \bibinfo {author} {\bibfnamefont {S.}~\bibnamefont
  {Pirandola}}, \bibinfo {author} {\bibfnamefont {R.}~\bibnamefont
  {Garc{\'\i}a-Patr{\'o}n}}, \bibinfo {author} {\bibfnamefont {N.~J.}\
  \bibnamefont {Cerf}}, \bibinfo {author} {\bibfnamefont {T.~C.}\ \bibnamefont
  {Ralph}}, \bibinfo {author} {\bibfnamefont {J.~H.}\ \bibnamefont {Shapiro}},\
  and\ \bibinfo {author} {\bibfnamefont {S.}~\bibnamefont {Lloyd}},\ }\bibfield
   {title} {\bibinfo {title} {Gaussian quantum information},\ }\href@noop {}
  {\bibfield  {journal} {\bibinfo  {journal} {Rev. Mod. Phys.}\ }\textbf
  {\bibinfo {volume} {84}},\ \bibinfo {pages} {621} (\bibinfo {year}
  {2012})}\BibitemShut {NoStop}%
\bibitem [{\citenamefont {Escher}\ \emph {et~al.}(2011)\citenamefont {Escher},
  \citenamefont {de~Matos~Filho},\ and\ \citenamefont
  {Davidovich}}]{Escher_2011}%
  \BibitemOpen
  \bibfield  {author} {\bibinfo {author} {\bibfnamefont {B.~M.}\ \bibnamefont
  {Escher}}, \bibinfo {author} {\bibfnamefont {R.~L.}\ \bibnamefont
  {de~Matos~Filho}},\ and\ \bibinfo {author} {\bibfnamefont {L.}~\bibnamefont
  {Davidovich}},\ }\bibfield  {title} {\bibinfo {title} {General framework for
  estimating the ultimate precision limit in noisy quantum-enhanced
  metrology},\ }\href@noop {} {\bibfield  {journal} {\bibinfo  {journal} {Nat
  Phys}\ }\textbf {\bibinfo {volume} {7}},\ \bibinfo {pages} {406} (\bibinfo
  {year} {2011})}\BibitemShut {NoStop}%
\bibitem [{\citenamefont {Collaboration}(2016)}]{abbott2016observation}%
  \BibitemOpen
  \bibfield  {author} {\bibinfo {author} {\bibfnamefont {L.~S.}\ \bibnamefont
  {Collaboration}},\ }\bibfield  {title} {\bibinfo {title} {Observation of
  gravitational waves from a binary black hole merger},\ }\href@noop {}
  {\bibfield  {journal} {\bibinfo  {journal} {Phys. Rev. Lett.}\ }\textbf
  {\bibinfo {volume} {116}},\ \bibinfo {pages} {061102} (\bibinfo {year}
  {2016})}\BibitemShut {NoStop}%
\bibitem [{\citenamefont {Shi}\ \emph {et~al.}(2020{\natexlab{a}})\citenamefont
  {Shi}, \citenamefont {Zhang},\ and\ \citenamefont
  {Zhuang}}]{shi2020practical}%
  \BibitemOpen
  \bibfield  {author} {\bibinfo {author} {\bibfnamefont {H.}~\bibnamefont
  {Shi}}, \bibinfo {author} {\bibfnamefont {Z.}~\bibnamefont {Zhang}},\ and\
  \bibinfo {author} {\bibfnamefont {Q.}~\bibnamefont {Zhuang}},\ }\bibfield
  {title} {\bibinfo {title} {Practical route to entanglement-assisted
  communication over noisy bosonic channels},\ }\href
  {https://doi.org/10.1103/PhysRevApplied.13.034029} {\bibfield  {journal}
  {\bibinfo  {journal} {Phys. Rev. Applied}\ }\textbf {\bibinfo {volume}
  {13}},\ \bibinfo {pages} {034029} (\bibinfo {year}
  {2020}{\natexlab{a}})}\BibitemShut {NoStop}%
\bibitem [{\citenamefont {Nair}\ and\ \citenamefont
  {Gu}(2020)}]{nair2020fundamental}%
  \BibitemOpen
  \bibfield  {author} {\bibinfo {author} {\bibfnamefont {R.}~\bibnamefont
  {Nair}}\ and\ \bibinfo {author} {\bibfnamefont {M.}~\bibnamefont {Gu}},\
  }\bibfield  {title} {\bibinfo {title} {Fundamental limits of quantum
  illumination},\ }\href@noop {} {\bibfield  {journal} {\bibinfo  {journal}
  {Optica}\ }\textbf {\bibinfo {volume} {7}},\ \bibinfo {pages} {771} (\bibinfo
  {year} {2020})}\BibitemShut {NoStop}%
\bibitem [{Note1()}]{Note1}%
  \BibitemOpen
  \bibinfo {note} {Similar to previous works~\cite
  {tan2008quantum,zhuang2017optimum,zhuang2021quantum}, we have chosen the
  $1-\kappa $ factor so that passive signature are not present.}\BibitemShut
  {Stop}%
\bibitem [{\citenamefont {Genoni}\ \emph {et~al.}(2016)\citenamefont {Genoni},
  \citenamefont {Lami},\ and\ \citenamefont
  {Serafini}}]{genoni2016conditional}%
  \BibitemOpen
  \bibfield  {author} {\bibinfo {author} {\bibfnamefont {M.~G.}\ \bibnamefont
  {Genoni}}, \bibinfo {author} {\bibfnamefont {L.}~\bibnamefont {Lami}},\ and\
  \bibinfo {author} {\bibfnamefont {A.}~\bibnamefont {Serafini}},\ }\bibfield
  {title} {\bibinfo {title} {Conditional and unconditional gaussian quantum
  dynamics},\ }\href@noop {} {\bibfield  {journal} {\bibinfo  {journal}
  {Contemp. Phys.}\ }\textbf {\bibinfo {volume} {57}},\ \bibinfo {pages} {331}
  (\bibinfo {year} {2016})}\BibitemShut {NoStop}%
\bibitem [{\citenamefont {Cook}\ \emph {et~al.}(2007)\citenamefont {Cook},
  \citenamefont {Martin},\ and\ \citenamefont {Geremia}}]{cook2007}%
  \BibitemOpen
  \bibfield  {author} {\bibinfo {author} {\bibfnamefont {R.~L.}\ \bibnamefont
  {Cook}}, \bibinfo {author} {\bibfnamefont {P.~J.}\ \bibnamefont {Martin}},\
  and\ \bibinfo {author} {\bibfnamefont {J.~M.}\ \bibnamefont {Geremia}},\
  }\bibfield  {title} {\bibinfo {title} {Optical coherent state discrimination
  using a closed-loop quantum measurement},\ }\href@noop {} {\bibfield
  {journal} {\bibinfo  {journal} {Nature}\ }\textbf {\bibinfo {volume} {446}},\
  \bibinfo {pages} {774} (\bibinfo {year} {2007})}\BibitemShut {NoStop}%
\bibitem [{\citenamefont {Angeletti}\ \emph {et~al.}(2023)\citenamefont
  {Angeletti}, \citenamefont {Shi}, \citenamefont {Lakshmanan}, \citenamefont
  {Vitali},\ and\ \citenamefont {Zhuang}}]{angeletti2023}%
  \BibitemOpen
  \bibfield  {author} {\bibinfo {author} {\bibfnamefont {J.}~\bibnamefont
  {Angeletti}}, \bibinfo {author} {\bibfnamefont {H.}~\bibnamefont {Shi}},
  \bibinfo {author} {\bibfnamefont {T.}~\bibnamefont {Lakshmanan}}, \bibinfo
  {author} {\bibfnamefont {D.}~\bibnamefont {Vitali}},\ and\ \bibinfo {author}
  {\bibfnamefont {Q.}~\bibnamefont {Zhuang}},\ }\bibfield  {title} {\bibinfo
  {title} {Microwave quantum illumination with correlation-to-displacement
  conversion},\ }\href@noop {} {\bibfield  {journal} {\bibinfo  {journal}
  {arXiv:2303.18206}\ } (\bibinfo {year} {2023})}\BibitemShut {NoStop}%
\bibitem [{\citenamefont {Reichert}\ \emph {et~al.}(2023)\citenamefont
  {Reichert}, \citenamefont {Zhuang}, \citenamefont {Shapiro},\ and\
  \citenamefont {Di~Candia}}]{reichert2023}%
  \BibitemOpen
  \bibfield  {author} {\bibinfo {author} {\bibfnamefont {M.}~\bibnamefont
  {Reichert}}, \bibinfo {author} {\bibfnamefont {Q.}~\bibnamefont {Zhuang}},
  \bibinfo {author} {\bibfnamefont {J.~H.}\ \bibnamefont {Shapiro}},\ and\
  \bibinfo {author} {\bibfnamefont {R.}~\bibnamefont {Di~Candia}},\ }\bibfield
  {title} {\bibinfo {title} {Quantum illumination with a hetero-homodyne
  receiver and sequential detection},\ }\href@noop {} {\bibfield  {journal}
  {\bibinfo  {journal} {arXiv:2303.18207}\ } (\bibinfo {year}
  {2023})}\BibitemShut {NoStop}%
\bibitem [{\citenamefont {Otterstrom}\ \emph {et~al.}(2021)\citenamefont
  {Otterstrom}, \citenamefont {Gertler}, \citenamefont {Kittlaus},
  \citenamefont {Gehl}, \citenamefont {Starbuck}, \citenamefont {Dallo},
  \citenamefont {Pomerene}, \citenamefont {Trotter}, \citenamefont {Rakich},
  \citenamefont {Davids} \emph {et~al.}}]{otterstrom2021}%
  \BibitemOpen
  \bibfield  {author} {\bibinfo {author} {\bibfnamefont {N.~T.}\ \bibnamefont
  {Otterstrom}}, \bibinfo {author} {\bibfnamefont {S.}~\bibnamefont {Gertler}},
  \bibinfo {author} {\bibfnamefont {E.~A.}\ \bibnamefont {Kittlaus}}, \bibinfo
  {author} {\bibfnamefont {M.}~\bibnamefont {Gehl}}, \bibinfo {author}
  {\bibfnamefont {A.~L.}\ \bibnamefont {Starbuck}}, \bibinfo {author}
  {\bibfnamefont {C.~M.}\ \bibnamefont {Dallo}}, \bibinfo {author}
  {\bibfnamefont {A.~T.}\ \bibnamefont {Pomerene}}, \bibinfo {author}
  {\bibfnamefont {D.~C.}\ \bibnamefont {Trotter}}, \bibinfo {author}
  {\bibfnamefont {P.~T.}\ \bibnamefont {Rakich}}, \bibinfo {author}
  {\bibfnamefont {P.~S.}\ \bibnamefont {Davids}}, \emph {et~al.},\ }\bibfield
  {title} {\bibinfo {title} {Nonreciprocal frequency domain beam splitter},\
  }\href@noop {} {\bibfield  {journal} {\bibinfo  {journal} {Phys. Rev. Lett.}\
  }\textbf {\bibinfo {volume} {127}},\ \bibinfo {pages} {253603} (\bibinfo
  {year} {2021})}\BibitemShut {NoStop}%
\bibitem [{\citenamefont {Li}\ \emph {et~al.}(2016)\citenamefont {Li},
  \citenamefont {Davan{\c{c}}o},\ and\ \citenamefont {Srinivasan}}]{li2016}%
  \BibitemOpen
  \bibfield  {author} {\bibinfo {author} {\bibfnamefont {Q.}~\bibnamefont
  {Li}}, \bibinfo {author} {\bibfnamefont {M.}~\bibnamefont {Davan{\c{c}}o}},\
  and\ \bibinfo {author} {\bibfnamefont {K.}~\bibnamefont {Srinivasan}},\
  }\bibfield  {title} {\bibinfo {title} {Efficient and low-noise
  single-photon-level frequency conversion interfaces using silicon
  nanophotonics},\ }\href@noop {} {\bibfield  {journal} {\bibinfo  {journal}
  {Nature Photon.}\ }\textbf {\bibinfo {volume} {10}},\ \bibinfo {pages} {406}
  (\bibinfo {year} {2016})}\BibitemShut {NoStop}%
\bibitem [{\citenamefont {McGuinness}\ \emph {et~al.}(2010)\citenamefont
  {McGuinness}, \citenamefont {Raymer}, \citenamefont {McKinstrie},\ and\
  \citenamefont {Radic}}]{mcguinness2010}%
  \BibitemOpen
  \bibfield  {author} {\bibinfo {author} {\bibfnamefont {H.~J.}\ \bibnamefont
  {McGuinness}}, \bibinfo {author} {\bibfnamefont {M.~G.}\ \bibnamefont
  {Raymer}}, \bibinfo {author} {\bibfnamefont {C.~J.}\ \bibnamefont
  {McKinstrie}},\ and\ \bibinfo {author} {\bibfnamefont {S.}~\bibnamefont
  {Radic}},\ }\bibfield  {title} {\bibinfo {title} {Quantum frequency
  translation of single-photon states in a photonic crystal fiber},\
  }\href@noop {} {\bibfield  {journal} {\bibinfo  {journal} {Phys. Rev. Lett.}\
  }\textbf {\bibinfo {volume} {105}},\ \bibinfo {pages} {093604} (\bibinfo
  {year} {2010})}\BibitemShut {NoStop}%
\bibitem [{\citenamefont {Grosshans}\ and\ \citenamefont
  {Grangier}(2002)}]{grosshans2002}%
  \BibitemOpen
  \bibfield  {author} {\bibinfo {author} {\bibfnamefont {F.}~\bibnamefont
  {Grosshans}}\ and\ \bibinfo {author} {\bibfnamefont {P.}~\bibnamefont
  {Grangier}},\ }\bibfield  {title} {\bibinfo {title} {Continuous variable
  quantum cryptography using coherent states},\ }\href@noop {} {\bibfield
  {journal} {\bibinfo  {journal} {Phys. Rev. Lett.}\ }\textbf {\bibinfo
  {volume} {88}},\ \bibinfo {pages} {057902} (\bibinfo {year}
  {2002})}\BibitemShut {NoStop}%
\bibitem [{\citenamefont {Garc{\'\i}a-Patr{\'o}n}\ and\ \citenamefont
  {Cerf}(2006)}]{garcia2006}%
  \BibitemOpen
  \bibfield  {author} {\bibinfo {author} {\bibfnamefont {R.}~\bibnamefont
  {Garc{\'\i}a-Patr{\'o}n}}\ and\ \bibinfo {author} {\bibfnamefont {N.~J.}\
  \bibnamefont {Cerf}},\ }\bibfield  {title} {\bibinfo {title} {Unconditional
  optimality of gaussian attacks against continuous-variable quantum key
  distribution},\ }\href@noop {} {\bibfield  {journal} {\bibinfo  {journal}
  {Phys. Rev. Lett.}\ }\textbf {\bibinfo {volume} {97}},\ \bibinfo {pages}
  {190503} (\bibinfo {year} {2006})}\BibitemShut {NoStop}%
\bibitem [{\citenamefont {Navascu{\'e}s}\ \emph {et~al.}(2006)\citenamefont
  {Navascu{\'e}s}, \citenamefont {Grosshans},\ and\ \citenamefont
  {Acin}}]{navascues2006}%
  \BibitemOpen
  \bibfield  {author} {\bibinfo {author} {\bibfnamefont {M.}~\bibnamefont
  {Navascu{\'e}s}}, \bibinfo {author} {\bibfnamefont {F.}~\bibnamefont
  {Grosshans}},\ and\ \bibinfo {author} {\bibfnamefont {A.}~\bibnamefont
  {Acin}},\ }\bibfield  {title} {\bibinfo {title} {Optimality of gaussian
  attacks in continuous-variable quantum cryptography},\ }\href@noop {}
  {\bibfield  {journal} {\bibinfo  {journal} {Phys. Rev. Lett.}\ }\textbf
  {\bibinfo {volume} {97}},\ \bibinfo {pages} {190502} (\bibinfo {year}
  {2006})}\BibitemShut {NoStop}%
\bibitem [{\citenamefont {Helstrom}(1969)}]{Helstrom1969}%
  \BibitemOpen
  \bibfield  {author} {\bibinfo {author} {\bibfnamefont {C.~W.}\ \bibnamefont
  {Helstrom}},\ }\bibfield  {title} {\bibinfo {title} {Quantum detection and
  estimation theory},\ }\href {https://doi.org/10.1007/BF01007479} {\bibfield
  {journal} {\bibinfo  {journal} {J. Stat. Phys.}\ }\textbf {\bibinfo {volume}
  {1}},\ \bibinfo {pages} {231} (\bibinfo {year} {1969})}\BibitemShut {NoStop}%
\bibitem [{\citenamefont {Helstrom}(1967)}]{Helstrom_1967}%
  \BibitemOpen
  \bibfield  {author} {\bibinfo {author} {\bibfnamefont {C.}~\bibnamefont
  {Helstrom}},\ }\bibfield  {title} {\bibinfo {title} {Minimum mean-squared
  error of estimates in quantum statistics},\ }\href@noop {} {\bibfield
  {journal} {\bibinfo  {journal} {Phys. Lett. A}\ }\textbf {\bibinfo {volume}
  {25}},\ \bibinfo {pages} {101} (\bibinfo {year} {1967})}\BibitemShut
  {NoStop}%
\bibitem [{\citenamefont {Helstrom}(1976)}]{Helstrom_1976}%
  \BibitemOpen
  \bibfield  {author} {\bibinfo {author} {\bibfnamefont {C.}~\bibnamefont
  {Helstrom}},\ }\href {https://books.google.com/books?id=fv9SAAAAMAAJ} {\emph
  {\bibinfo {title} {Quantum Detection and Estimation Theory}}},\ Mathematics
  in Science and Engineering : a series of monographs and textbooks\ (\bibinfo
  {publisher} {Academic Press},\ \bibinfo {year} {1976})\BibitemShut {NoStop}%
\bibitem [{\citenamefont {Gagatsos}\ \emph {et~al.}(2017)\citenamefont
  {Gagatsos}, \citenamefont {Bash}, \citenamefont {Guha},\ and\ \citenamefont
  {Datta}}]{gagatsos2017bounding}%
  \BibitemOpen
  \bibfield  {author} {\bibinfo {author} {\bibfnamefont {C.~N.}\ \bibnamefont
  {Gagatsos}}, \bibinfo {author} {\bibfnamefont {B.~A.}\ \bibnamefont {Bash}},
  \bibinfo {author} {\bibfnamefont {S.}~\bibnamefont {Guha}},\ and\ \bibinfo
  {author} {\bibfnamefont {A.}~\bibnamefont {Datta}},\ }\bibfield  {title}
  {\bibinfo {title} {Bounding the quantum limits of precision for phase
  estimation with loss and thermal noise},\ }\href@noop {} {\bibfield
  {journal} {\bibinfo  {journal} {Phys. Rev. A}\ }\textbf {\bibinfo {volume}
  {96}},\ \bibinfo {pages} {062306} (\bibinfo {year} {2017})}\BibitemShut
  {NoStop}%
\bibitem [{\citenamefont {Holevo}(1973)}]{holevo1973bounds}%
  \BibitemOpen
  \bibfield  {author} {\bibinfo {author} {\bibfnamefont {A.~S.}\ \bibnamefont
  {Holevo}},\ }\bibfield  {title} {\bibinfo {title} {Bounds for the quantity of
  information transmitted by a quantum communication channel},\ }\href@noop {}
  {\bibfield  {journal} {\bibinfo  {journal} {Problemy Peredachi Informatsii}\
  }\textbf {\bibinfo {volume} {9}},\ \bibinfo {pages} {3} (\bibinfo {year}
  {1973})}\BibitemShut {NoStop}%
\bibitem [{\citenamefont {Giovannetti}\ \emph {et~al.}(2014)\citenamefont
  {Giovannetti}, \citenamefont {Garcia-Patron}, \citenamefont {Cerf},\ and\
  \citenamefont {Holevo}}]{giovannetti2014ultimate}%
  \BibitemOpen
  \bibfield  {author} {\bibinfo {author} {\bibfnamefont {V.}~\bibnamefont
  {Giovannetti}}, \bibinfo {author} {\bibfnamefont {R.}~\bibnamefont
  {Garcia-Patron}}, \bibinfo {author} {\bibfnamefont {N.~J.}\ \bibnamefont
  {Cerf}},\ and\ \bibinfo {author} {\bibfnamefont {A.~S.}\ \bibnamefont
  {Holevo}},\ }\bibfield  {title} {\bibinfo {title} {Ultimate classical
  communication rates of quantum optical channels},\ }\href@noop {} {\bibfield
  {journal} {\bibinfo  {journal} {Nature Photonics}\ }\textbf {\bibinfo
  {volume} {8}},\ \bibinfo {pages} {796} (\bibinfo {year} {2014})}\BibitemShut
  {NoStop}%
\bibitem [{\citenamefont {Bash}\ \emph {et~al.}(2015)\citenamefont {Bash},
  \citenamefont {Gheorghe}, \citenamefont {Patel}, \citenamefont {Habif},
  \citenamefont {Goeckel}, \citenamefont {Towsley},\ and\ \citenamefont
  {Guha}}]{bash2015quantum}%
  \BibitemOpen
  \bibfield  {author} {\bibinfo {author} {\bibfnamefont {B.~A.}\ \bibnamefont
  {Bash}}, \bibinfo {author} {\bibfnamefont {A.~H.}\ \bibnamefont {Gheorghe}},
  \bibinfo {author} {\bibfnamefont {M.}~\bibnamefont {Patel}}, \bibinfo
  {author} {\bibfnamefont {J.~L.}\ \bibnamefont {Habif}}, \bibinfo {author}
  {\bibfnamefont {D.}~\bibnamefont {Goeckel}}, \bibinfo {author} {\bibfnamefont
  {D.}~\bibnamefont {Towsley}},\ and\ \bibinfo {author} {\bibfnamefont
  {S.}~\bibnamefont {Guha}},\ }\bibfield  {title} {\bibinfo {title}
  {Quantum-secure covert communication on bosonic channels},\ }\href@noop {}
  {\bibfield  {journal} {\bibinfo  {journal} {Nat. Commun.}\ }\textbf {\bibinfo
  {volume} {6}},\ \bibinfo {pages} {1} (\bibinfo {year} {2015})}\BibitemShut
  {NoStop}%
\bibitem [{\citenamefont {Shi}\ \emph {et~al.}(2020{\natexlab{b}})\citenamefont
  {Shi}, \citenamefont {Zhang}, \citenamefont {Pirandola},\ and\ \citenamefont
  {Zhuang}}]{shi2020entanglement}%
  \BibitemOpen
  \bibfield  {author} {\bibinfo {author} {\bibfnamefont {H.}~\bibnamefont
  {Shi}}, \bibinfo {author} {\bibfnamefont {Z.}~\bibnamefont {Zhang}}, \bibinfo
  {author} {\bibfnamefont {S.}~\bibnamefont {Pirandola}},\ and\ \bibinfo
  {author} {\bibfnamefont {Q.}~\bibnamefont {Zhuang}},\ }\bibfield  {title}
  {\bibinfo {title} {Entanglement-assisted absorption spectroscopy},\ }\href
  {https://doi.org/10.1103/PhysRevLett.125.180502} {\bibfield  {journal}
  {\bibinfo  {journal} {Phys. Rev. Lett.}\ }\textbf {\bibinfo {volume} {125}},\
  \bibinfo {pages} {180502} (\bibinfo {year} {2020}{\natexlab{b}})}\BibitemShut
  {NoStop}%
\bibitem [{\citenamefont {Zhuang}\ and\ \citenamefont
  {Pirandola}(2020)}]{zhuang2020entanglement}%
  \BibitemOpen
  \bibfield  {author} {\bibinfo {author} {\bibfnamefont {Q.}~\bibnamefont
  {Zhuang}}\ and\ \bibinfo {author} {\bibfnamefont {S.}~\bibnamefont
  {Pirandola}},\ }\bibfield  {title} {\bibinfo {title} {Entanglement-enhanced
  testing of multiple quantum hypotheses},\ }\href@noop {} {\bibfield
  {journal} {\bibinfo  {journal} {Commun. Phys.}\ }\textbf {\bibinfo {volume}
  {3}},\ \bibinfo {pages} {1} (\bibinfo {year} {2020})}\BibitemShut {NoStop}%
\bibitem [{\citenamefont {Banchi}\ \emph {et~al.}(2020)\citenamefont {Banchi},
  \citenamefont {Zhuang},\ and\ \citenamefont {Pirandola}}]{banchi2020quantum}%
  \BibitemOpen
  \bibfield  {author} {\bibinfo {author} {\bibfnamefont {L.}~\bibnamefont
  {Banchi}}, \bibinfo {author} {\bibfnamefont {Q.}~\bibnamefont {Zhuang}},\
  and\ \bibinfo {author} {\bibfnamefont {S.}~\bibnamefont {Pirandola}},\
  }\bibfield  {title} {\bibinfo {title} {Quantum-enhanced barcode decoding and
  pattern recognition},\ }\href@noop {} {\bibfield  {journal} {\bibinfo
  {journal} {Phys. Rev. Applied}\ }\textbf {\bibinfo {volume} {14}},\ \bibinfo
  {pages} {064026} (\bibinfo {year} {2020})}\BibitemShut {NoStop}%
\bibitem [{\citenamefont {Tsujino}\ \emph {et~al.}(2011)\citenamefont
  {Tsujino}, \citenamefont {Fukuda}, \citenamefont {Fujii}, \citenamefont
  {Inoue}, \citenamefont {Fujiwara}, \citenamefont {Takeoka},\ and\
  \citenamefont {Sasaki}}]{tsujino2011quantum}%
  \BibitemOpen
  \bibfield  {author} {\bibinfo {author} {\bibfnamefont {K.}~\bibnamefont
  {Tsujino}}, \bibinfo {author} {\bibfnamefont {D.}~\bibnamefont {Fukuda}},
  \bibinfo {author} {\bibfnamefont {G.}~\bibnamefont {Fujii}}, \bibinfo
  {author} {\bibfnamefont {S.}~\bibnamefont {Inoue}}, \bibinfo {author}
  {\bibfnamefont {M.}~\bibnamefont {Fujiwara}}, \bibinfo {author}
  {\bibfnamefont {M.}~\bibnamefont {Takeoka}},\ and\ \bibinfo {author}
  {\bibfnamefont {M.}~\bibnamefont {Sasaki}},\ }\bibfield  {title} {\bibinfo
  {title} {Quantum receiver beyond the standard quantum limit of coherent
  optical communication},\ }\href@noop {} {\bibfield  {journal} {\bibinfo
  {journal} {Phys. Rev. Lett.}\ }\textbf {\bibinfo {volume} {106}},\ \bibinfo
  {pages} {250503} (\bibinfo {year} {2011})}\BibitemShut {NoStop}%
\bibitem [{\citenamefont {Chen}\ \emph {et~al.}(2012)\citenamefont {Chen},
  \citenamefont {Habif}, \citenamefont {Dutton}, \citenamefont {Lazarus},\ and\
  \citenamefont {Guha}}]{chen2012optical}%
  \BibitemOpen
  \bibfield  {author} {\bibinfo {author} {\bibfnamefont {J.}~\bibnamefont
  {Chen}}, \bibinfo {author} {\bibfnamefont {J.~L.}\ \bibnamefont {Habif}},
  \bibinfo {author} {\bibfnamefont {Z.}~\bibnamefont {Dutton}}, \bibinfo
  {author} {\bibfnamefont {R.}~\bibnamefont {Lazarus}},\ and\ \bibinfo {author}
  {\bibfnamefont {S.}~\bibnamefont {Guha}},\ }\bibfield  {title} {\bibinfo
  {title} {Optical codeword demodulation with error rates below the standard
  quantum limit using a conditional nulling receiver},\ }\href@noop {}
  {\bibfield  {journal} {\bibinfo  {journal} {Nat. Photonics}\ }\textbf
  {\bibinfo {volume} {6}},\ \bibinfo {pages} {374} (\bibinfo {year}
  {2012})}\BibitemShut {NoStop}%
\bibitem [{\citenamefont {Becerra}\ \emph {et~al.}(2013)\citenamefont
  {Becerra}, \citenamefont {Fan}, \citenamefont {Baumgartner}, \citenamefont
  {Goldhar}, \citenamefont {Kosloski},\ and\ \citenamefont
  {Migdall}}]{becerra2013experimental}%
  \BibitemOpen
  \bibfield  {author} {\bibinfo {author} {\bibfnamefont {F.}~\bibnamefont
  {Becerra}}, \bibinfo {author} {\bibfnamefont {J.}~\bibnamefont {Fan}},
  \bibinfo {author} {\bibfnamefont {G.}~\bibnamefont {Baumgartner}}, \bibinfo
  {author} {\bibfnamefont {J.}~\bibnamefont {Goldhar}}, \bibinfo {author}
  {\bibfnamefont {J.}~\bibnamefont {Kosloski}},\ and\ \bibinfo {author}
  {\bibfnamefont {A.}~\bibnamefont {Migdall}},\ }\bibfield  {title} {\bibinfo
  {title} {Experimental demonstration of a receiver beating the standard
  quantum limit for multiple nonorthogonal state discrimination},\ }\href@noop
  {} {\bibfield  {journal} {\bibinfo  {journal} {Nat. Photonics}\ }\textbf
  {\bibinfo {volume} {7}},\ \bibinfo {pages} {147} (\bibinfo {year}
  {2013})}\BibitemShut {NoStop}%
\bibitem [{\citenamefont {Becerra}\ \emph {et~al.}(2015)\citenamefont
  {Becerra}, \citenamefont {Fan},\ and\ \citenamefont
  {Migdall}}]{becerra2015photon}%
  \BibitemOpen
  \bibfield  {author} {\bibinfo {author} {\bibfnamefont {F.}~\bibnamefont
  {Becerra}}, \bibinfo {author} {\bibfnamefont {J.}~\bibnamefont {Fan}},\ and\
  \bibinfo {author} {\bibfnamefont {A.}~\bibnamefont {Migdall}},\ }\bibfield
  {title} {\bibinfo {title} {Photon number resolution enables quantum receiver
  for realistic coherent optical communications},\ }\href@noop {} {\bibfield
  {journal} {\bibinfo  {journal} {Nat. Photonics}\ }\textbf {\bibinfo {volume}
  {9}},\ \bibinfo {pages} {48} (\bibinfo {year} {2015})}\BibitemShut {NoStop}%
\bibitem [{\citenamefont {Ferdinand}\ \emph {et~al.}(2017)\citenamefont
  {Ferdinand}, \citenamefont {DiMario},\ and\ \citenamefont
  {Becerra}}]{ferdinand2017multi}%
  \BibitemOpen
  \bibfield  {author} {\bibinfo {author} {\bibfnamefont {A.}~\bibnamefont
  {Ferdinand}}, \bibinfo {author} {\bibfnamefont {M.}~\bibnamefont {DiMario}},\
  and\ \bibinfo {author} {\bibfnamefont {F.}~\bibnamefont {Becerra}},\
  }\bibfield  {title} {\bibinfo {title} {Multi-state discrimination below the
  quantum noise limit at the single-photon level},\ }\href@noop {} {\bibfield
  {journal} {\bibinfo  {journal} {npj Quantum Inf.}\ }\textbf {\bibinfo
  {volume} {3}},\ \bibinfo {pages} {1} (\bibinfo {year} {2017})}\BibitemShut
  {NoStop}%
\bibitem [{\citenamefont {Burenkov}\ \emph {et~al.}(2018)\citenamefont
  {Burenkov}, \citenamefont {Tikhonova},\ and\ \citenamefont
  {Polyakov}}]{burenkov2018quantum}%
  \BibitemOpen
  \bibfield  {author} {\bibinfo {author} {\bibfnamefont {I.}~\bibnamefont
  {Burenkov}}, \bibinfo {author} {\bibfnamefont {O.}~\bibnamefont
  {Tikhonova}},\ and\ \bibinfo {author} {\bibfnamefont {S.}~\bibnamefont
  {Polyakov}},\ }\bibfield  {title} {\bibinfo {title} {Quantum receiver for
  large alphabet communication},\ }\href@noop {} {\bibfield  {journal}
  {\bibinfo  {journal} {Optica}\ }\textbf {\bibinfo {volume} {5}},\ \bibinfo
  {pages} {227} (\bibinfo {year} {2018})}\BibitemShut {NoStop}%
\bibitem [{\citenamefont {Izumi}\ \emph {et~al.}(2020)\citenamefont {Izumi},
  \citenamefont {Neergaard-Nielsen}, \citenamefont {Miki}, \citenamefont
  {Terai},\ and\ \citenamefont {Andersen}}]{izumi2020experimental}%
  \BibitemOpen
  \bibfield  {author} {\bibinfo {author} {\bibfnamefont {S.}~\bibnamefont
  {Izumi}}, \bibinfo {author} {\bibfnamefont {J.~S.}\ \bibnamefont
  {Neergaard-Nielsen}}, \bibinfo {author} {\bibfnamefont {S.}~\bibnamefont
  {Miki}}, \bibinfo {author} {\bibfnamefont {H.}~\bibnamefont {Terai}},\ and\
  \bibinfo {author} {\bibfnamefont {U.~L.}\ \bibnamefont {Andersen}},\
  }\bibfield  {title} {\bibinfo {title} {Experimental demonstration of a
  quantum receiver beating the standard quantum limit at telecom wavelength},\
  }\href@noop {} {\bibfield  {journal} {\bibinfo  {journal} {Phys. Rev. Appl.}\
  }\textbf {\bibinfo {volume} {13}},\ \bibinfo {pages} {054015} (\bibinfo
  {year} {2020})}\BibitemShut {NoStop}%
\bibitem [{\citenamefont {Guha}\ and\ \citenamefont {Erkmen}(2009)}]{Guha2009}%
  \BibitemOpen
  \bibfield  {author} {\bibinfo {author} {\bibfnamefont {S.}~\bibnamefont
  {Guha}}\ and\ \bibinfo {author} {\bibfnamefont {B.~I.}\ \bibnamefont
  {Erkmen}},\ }\bibfield  {title} {\bibinfo {title} {Gaussian-state
  quantum-illumination receivers for target detection},\ }\href
  {https://doi.org/10.1103/PhysRevA.80.052310} {\bibfield  {journal} {\bibinfo
  {journal} {Phys. Rev. A}\ }\textbf {\bibinfo {volume} {80}},\ \bibinfo
  {pages} {052310} (\bibinfo {year} {2009})}\BibitemShut {NoStop}%
\bibitem [{\citenamefont {Wilde}\ \emph {et~al.}(2012)\citenamefont {Wilde},
  \citenamefont {Guha}, \citenamefont {Tan},\ and\ \citenamefont
  {Lloyd}}]{wilde2012explicit}%
  \BibitemOpen
  \bibfield  {author} {\bibinfo {author} {\bibfnamefont {M.~M.}\ \bibnamefont
  {Wilde}}, \bibinfo {author} {\bibfnamefont {S.}~\bibnamefont {Guha}},
  \bibinfo {author} {\bibfnamefont {S.-H.}\ \bibnamefont {Tan}},\ and\ \bibinfo
  {author} {\bibfnamefont {S.}~\bibnamefont {Lloyd}},\ }\bibfield  {title}
  {\bibinfo {title} {Explicit capacity-achieving receivers for optical
  communication and quantum reading},\ }in\ \href@noop {} {\emph {\bibinfo
  {booktitle} {2012 IEEE International Symposium on Information Theory
  Proceedings}}}\ (\bibinfo {organization} {IEEE},\ \bibinfo {year} {2012})\
  pp.\ \bibinfo {pages} {551--555}\BibitemShut {NoStop}%
\bibitem [{\citenamefont {Guha}\ \emph {et~al.}(2020)\citenamefont {Guha},
  \citenamefont {Zhuang},\ and\ \citenamefont {Bash}}]{guha2020infinite}%
  \BibitemOpen
  \bibfield  {author} {\bibinfo {author} {\bibfnamefont {S.}~\bibnamefont
  {Guha}}, \bibinfo {author} {\bibfnamefont {Q.}~\bibnamefont {Zhuang}},\ and\
  \bibinfo {author} {\bibfnamefont {B.~A.}\ \bibnamefont {Bash}},\ }\bibfield
  {title} {\bibinfo {title} {Infinite-fold enhancement in communications
  capacity using pre-shared entanglement},\ }in\ \href@noop {} {\emph {\bibinfo
  {booktitle} {2020 IEEE International Symposium on Information Theory
  (ISIT)}}}\ (\bibinfo {organization} {IEEE},\ \bibinfo {year} {2020})\ pp.\
  \bibinfo {pages} {1835--1839}\BibitemShut {NoStop}%
\bibitem [{\citenamefont {Shi}\ \emph {et~al.}(2022)\citenamefont {Shi},
  \citenamefont {Zhang},\ and\ \citenamefont {Zhuang}}]{shi2022fulfilling}%
  \BibitemOpen
  \bibfield  {author} {\bibinfo {author} {\bibfnamefont {H.}~\bibnamefont
  {Shi}}, \bibinfo {author} {\bibfnamefont {B.}~\bibnamefont {Zhang}},\ and\
  \bibinfo {author} {\bibfnamefont {Q.}~\bibnamefont {Zhuang}},\ }\bibfield
  {title} {\bibinfo {title} {Fulfilling entanglement's benefit via converting
  correlation to coherence},\ }\href@noop {} {\bibfield  {journal} {\bibinfo
  {journal} {arXiv:2207.06609}\ } (\bibinfo {year} {2022})}\BibitemShut
  {NoStop}%
\bibitem [{\citenamefont
  {Zhuang}(2021{\natexlab{b}})}]{zhuang2021quantum-enabled}%
  \BibitemOpen
  \bibfield  {author} {\bibinfo {author} {\bibfnamefont {Q.}~\bibnamefont
  {Zhuang}},\ }\bibfield  {title} {\bibinfo {title} {Quantum-enabled
  communication without a phase reference},\ }\href@noop {} {\bibfield
  {journal} {\bibinfo  {journal} {Phys. Rev. Lett.}\ }\textbf {\bibinfo
  {volume} {126}},\ \bibinfo {pages} {060502} (\bibinfo {year}
  {2021}{\natexlab{b}})}\BibitemShut {NoStop}%
\bibitem [{\citenamefont {Chen}\ and\ \citenamefont {Zhuang}(2022)}]{chen2022}%
  \BibitemOpen
  \bibfield  {author} {\bibinfo {author} {\bibfnamefont {X.}~\bibnamefont
  {Chen}}\ and\ \bibinfo {author} {\bibfnamefont {Q.}~\bibnamefont {Zhuang}},\
  }\href@noop {} {\bibinfo {title} {{Entanglement-assisted detection of fading
  targets via correlation-to-coherence conversion}}} (\bibinfo {year}
  {2022})\BibitemShut {NoStop}%
\bibitem [{\citenamefont {Gao}\ and\ \citenamefont
  {Lee}(2014)}]{gao2014bounds}%
  \BibitemOpen
  \bibfield  {author} {\bibinfo {author} {\bibfnamefont {Y.}~\bibnamefont
  {Gao}}\ and\ \bibinfo {author} {\bibfnamefont {H.}~\bibnamefont {Lee}},\
  }\bibfield  {title} {\bibinfo {title} {Bounds on quantum multiple-parameter
  estimation with gaussian state},\ }\href@noop {} {\bibfield  {journal}
  {\bibinfo  {journal} {Eur. Phys. J. D}\ }\textbf {\bibinfo {volume} {68}},\
  \bibinfo {pages} {1} (\bibinfo {year} {2014})}\BibitemShut {NoStop}%
\bibitem [{\citenamefont {Calsamiglia}\ \emph {et~al.}(2010)\citenamefont
  {Calsamiglia}, \citenamefont {de~Vicente}, \citenamefont {Mu\~noz Tapia},\
  and\ \citenamefont {Bagan}}]{LOCC_NO_GO}%
  \BibitemOpen
  \bibfield  {author} {\bibinfo {author} {\bibfnamefont {J.}~\bibnamefont
  {Calsamiglia}}, \bibinfo {author} {\bibfnamefont {J.~I.}\ \bibnamefont
  {de~Vicente}}, \bibinfo {author} {\bibfnamefont {R.}~\bibnamefont {Mu\~noz
  Tapia}},\ and\ \bibinfo {author} {\bibfnamefont {E.}~\bibnamefont {Bagan}},\
  }\bibfield  {title} {\bibinfo {title} {Local discrimination of mixed
  states},\ }\href {https://doi.org/10.1103/PhysRevLett.105.080504} {\bibfield
  {journal} {\bibinfo  {journal} {Phys. Rev. Lett.}\ }\textbf {\bibinfo
  {volume} {105}},\ \bibinfo {pages} {080504} (\bibinfo {year}
  {2010})}\BibitemShut {NoStop}%
\bibitem [{\citenamefont {Cheng}\ \emph {et~al.}(2021)\citenamefont {Cheng},
  \citenamefont {Winter},\ and\ \citenamefont {Yu}}]{cheng2021discrimination}%
  \BibitemOpen
  \bibfield  {author} {\bibinfo {author} {\bibfnamefont {H.-C.}\ \bibnamefont
  {Cheng}}, \bibinfo {author} {\bibfnamefont {A.}~\bibnamefont {Winter}},\ and\
  \bibinfo {author} {\bibfnamefont {N.}~\bibnamefont {Yu}},\ }\bibfield
  {title} {\bibinfo {title} {Discrimination of quantum states under locality
  constraints in the many-copy setting},\ }in\ \href@noop {} {\emph {\bibinfo
  {booktitle} {2021 IEEE International Symposium on Information Theory
  (ISIT)}}}\ (\bibinfo {organization} {IEEE},\ \bibinfo {year} {2021})\ pp.\
  \bibinfo {pages} {1188--1193}\BibitemShut {NoStop}%
\bibitem [{\citenamefont {Bradshaw}\ \emph {et~al.}(2017)\citenamefont
  {Bradshaw}, \citenamefont {Assad}, \citenamefont {Haw}, \citenamefont {Tan},
  \citenamefont {Lam},\ and\ \citenamefont {Gu}}]{bradshaw2017overarching}%
  \BibitemOpen
  \bibfield  {author} {\bibinfo {author} {\bibfnamefont {M.}~\bibnamefont
  {Bradshaw}}, \bibinfo {author} {\bibfnamefont {S.~M.}\ \bibnamefont {Assad}},
  \bibinfo {author} {\bibfnamefont {J.~Y.}\ \bibnamefont {Haw}}, \bibinfo
  {author} {\bibfnamefont {S.-H.}\ \bibnamefont {Tan}}, \bibinfo {author}
  {\bibfnamefont {P.~K.}\ \bibnamefont {Lam}},\ and\ \bibinfo {author}
  {\bibfnamefont {M.}~\bibnamefont {Gu}},\ }\bibfield  {title} {\bibinfo
  {title} {Overarching framework between gaussian quantum discord and gaussian
  quantum illumination},\ }\href@noop {} {\bibfield  {journal} {\bibinfo
  {journal} {Phys. Rev. A}\ }\textbf {\bibinfo {volume} {95}},\ \bibinfo
  {pages} {022333} (\bibinfo {year} {2017})}\BibitemShut {NoStop}%
\bibitem [{\citenamefont {Giovannetti}\ \emph {et~al.}(2004)\citenamefont
  {Giovannetti}, \citenamefont {Lloyd},\ and\ \citenamefont
  {Maccone}}]{giovannetti2004}%
  \BibitemOpen
  \bibfield  {author} {\bibinfo {author} {\bibfnamefont {V.}~\bibnamefont
  {Giovannetti}}, \bibinfo {author} {\bibfnamefont {S.}~\bibnamefont {Lloyd}},\
  and\ \bibinfo {author} {\bibfnamefont {L.}~\bibnamefont {Maccone}},\
  }\bibfield  {title} {\bibinfo {title} {Quantum-enhanced measurements: beating
  the standard quantum limit},\ }\href@noop {} {\bibfield  {journal} {\bibinfo
  {journal} {Science}\ }\textbf {\bibinfo {volume} {306}},\ \bibinfo {pages}
  {1330} (\bibinfo {year} {2004})}\BibitemShut {NoStop}%
\bibitem [{\citenamefont {Kennedy}(1972)}]{Kennedy_1972}%
  \BibitemOpen
  \bibfield  {author} {\bibinfo {author} {\bibfnamefont {R.~S.}\ \bibnamefont
  {Kennedy}},\ }\href@noop {} {}\bibinfo {type} {Technical {Report}}\ (\bibinfo
   {institution} {Research Laboratory of Electronics (RLE) at the Massachusetts
  Institute of Technology (MIT)},\ \bibinfo {year} {1972})\BibitemShut
  {NoStop}%
\bibitem [{\citenamefont {Nielsen}\ and\ \citenamefont
  {Chuang}(2002)}]{nielsen2002quantum}%
  \BibitemOpen
  \bibfield  {author} {\bibinfo {author} {\bibfnamefont {M.~A.}\ \bibnamefont
  {Nielsen}}\ and\ \bibinfo {author} {\bibfnamefont {I.}~\bibnamefont
  {Chuang}},\ }\href@noop {} {\bibinfo {title} {Quantum computation and quantum
  information}} (\bibinfo {year} {2002})\BibitemShut {NoStop}%
\bibitem [{\citenamefont {Stewart}(1990)}]{stewart1990matrix}%
  \BibitemOpen
  \bibfield  {author} {\bibinfo {author} {\bibfnamefont {G.~W.}\ \bibnamefont
  {Stewart}},\ }\href@noop {} {\emph {\bibinfo {title} {Matrix perturbation
  theory}}}\ (\bibinfo  {publisher} {Citeseer},\ \bibinfo {year}
  {1990})\BibitemShut {NoStop}%
\bibitem [{\citenamefont {Lachs}(1965)}]{Lachs_1965}%
  \BibitemOpen
  \bibfield  {author} {\bibinfo {author} {\bibfnamefont {G.}~\bibnamefont
  {Lachs}},\ }\bibfield  {title} {\bibinfo {title} {Theoretical aspects of
  mixtures of thermal and coherent radiation},\ }\href
  {https://doi.org/10.1103/PhysRev.138.B1012} {\bibfield  {journal} {\bibinfo
  {journal} {Phys. Rev.}\ }\textbf {\bibinfo {volume} {138}},\ \bibinfo {pages}
  {B1012} (\bibinfo {year} {1965})}\BibitemShut {NoStop}%
\end{thebibliography}

%

\end{document}